\documentclass[a4paper,11pt]{article}
\pdfoutput=1 

\usepackage[usenames,dvipsnames,svgnames,table]{xcolor} 
\usepackage{jcapmod}
\usepackage{verbatim}
\usepackage[T1]{fontenc} 
\usepackage[latin9]{inputenc}
\definecolor{lightgray}{gray}{0.9}
\definecolor{Amber}{rgb}{1.0, 0.75, 0.0}
\definecolor{blizzardblue}{rgb}{0.67, 0.9, 0.93}
\usepackage{float}
\usepackage{graphicx}
\usepackage{hyperref}
\usepackage{amsmath}
\usepackage{xcolor}
\usepackage{tcolorbox}
\usepackage{amssymb}
\usepackage{physics}
\usepackage{microtype}
\usepackage[shortlabels]{enumitem}
\usepackage{cancel}
\usepackage{amsbsy}
\usepackage{color}
\usepackage[capitalise]{cleveref}
\usepackage{breakurl}
\usepackage{mathtools}
\usepackage{comment}
\usepackage{csquotes}
\usepackage{lmodern}
\usepackage{bm}
\usepackage{dsfont}
\usepackage{bbold}
\usepackage{empheq}
\usepackage{soul}
\usepackage{caption}
\usepackage{subcaption}
\usepackage[normalem]{ulem}

\pdfpageheight\paperheight
\pdfpagewidth\paperwidth

\usepackage[textwidth=0.71\paperwidth, textheight=0.8\paperheight]{geometry}

\renewcommand*{\vec}[1]{\bm{#1}}
\newcommand*{\unitvec}[1]{\vec{\hat{#1}}}
\newcommand*{\mat}[1]{\bm{\mathsf{#1}}}
\DeclareMathOperator{\diag}{diag}

\let\field\phi
\let\phi\varphi

\newcommand{\identity}{\ensuremath{\mathds{1}}}
\newcommand*{\transpose}[1]{\ensuremath{{#1}^{T}}}

\newcommand{\integers}{\mathbb{Z}}

\newcommand{\LSS}{L_{\mathrm{LSS}}}

\newcommand{\tensorsym}{\mathrm{t}}
\newcommand{\Tpower}{\mathcal{P}^{\tensorsym}} 
\newcommand*{\E}[1]{\texorpdfstring{\ensuremath{E_{#1}}}{E#1}}

\newcommand*{\A}[1]{\texorpdfstring{\ensuremath{A_{#1}}}{A#1}}
\newcommand{\slabh}{\texorpdfstring{\ensuremath{\E{16}^{(\mathrm{h})}}}{E16h}}
\newcommand{\slabi}
{\texorpdfstring{\ensuremath{\E{16}^{(\mathrm{i})}}}{E16i}}
\newcommand{\Kdelta}{\delta^{(\textrm{K})}}
\newcommand{\Ddelta}{\delta^{(\textrm{D})}}
\newcommand{\setN}{\mathcal{N}}
\newcommand*{\Mod}[1]{\,(\mathrm{mod}\ #1)}

\newcommand{\DeltaYstar}{\Delta^{Y*}}
\newcommand{\vecnp}{\smash{\vec{n}'}}

\usepackage[table]{xcolor}

\allowdisplaybreaks

\makeatletter
\DeclareRobustCommand{\rcite}[1]{%
  \rcite@aux#1,\@nil{#1}%
}
\def\rcite@aux#1,#2\@nil#3{%
  \if\relax#2\relax
    Ref.~\cite{#3}%
  \else
    Refs.~\cite{#3}%
  \fi
}
\makeatother

\subheader{IFT-UAM/CSIC-25-25}

\title{Cosmic topology. Part IIIb. Eigenmodes and correlation matrices of spin-2 perturbations in orientable Euclidean manifolds}

\author[a]{Amirhossein Samandar,}
\author[b]{Javier Carr\'on Duque,}
\author[a]{Craig J. Copi,}
\author[b]{Mikel Martin Barandiaran,}
\author[a]{Deyan P. Mihaylov,}
\author[a]{Glenn D. Starkman,}
\author[b,a,c]{Yashar Akrami,}
\author[d,e,f]{Stefano Anselmi,}
\author[a,g]{Fernando Cornet-Gomez,}
\author[h]{Johannes R. Eskilt,}
\author[c]{Andrew H. Jaffe,}
\author[i]{Arthur Kosowsky,}
\author[a]{Anna Negro,}           \author[b,j]{Joline Noltmann,}
\author[k,l]{Thiago S. Pereira,}
\author[b,a]{Andrius Tamosiunas}

\collaboration{(COMPACT Collaboration)}
\collaborationImg{\includegraphics[width = 0.1\textwidth]{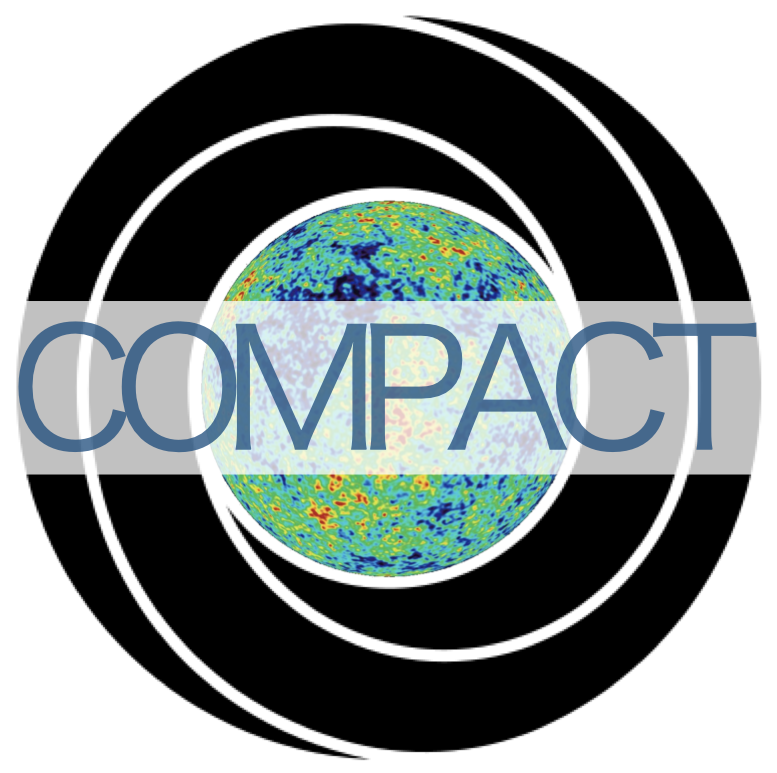}}

\affiliation[a]{CERCA/ISO, Department of Physics, Case Western Reserve University, 10900 Euclid Avenue, Cleveland, OH 44106, USA}
\affiliation[b]{Instituto de F\'isica Te\'orica (IFT) UAM-CSIC, C/ Nicol\'as Cabrera 13-15, Campus de Cantoblanco UAM, 28049 Madrid, Spain}
\affiliation[c]{Astrophysics Group \& Imperial Centre for Inference and Cosmology, Department of Physics, Imperial College London, Blackett Laboratory, Prince Consort Road, London SW7 2AZ, United Kingdom}
\affiliation[d]{INFN, Sezione di Padova, via Marzolo 8, I-35131 Padova, Italy}
\affiliation[e]{Dipartimento di Fisica e Astronomia ``G. Galilei'', Universit\`a degli Studi di Padova, via Marzolo 8, I-35131 Padova, Italy}
\affiliation[f]{Laboratoire Univers et Th\'eories, Observatoire de Paris, Universit\'e PSL, Universit\'e Paris Cit\'e, CNRS, F-92190 Meudon, France}
\affiliation[g]{Departamento de F\'isica, Universidad de C\'ordoba, Campus Universitario de Rabanales, Ctra. N-IV Km. 396, E-14071 C\'ordoba, Spain}
\affiliation[h]{Institute of Theoretical Astrophysics, University of Oslo, P.O. Box 1029 Blindern, N-0315 Oslo, Norway}
\affiliation[i]{Department of Physics and Astronomy, University of Pittsburgh, Pittsburgh, PA 15260, USA}
\affiliation[j]{Institute for Theoretical Particle Physics and Cosmology, RWTH Aachen University, Templergraben 55, 52062 Aachen, Germany}
\affiliation[k]{Departamento de F\'{i}sica, Universidade Estadual de Londrina, Rod. Celso Garcia Cid, Km 380, 86057-970, Londrina, Paran\'{a}, Brazil}
\affiliation[l]{Instituto de F\'{i}sica, Universidade Federal do Rio de Janeiro, 21941-972, Rio de Janeiro, RJ, Brazil}

\emailAdd{amirhossein.samandar@case.edu}
\emailAdd{javier.carron@csic.es}
\emailAdd{craig.copi@case.edu}
\emailAdd{mikel.martin@uam.es}
\emailAdd{deyan.mihaylov@case.edu}
\emailAdd{glenn.starkman@case.edu}
\emailAdd{yashar.akrami@csic.es}
\emailAdd{stefano.anselmi@pd.infn.it}
\emailAdd{fernando.cornetgomez@case.edu}
\emailAdd{j.r.eskilt@astro.uio.no}
\emailAdd{a.jaffe@imperial.ac.uk}
\emailAdd{kosowsky@pitt.edu}
\emailAdd{anna.negro@case.edu}
\emailAdd{joline.noltmann@rwth-aachen.de}
\emailAdd{tspereira@uel.br}
\emailAdd{andrius.tamosiunas@case.edu}
\date{\today}

\abstract{
We study the eigenmodes of the spin-2 Laplacian in orientable Euclidean manifolds and their implications for the tensor-induced part of the cosmic microwave background (CMB) temperature and polarization anisotropies. 
We provide analytic expressions for the correlation matrices of Fourier-mode amplitudes and of spherical harmonic coefficients. We demonstrate that non-trivial spatial topology alters the statistical properties of CMB tensor anisotropies, inducing correlations between harmonic coefficients of differing $\ell$ and $m$ and across every possible pair of temperature and $E$- and $B$-modes of polarization.
This includes normally forbidden $TB$ and $EB$ correlations.
We compute the Kullback-Leibler (KL) divergence between the pure tensor-induced CMB fluctuations in the usual infinite covering space and those in each of the non-trivial manifolds under consideration, varying both the size of the manifolds and the location of the observer. 
We find that the amount of information about the topology of the Universe contained in tensor-induced anisotropies does not saturate as fast as its scalar counterpart; indeed, the KL divergence continues to grow with the inclusion of higher multipoles up to the largest $\ell$ we have computed.
Our results suggest that CMB polarization measurements from upcoming experiments can provide new avenues for detecting signatures of cosmic topology, motivating a full analysis where scalar and tensor perturbations are combined and noise is included.
}


\begin{document}
\maketitle
\flushbottom

\section{Introduction}
\label{secn:intro}

The topology of the Universe has been a matter of scientific interest since before \cite{schwarzschild1998permissible} the discovery of general relativity \cite{Einstein:1916vd}, and there is a long history of efforts to determine that topology observationally \cite{Cornish:1996kv, Cornish:1997ab, Cornish:1997hz, Cornish:1997rp, deOliveira-Costa:2003utu, Cornish:2003db, Luminet2006brz, ShapiroKey:2006hm, Mota:2010jb, Bielewicz:2010bh, Cornish:2011ys, Bielewicz:2011jz, Vaudrevange:2012da, Aurich:2013fwa, Starkman_Priv_Comm, Planck:2013okc, Planck:2015gmu, Luminet2016:uni}.
These efforts have mostly fallen into two distinct classes. 
The first class comprises searches for objects that are visible in two or more directions on the sky corresponding to multiple null geodesics connecting us to the past worldline of those objects. 
This has been called ``cosmic crystallography'' \cite{Lehoucq:1996qe, Fujii2011}. 
The second class searches for evidence of topology in the fluctuations of the cosmic microwave background (CMB) temperature and polarization \cite{Cornish:1997ab, Cornish:2003db, Mota:2010jb, Vaudrevange:2012da, Planck:2013okc, Aslanyan:2013lsa,  Planck:2015gmu,  COMPACT:2022gbl, COMPACT:2022nsu, Mihaylov_2024, COMPACT:2023rkp, COMPACT:2024dqe, COMPACT:2024cud, COMPACT:2024qni}.

By considering the temperature data from the Wilkinson Microwave
Anisotropy Probe (WMAP) and {\it Planck} CMB satellite missions, cosmologists have been able to place reasonably strong limits on the cosmic topology. 
In a topologically non-trivial universe, if any closed spatial loop through the observer is shorter than the diameter of the last-scattering surface (LSS) of the CMB, the LSS self-intersects \cite{Cornish:1996kv,Cornish:1997ab,  Cornish:1997hz,Cornish:1997rp}, resulting in a thin spherical annulus centered on the observer being visible in two distinct directions on the sky. Checking for this ``circles-in-the-sky'' signature in WMAP and {\it Planck} data, it has been possible to show that the shortest closed loop through us is longer than $98.5\%$ of the diameter of the LSS.
A complementary Bayesian method comparing the pixel-pixel correlations in the observed CMB temperature map to those expected to be induced in manifolds with non-trivial topology  and to those expected in the covering space, found no evidence in {\it Planck} temperature data for non-trivial topology \cite{Souradeep1998arx, Planck:2013okc, Planck:2015gmu} for the specific subset of possible topologies that were considered.
These methods were extended to include $E$-mode polarization in 2015 {\it Planck} team search for topology, with similar results \cite{Planck:2015gmu}. 

In vanilla $\Lambda$ Cold Dark Matter ($\Lambda$CDM) cosmology, i.e., assuming that we live in the covering space of one of the three Friedmann-Lema\^itre-Robertson-Walker (FLRW) geometries, CMB polarization is sourced by both scalar and tensor fluctuations.  
A non-trivial topology of the Universe does not change the differential structure of the fluctuation mode equations, so most, though not all, of the conventional wisdom remains valid.
Scalar curvature modes source temperature and $E$-mode polarization fluctuations.
The adiabatic vector modes in the covering spaces of the FLRW geometries have no growing modes so are generally ignored. Tensor modes source temperature and both $E$- and $B$-mode polarization fluctuations. 

So far, all analyses of topology-induced effects have focused on primordial scalar (spin-0) fluctuations.
However, there is considerable activity searching for direct evidence of spin-2 (tensor-mode) fluctuations from the early Universe in $B$-mode CMB polarization---mostly as a smoking gun for inflation.
Both recent/current (BICEP \cite{BICEP2:2018kqh} and Simons Observatory \cite{SimonsObservatory:2018koc}) and future (TauRUS \cite{Taurus:2024dyi}, CMB-S4 \cite{Abazajian:2019eic}, and LiteBIRD \cite{LiteBIRD:2022cnt}) CMB observational programs have this as a primary science goal.
Just as topology, by breaking statistical isotropy, changes the properties of scalar eigenmodes, and thus the $TT$, $TE$, and $EE$ CMB correlations that scalar fluctuations induce, so too topology alters tensor eigenmodes, affecting all of $TT$, $TE$, $EE$, $TB$, $EB$, and $BB$ correlations.

The COMPACT collaboration~\cite{COMPACT:2022gbl} is working to create a publicly available code and comprehensive compendium of the scalar and tensor eigenmodes of topologically non-trivial manifolds with the most general set of parameters describing them and the position and orientation of an observer in them.
In addition, we aim to provide correlation matrices of both Fourier modes (of the underlying scalar and tensor fluctuations) and spherical harmonic coefficients ($a_{\ell m}$) of $T$, $E$, and $B$.
These will be presented under the conventional assumption that the amplitudes of the eigenmodes are statistically independent Gaussian random variables of zero mean.

 Non-trivial topology has important effects on the set of eigenmodes of the Laplacian operator, which are ultimately at the root of the potential detectability of non-trivial topology even when the shortest closed loop through us is longer than the diameter of the LSS\@. The eigenmodes of the Laplacian operator in topologically non-trivial manifolds (which form a basis in which the fields defined on the manifold are expanded) can be written as linear combinations of different covering-space eigenmodes with the same eigenvalue. Depending on the number of compact dimensions of the manifold under consideration, the eigenspectrum may be fully or partially discretized, meaning that the parametrization of the eigenmode basis will necessarily include one or more discrete parameters alongside any continuous ones. See \cref{app:topology_details} for a more detailed discussion.

In \rcite{COMPACT:2023rkp}, we presented the scalar eigenmodes of the 9 topologically non-trivial orientable Euclidean manifolds: \E{1} (the 3-torus), \E{2} (known as the half-turn space), \E{3} (the quarter-turn space), \E{4} (the third-turn space), \E{5} (the sixth-turn space), \E{6} (the Hantzsche-Wendt space), \E{11} (the chimney space), \E{12} (the chimney space with half-turn), and \E{16} (the slab space, including the previously neglected possibility of rotation), as well as the well-known results for the covering space, \E{18}.
We also presented the correlation matrices for CMB temperature, and used them to calculate the Kullback-Leibler (KL) divergences \cite{kullback1951, kullback1959information} for CMB temperature between the non-trivial manifolds and the covering space as a function of certain parameters of the topology.\footnote{
    The equations in the paper and the software we will  make available in the near future will allow these correlation matrices and the KL divergences to be calculated for arbitrary topological parameter values, within reasonable ranges, and for arbitrary values of the cosmological parameters, within reasonable ranges. 
} 
This KL divergence quantifies the expected power to distinguish between the two models for a typical observation (more technically, the relative entropy between the probability distributions of the CMB under both models).
In a future publication~\cite{COMPACT3c}, we will extend this to the auto-correlation matrices of CMB $E$-mode polarization, and to the cross-correlations between $T$ and $E$. 
As pointed out in a recent COMPACT paper \cite{COMPACT:2024cud}, the fact that topological boundary conditions ``spontaneously'' break isotropy means that the correlation matrices of $T$ and $E$ are not diagonal and are far from sparse.
In an upcoming paper \cite{Copi:2023tbd}, we will extend the presentation of scalar eigenmodes and the resulting $TT$, $TE$, and $EE$ correlation matrices to the 8 non-orientable Euclidean manifolds:
\E{7} (the Klein space), \E{8} (the Klein space with a horizontal flip), \E{9} (the Klein space with a vertical flip), \E{10} (the Klein space with a half-turn), \E{13} (the chimney space with a vertical flip), \E{14} (the chimney space with a horizontal flip), \E{15} (the chimney space with a half-turn and a flip) and \E{17} (the slab space with a flip). 

In this paper, we present the spin-2 eigenmodes for the 9 orientable Euclidean non-trivial manifolds,\footnote{
    In a future paper \cite{Martin:2025tbd}, we will extend the work of others \cite{Lehoucq:2002wy,Lachieze-Rey:2005nyz} on eigenmodes of $S^3$ manifolds to full generality.
    The task of extending this to $H^3$ manifolds is daunting because there are infinitely many of them and, unlike for $S^3$, they are not classified.  We will consider the anisotropic Thurston geometries and manifolds with boundaries as time and interest dictate and permit.
    }
as well as the consequent covering-space eigenmodes (i.e., polarized plane waves)  correlation matrices, and the resulting CMB $XY$ harmonic coefficient correlation matrices 
\begin{equation}
    C^{XY}_{\ell m\ell'm'} \equiv \left< a^X_{\ell m} a^{Y*}_{\ell'm'}\right> \,,
\end{equation}
where $X,Y \in \{T,E,B\}$, so that $XY \in \{TT, TE, TB, EE, EB, BB\}$. (We refer to these below as $TEB$ correlation matrices.)
In upcoming papers, we will extend this to the non-orientable Euclidean manifolds, and also treat the scalar and tensor eigenmodes (first for orientable Euclidean manifolds \cite{COMPACT3c}) jointly, so as to study the sensitivity to tensor modes (i.e., to the tensor-to-scalar ratio $r$). 

This paper extends the work in \rcite{COMPACT:2024cud}, where it was demonstrated how the violation of statistical isotropy by topological boundary conditions radically changed the nature of CMB correlations.   
Statistical isotropy of the Universe, not just microphysical parity conservation, is responsible in standard covering-space $\Lambda$CDM for the sparseness of the CMB correlation matrices, and in particular for the absence of correlations between \emph{B}-mode polarization and \emph{T} or \emph{E}. 
It was also pointed out that in addition to breaking statistical isotropy, the boundary conditions of most of the Euclidean topologies (i.e., except \E{1}, \E{11}, unrotated \E{16}, and \E{18}) break statistical homogeneity, by establishing special axes of rotation or planes of reflection in the non-orientable cases.  
Thus, except at special locations of accidental symmetry (e.g., on the axes of rotation or the planes of reflection), parity is not a good symmetry for the generic observer. 
Without statistical isotropy or parity, essentially all elements of the $T$, $E$, and $B$ auto- and cross-correlation matrices of the CMB are non-zero. 
The effects of tensor modes thus differ in an important way from those of scalar modes, which, even in a topologically non-trivial manifold, induce no polarization $B$ modes and thus lead to no correlations between  $B$ and either $T$ or $E$. 

 As a first step toward far more detailed exploration of the observable consequences of tensor modes in topologically non-trivial spaces, we use the tensor-only $TEB$ correlation matrices to calculate the KL divergence between topologically non-trivial manifolds and the covering space.
We find that there is potentially more topological information than previously thought in polarization.
In particular, we find that even when the shortest distance around the manifold through the observer is greater than the LSS diameter, KL divergences of $TEB$ correlations can continue to grow with the maximum spherical harmonic angular-momentum multipole $\ell_{\mathrm{max}}$ as far as we track it due to computational constraints. 
This is unlike the scalar $TT$ correlations reported on in \rcite{Eskilt:2022cff}, which saturated around $\ell\simeq30$.
It is also unlike what had been reported for scalar \emph{E}-mode correlations in the past \cite{Fabre:2013wia}, which was at least in part responsible for discouraging consideration of polarization in CMB topology analysis.  
We find that the (pure tensor) KL divergence can continue to be large for much larger manifolds than previous results had led us to expect.
We speculate that this is also an optimistic indication for the detectability of tensor modes in the presence of topology, but acknowledge that both noise and the much-larger scalar signal may confound this optimism.

The balance of the paper is structured as follows.
In \cref{secn:eigenmodes}, we present analytic expressions for the spin-2 eigenmodes of \E{18}, the universal covering space of Euclidean geometry (known to cosmologists as $E^3$ or $\mathbb{R}^3$), and of the 9 orientable topologically non-trivial Euclidean manifolds: \E{1}--\E{6}, \E{11}, \E{12}, \E{16}.
In \cref{sec:correlations} we also present analytic expressions for the correlation matrices of covering-space tensor eigenmodes (plane gravitational waves), and of the tensor $T$, $E$, and $B$ signals in the CMB.
In \cref{secn:numerical_results}, we present numerically calculated correlation matrices of CMB $T$, $E$, and $B$, and study the KL divergence between each \E{i} and the covering space for tensor-induced CMB fluctuations as a function of size parameters of the \E{i} manifolds.
Finally, in \cref{secn:conclusions} we draw our conclusions.

The GitHub repositories for COMPACT are  publicly available at \url{https://github.com/CompactCollaboration}. Codes associated with this study will be deposited there as publicly usable versions become available.

\section{General considerations for tensor perturbations}
\label{secn:eigenmodes}

It is often convenient to decompose metric and stress-energy perturbations around their background values into scalar (spin-0), vector (spin-1), and tensor (spin-2) representations of the proper rotation group $SO(3)$. 
Because the FLRW cosmological metric is homogeneous and isotropic, these scalar, vector, and tensor perturbations decouple and can be studied independently of one another.
The scalar perturbations play the most prominent role in the observable Universe since they are the primary source for temperature and polarization fluctuations of the CMB and for density fluctuations and other similar physical phenomena. 
However, the tensor  perturbations are of great interest as well.
In inflationary $\Lambda$CDM, they are the sole source of primordial $B$-type polarization of the CMB and are considered the key signature of inflation---gravitational waves produced during the inflationary epoch. Regardless of their origin, detecting tensor modes would offer an independent probe of the early Universe.

\subsection{Spin-2 Laplacian eigenmodes in the covering space \E{18}}
Plane waves $\mathrm{e}^{i\vec{k}\cdot(\vec{x}-\vec{x}_0)}$, i.e., Fourier modes, are well known to be a complete basis of scalar eigenmodes of the Laplacian in the Euclidean covering space, \E{18}.
In these modes $\vec{x}_0$ represents the position of an arbitrary origin relative to the observer's coordinates\footnote{
    The inclusion of $\vec{x}_0$ is inconsequential in the covering space, for \E{1}, \E{11}, or \slabh. However, it plays a crucial role in the analysis for \E{2}--\E{6}, \E{12}, and \slabi.
    }
and $\vec{k}=\transpose{(k_x,k_y,k_z)}$ is the wavevector, with its magnitude $k$ referred to as the wavenumber.
Tensor eigenmodes follow a similarly elegant structure: 
\begin{equation}
    \label{eqn:EuclideanFourierBasis}
    \Upsilon^{\E{18}}_{ij,\vec{k}}(\vec{x},\lambda) = e_{ij}(\unitvec{k},\lambda) \ \mathrm{e}^{i\vec{k}\cdot(\vec{x}-\vec{x}_0)},
\end{equation}
where $e_{ij}$ is a polarization tensor that is symmetric, traceless, and transverse (i.e., orthogonal to $\unitvec{k}$). 
There are two independent polarization states, which may be labeled by $\lambda=\pm2$ for the two possible helicity states. 
For details of our choice of conventions for polarization helicity tensors $e_{ij}(\unitvec{k},\lambda)$, see \cref{app:polzn}.
These tensor eigenmodes satisfy the eigenvalue equation
\begin{equation}
   \nabla^2 \Upsilon^{\E{18}}_{ij,\vec{k}} (\vec{x}, \lambda) = -\vert\vec{k}\vert^2 \Upsilon^{\E{18}}_{ij,\vec{k}} (\vec{x}, \lambda) ,
\end{equation}
where the eigenvalue associated with $\Upsilon^{\E{18}}_{ij,\vec{k}} (\vec{x}, \lambda) $ is $-\vert\vec{k}\vert^2\equiv -k^2$, which is independent of the helicity $\lambda$.

The adiabatic tensor perturbation field around the FLRW background at time $t$ can be fully characterized \cite{Tsagas:2007yx, Brechet:2009fa} by the symmetric, traceless, and transverse tensor $\mathcal{D}_{ij}(\vec{x}, t)$, which can be represented in terms of the helicity $\pm 2$ eigenmodes \cite{Weinberg:2008}.
For \E{18},
\begin{equation}
\mathcal{D}^{\E{18}}_{ij}(\vec{x},t)=\sum_{\lambda=\pm 2} \int \frac{\mathrm{d}^3 k}{(2\pi)^3}  \mathcal{D}(\vec{k},\lambda,t) \Upsilon^{\E{18}}_{ij, \vec{k}} (\vec{x}, \lambda).
\end{equation}
Here $\mathcal{D}(\vec{k}, \lambda, t)$ is the time-dependent amplitude for wavevector $\vec{k}$ and helicity $\lambda$ corresponding to the eigenmode $\Upsilon^{\E{18}}_{ij, \vec{k}} (\vec{x}, \lambda)$ in the covering space. 

Translation invariance of the covering space ensures the statistical independence of modes of different $\vec{k}$, except for $\vec{k}$ and $-\vec{k}$, which are conjugate due to the reality of the field.
The correlation of $\mathcal{D}(\vec{k},\lambda,t)$ can be calculated at any time $t$.
In particular, the primordial ($t=0$) correlation is
\begin{equation}
    \label{eqn:tensor2pcf}
    \langle \mathcal{D}(\vec{k},\lambda, 0) \: \mathcal{D}^*(\vec{k}',\lambda', 0)\rangle = \Kdelta_{\lambda \lambda'} \frac{\pi^2 \Tpower(k)}{2 k^3}  (2\pi)^3 \Ddelta(\vec{k}-\vec{k}')\, , 
\end{equation}
where $\Tpower (k)$ represents the primordial power spectrum of tensor modes. 

As we did for scalar fluctuations in \rcite{COMPACT:2023rkp}, we assume that the isometries of the geometry are preserved by the microphysical processes responsible for generating fluctuations.
The Dirac delta function, $\Ddelta(\vec{k} - \vec{k}')$, reflects the translational invariance of the $E^3$ isometry group, ensuring that different Fourier modes are uncorrelated.
The term $\Kdelta_{\lambda \lambda'}$ arises from the orientability of \E{18}, while the dependence of $\Tpower(k)$ solely on the magnitude of $\vec{k}$ reflects the rotational invariance of the Laplacian operator. Consequently, the primordial power spectrum depends only on the eigenvalues of this operator, $-k^2$. Parity invariance of the Laplacian ensures that the power spectrum is independent of $\lambda$.

We will be interested in any tensor fields $\field_{ij}^X$ that are linearly related to $\mathcal{D}(\vec{k},\lambda,t)$
by a transfer function $\Delta^{X}(k)$.
The transfer function only depends on the magnitude of $\vec{k}$ again because we assume that the microphysics that is responsible for the transfer function respects all the symmetries of $E^3$.
The expectation value of any pair of $\field_{ij}^X(\vec{k}, \lambda)$ is\footnote{The overall normalization of \eqref{eqn:CE18XY} follows the convention of the \texttt{CAMB} software package \cite{Lewis:1999bs}}
\begin{empheq}[box=\fbox]{align}
    \label{eqn:CE18XY}
    C^{\E{18};XY}_{iji'j', \vec{k} \vec{k}', \lambda \lambda'} 
        &\equiv \langle \field_{ij}^X(\vec{k}, \lambda) \field_{i'j'}^{Y *}(\vec{k}', \lambda')\rangle \nonumber\\
        &= (2\pi)^3\frac{\pi^2}{2 k^3} \Tpower(k) \Delta^{X}(k) \DeltaYstar(k)
        \Ddelta(\vec{k}-\vec{k}') \Kdelta_{\lambda \lambda'} \mathcal{E}_{iji'j'}(\unitvec{k}, \unitvec{k}', \lambda)\; ,
\end{empheq}
where
\begin{empheq}[box=\fbox]{align}
    \label{eqn:chi_tensor}
    \mathcal{E}_{iji'j'}(\unitvec{k}, \unitvec{k}', \lambda) \equiv e_{ij}(\unitvec{k},\lambda) e^*_{i'j'}(\unitvec{k}',\lambda)\,.
\end{empheq}
In an orientable space like \E{18}, correlations are diagonal in $\lambda$, and their values are independent of $\lambda$.
It is therefore more convenient to consider the helicity-summed correlations
\begin{empheq}[box=\fbox]{align}
C^{\E{18};XY}_{iji'j', \vec{k} \vec{k}'} 
        &\equiv \sum_{\lambda, \lambda' = \pm 2} \langle \field_{ij}^X(\vec{k}, \lambda) \field_{i'j'}^{Y*}(\vec{k}', \lambda')\rangle \nonumber\\
        &= (2\pi)^3\frac{\pi^2}{2 k^3} \Tpower(k) \Delta^{X}(k) \DeltaYstar(k')
        \Ddelta(\vec{k}-\vec{k}') \Pi_{ij,i'j'}(\unitvec{k})\; ,
\end{empheq}
where \cite{Weinberg:2008}
\begin{empheq}[box=\fbox]{align}
\Pi_{ij,i'j'}(\unitvec{k}) &\equiv \sum_{\lambda} e_{ij}(\unitvec{k}, \lambda) e^*_{i'j'}(\unitvec{k}, \lambda) \\
&= \Kdelta_{ii'}\Kdelta_{jj'} + \Kdelta_{ij'}\Kdelta_{ji'} - \Kdelta_{ij}\Kdelta_{i'j'} 
+ \Kdelta_{ij} \hat{k}_{i'} \hat{k}_{j'} + \Kdelta_{i'j'} \hat{k}_i \hat{k}_j \nonumber \\
&\quad - \Kdelta_{ii'} \hat{k}_j \hat{k}_{j'} - \Kdelta_{ij'} \hat{k}_j \hat{k}_{i'} 
- \Kdelta_{ji'} \hat{k}_i \hat{k}_{j'} - \Kdelta_{jj'} \hat{k}_i \hat{k}_{i'} 
+ \hat{k}_i \hat{k}_j \hat{k}_{i'} \hat{k}_{j'}\,. \nonumber
\end{empheq}

Contemporary CMB observations include both the CMB intensity (characterized by the equivalent blackbody temperature) and the CMB polarization as functions of position on the sky.
The (net) polarization of the CMB is due to Thomson scattering of the locally anisotropic photon distribution by free electrons. 
These electrons were present in sufficient number densities to lead to observable polarization both during recombination and during the later period of reionization triggered by ultraviolet light from the first generation of stars. 
There is interest in the polarization signals from both epochs.
The reionization signal can be used to characterize the Universe's reionization history, giving insight into early star formation and other energetic processes in the post-recombination Universe.
The recombination signal is useful for better characterizing primordial scalar perturbations, but essential for detecting and then characterizing any primordial tensor perturbations.

CMB fluctuations on the sky are characterized by  the intensity Stokes parameter ($I$), which is typically replaced by the temperature ($T$), and by the polarization Stokes parameters $Q$ and $U$. 
The third polarization Stokes parameter $V$, which characterizes circular polarization of the photon field, is not generated by Thomson scattering and is typically taken to vanish in the absence of any microphysical source of circular polarization in the early Universe.
In this work we will adopt that assumption.
The temperature and polarization Stokes parameters are conventionally expanded in appropriate eigenmodes on the sky as
\begin{align}
    \label{eqn:Tofnhat}
    \Delta T(\unitvec{n}) &= \sum_{\ell=0}^{\infty} \sum_{m=-\ell}^{\ell} a^T_{\ell m} Y_{\ell m} (\unitvec{n}),\\
    \label{eqn:QUofnhat}
    Q(\unitvec{n}) \pm i U(\unitvec{n}) &= - \sum_{\ell m} \left( a^{ E}_{\ell m} \pm i a^{ B}_{\ell m}\right) {}_{\pm 2}Y_{\ell m} (\unitvec{n}) \,.
\end{align}
Here $Y_{\ell m} (\unitvec{n})$ and ${}_{\pm 2}Y_{\ell m}(\unitvec{n})$ are the scalar (spin-0) and tensor (spin-2) spin-weighted spherical harmonics. More details about the ${}_{\pm 2}Y_{\ell m}(\unitvec{n})$ can be found in \cref{app:2Ylm}.

Locally the CMB polarization fluctuations are characterized by a scalar field on the sky, the \emph{E}-mode polarization, and a pseudo-scalar field on the sky, the \emph{B}-mode polarization.
Our notation for the coefficients of ${}_{\pm 2}Y_{\ell m} (\unitvec{n})$ in \eqref{eqn:QUofnhat} reflects that conventional decomposition.
$E(\unitvec{n})$  and $B(\unitvec{n})$ can be expressed in terms of ordinary (scalar) spherical harmonics as \cite{Bessada:2009np}
\begin{align}
\label{eqn:localEofnhat}
E(\unitvec{n}) &= \sum_{\ell=2}^{\infty} \sum_{m=-\ell}^{\ell} 
    \sqrt{\frac{(\ell + 2)!}{(\ell - 2)!}}
    a^E_{\ell m} Y_{\ell m} (\unitvec{n}),\\
\label{eqn:localBofnhat}
B(\unitvec{n}) &= \sum_{\ell=2}^{\infty}\sum_{m=-\ell }^\ell  
    \sqrt{\frac{(\ell + 2)!}{(\ell - 2)!}}
   a^B_{\ell m} Y_{\ell m} (\unitvec{n}).
\end{align}
We emphasize the importance of the $\ell$-dependent coefficients preceding $a^E_{\ell m}$ in \eqref{eqn:localEofnhat} and $a^B_{\ell m}$ in \eqref{eqn:localBofnhat}.  
With these coefficients \cite{Kosowsky:1994cy}, $E(\unitvec{n})$ and $B(\unitvec{n})$ are functions only of the measured polarization-tensor field and its derivatives at $\unitvec{n}$; without them, they would be non-local functions of the polarization tensor.

Under space inversion, \(\Delta T(-\unitvec{n}) \rightarrow \Delta T(\unitvec{n})\), \(E(-\unitvec{n}) \rightarrow E(\unitvec{n})\), and \(B(-\unitvec{n}) \rightarrow -B(\unitvec{n})\).
In other words, \emph{T} and \emph{E} are scalar fields (parity-even), while \emph{B} is a pseudo-scalar field (parity-odd).
Consequently, the multipole coefficients \(a^X_{\ell m}\), where \(X \in \{T, E, B\}\), transform under space inversion as \(a^{T}_{\ell m} \rightarrow (-1)^\ell a^{T}_{\ell m}\), \(a^{E}_{\ell m} \rightarrow (-1)^\ell a^{E}_{\ell m}\), and \(a^B_{\ell m} \rightarrow -(-1)^\ell a^B_{ \ell m}\).

The tensor-fluctuation  contributions to the CMB anisotropy multipole coefficients, \(a^{\tensorsym, X}_{ \ell m}\), for \(X \in \{T, E, B\}\) can be expressed as
\begin{equation}
\label{eqn:almE18}
a_{\ell m}^{\tensorsym, X} = \sum_{\lambda = \pm 2}  \int \frac{\mathrm{d}^3 k}{(2\pi)^3} \,  \mathcal{D}(\vec{k}, \lambda)\ \Delta^{\tensorsym, X}_{\ell} (k) \, \xi^{\tensorsym,\E{18};X, \unitvec{k}}_{k,\ell m,\lambda},
\end{equation}
where \( \mathcal{D}(\vec{k}, \lambda)\) is the amplitude of primordial tensor perturbations for each helicity. $\Delta^{\tensorsym, X}_{\ell}(k)$ is the tensor transfer function, which we again assume depends only on $\ell$ and the magnitude of $\vec{k}$, reflecting the rotational symmetry of the local geometry.
The \(\xi^{\tensorsym,\E{i};X,\unitvec{k}}_{{k},\ell m,\lambda}\) factors encapsulate the effects on the multipole coefficients of both the projection of the three-dimensional perturbation field onto the two-dimensional sphere of the sky and the correlations between Fourier modes in each of the topologies.
For \E{18} they are given by
\begin{empheq}[box=\fbox]{align}
    \label{eqn: E18coeffT}
  \xi^{\tensorsym,\E{18};T, \unitvec{k}}_{k,\ell m,\lambda} &\equiv i^{\ell} \sqrt{\frac{2\pi^2(\ell + 2)!}{(\ell - 2)!}} \ {}_{-\lambda}Y_{\ell m} (\unitvec{k})  \mathrm{e}^{-i \vec{k}\cdot \vec{x}_0},
\end{empheq}

\begin{empheq}[box=\fbox]{align}
    \label{eqn: E18coeffE}
  \xi^{\tensorsym,\E{18};E, \unitvec{k}}_{k,\ell m,\lambda} &\equiv -i^{\ell} \ \sqrt{2 \pi^2} \ {}_{-\lambda}Y_{\ell m} (\unitvec{k})\mathrm{e}^{-i \vec{k}\cdot \vec{x}_0},
\end{empheq}

\begin{empheq}[box=\fbox]{align}
    \label{eqn: E18coeffB}
  \xi^{\tensorsym,\E{18};B, \unitvec{k}}_{k,\ell m,\lambda} &= -\frac{\lambda}{2} \xi^{\tensorsym,\E{18};E, \unitvec{k}}_{k,\ell m,\lambda}\,.
\end{empheq}

Throughout this discussion, we have used the superscript ``$\tensorsym$'' to denote the tensor component of the CMB anisotropies. Since our focus in this paper is on tensor perturbations, henceforth, except on the power spectrum, we will omit the $\tensorsym$ superscript.

\subsection{From covering space to non-trivial topologies}

In our previous work \cite{COMPACT:2024cud}, we investigated how statistical isotropy and homogeneity affect spherical harmonic correlations of a random field on the sphere. Assuming a statistically rotationally invariant field, all correlations vanish except for the diagonal elements in $\ell$, commonly referred to as $C^{XY}_{\ell}$.
We also demonstrated that parity invariance alone does not eliminate correlations between observables of different parity (e.g., $EB$ and $TB$); rather, it constrains them to only have $\ell + \ell'$ odd correlations.
Only the combination of parity invariance and statistical isotropy forces all correlations between observables of opposite parity to vanish.
On the other hand, statistical anisotropy often, but not always, accompanies statistical inhomogeneity, which generically implies parity violation at specific locations.

As we discussed in \rcite{COMPACT:2023rkp}, briefly  above, and  in detail in \cref{app:topology_details}, non-trivial topological boundary conditions have several significant effects on the eigenmodes of the Laplacian \cite{courant2024methods,chavel1984eigenvalues,zee2016group}.  Specializing to the Euclidean case:
\begin{enumerate}
    \item The boundary conditions discretize the components of allowed wavevectors $\vec{k}$ for the Laplacian. For compact manifolds (\E{1}--\E{10}) all three components are discrete and the allowed $\vec{k}$ form a discrete lattice; in other manifolds, for each non-compact spatial dimension one component of $\vec{k}$ remains continuous. Notably, the discretization for the tensor Laplacian is different from the discretization for the scalar Laplacian for some topologies. The two helicities can even differ in their allowed wavevectors.
\item For compact topologies, the eigenspectrum becomes discrete and there are only a finite number of eigenmodes of each eigenvalue; consequently,
the correlator $\langle \field_{ij}^X(\vec{k}, \lambda) \field_{i'j'}^{Y*}(\vec{k}', \lambda') \rangle$ includes terms involving ${\Tpower}(k) \Kdelta_{\vec{k} \vec{k}'}$, where $\Kdelta$ represents a Kronecker delta rather than a Dirac delta. For the non-compact chimney and slab spaces, which include a mix of finite and infinite directions, the eigenspectrum is continuous and the correlator may exhibit a combination of Kronecker and Dirac delta terms.
    \item Except for \E{1}, \E{11}, and \slabh, the eigenmodes are not simply single plane-wave eigenmodes of the covering space. Instead, they are linear combinations of these modes, involving different $\vec{k}$ vectors with the same magnitude. This mixing introduces additional terms in the correlator, coupling $\vec{k}$ to its rotations under the generators of the topology. These couplings are expressed through a Kronecker or a Dirac delta function.
\end{enumerate}
These effects encode violations of statistical isotropy and violate the direct connection described earlier between the statistics of $\field_{ij}^X(\vec{k}, \lambda)$ and the multipole coefficients $a^X_{\ell m}$. For instance, instead of $C^{\E{18};XY}_{iji'j', \vec{k} \vec{k}', \lambda \lambda'} $being proportional to a Dirac delta function of $\vec{k}$ and $\vec{k}'$, the correlator vanishes for disallowed $\vec{k}$ values and, more generally, establishes correlations between all pairs of allowed $\vec{k}$. These correlations have equal magnitudes but location-dependent phases, reflecting the constraints imposed by the topology.

Applying these arguments to the auto- and cross-correlations of the multipole coefficients $a^{X}_{\ell m}$, where $X \in \{T, E, B\}$, we can classify the patterns of non-zero elements in the covariance matrices that arise due to the manifold with different topologies into three general categories:

\begin{enumerate}
\label{sec:symmetries_cat}
    \item If the topology of the manifold preserves the statistical isotropy, homogeneity, and parity invariance inherent to the underlying microphysics:
    \begin{align}
    \label{eqn:SICdiagonal}
        C^{\E{18}; XY}_{\ell m \ell' m'} \equiv 
    	\langle a^{\E{18}; X}_{\ell m} a^{\E{18}; Y*}_{\ell' m'} \rangle = 
        \begin{cases}  
        C^{XY}_\ell \Kdelta_{\ell \ell'}  \Kdelta_{m m'}, & \text{for } XY \in \{TT, EE, BB, TE\}, \\ 
        0, & \text{for } XY \in \{EB, TB\}.
        \end{cases}
    \end{align}
    We include the \E{18} index because this can only arise in the context of the covering space topology, while all the other, non-trivial topologies necessarily at least break statistical isotropy.
    \item If the topology of the manifold breaks statistical isotropy while preserving statistical homogeneity and statistical parity invariance inherent to the underlying microphysics:
    \begin{enumerate}
        \item For $XY \in \{TT, EE, BB, TE\}$:
        \begin{equation}
        C^{XY}_{\ell m \ell' m'} =
                \begin{cases}
                    \neq 0, & \ell + \ell' \text{ even}, \\ 
                    = 0, & \ell + \ell' \text{ odd}.
                \end{cases}
        \end{equation}
        \item For $XY \in \{EB, TB\}$:
        \begin{equation}
        C^{XY}_{\ell m \ell' m'} =
                \begin{cases}
                    \neq 0, & \ell + \ell' \text{ odd}, \\ 
                    = 0, & \ell + \ell' \text{ even}.
                \end{cases}
        \end{equation}
    \end{enumerate}
     \item If the topology of the manifold breaks statistical isotropy, statistical homogeneity, and statistical parity invariance inherent to the underlying microphysics, the covariance matrix $C^{XY}_{\ell m \ell' m'}$ can have non-zero elements across all $\ell + \ell'$ blocks, encompassing both even and odd cases, for all auto- and cross-correlations.
\end{enumerate}

In the following section, we present the tensor eigenmodes and eigenspectra of orientable Euclidean manifolds as functions of their topological parameters, for an observer located at an arbitrary position. Assuming that the amplitudes of these eigenmodes are Gaussian random variables with zero mean and a dimensionless power spectrum $\Tpower(k)$, we present the correlation for the Fourier-mode amplitudes, $C^{\E{i};XY}_{iji'j', \vec{k} \vec{k}', \lambda \lambda'}$, and the spherical-harmonic amplitudes, $C^{\E{i};XY}_{\ell m \ell' m'}$.

\section{Eigenmodes of tensor Laplacian and correlation matrices}\label{sec:correlations}
In each of the manifolds, the eigenmodes $\Upsilon^{\E{i}}_{ij,\vec{k}}(\vec{x},\lambda)$ of the tensor Laplacian, like the eigenmodes of the scalar Laplacian discussed in \rcite{COMPACT:2023rkp},  must be invariant under every possible group transformation $G_\alpha \in \Gamma^{\E{i}}$. 
Here $\Gamma^{\E{i}}$ is the isometry subgroup that yields such manifold: $\E{i}\cong \E{18}/\Gamma^{\E{i}}$. Therefore, with some abuse of notation, we must have:

\begin{equation}
    \label{eqn:generaleigenmodeinvariance}
  G_\alpha (\Upsilon^{\E{i}}_{ij,\vec{k}}(\vec{x},\lambda)) = \Upsilon^{\E{i}}_{ij,\vec{k}}(\vec{x},\lambda)\,.
\end{equation}

Formally, the solution is that $\Upsilon^{\E{i}}_{ij,\vec{k}}(\vec{x},\lambda)$ is a simple linear combination of all covering-space eigenmodes related by the group transformations:
\begin{equation}
   \Upsilon^{\E{i}}_{ij,\vec{k}}(\vec{x},\lambda) \propto \sum_{G_\alpha \in \Gamma^{\E{i}}} G_\alpha (\Upsilon^{\E{18}}_{ij,\vec{k}}(\vec{x},\lambda)) \,,
\end{equation}
with the additional caveat that only certain wavevectors $\vec{k}$ will be suitable. Again, this is analogous to the situation for scalar eigenmodes discussed in \rcite{COMPACT:2023rkp}, and again, 
more practically, we can limit the sum to a small, finite set of group elements $G_\alpha \in \mathcal{G}^{\E{i}}$. Taking into account the transformation property \eqref{eq:genericrotation} and normalizing accordingly, we have
\begin{equation}
\label{eqn:generaleigenmodeformula}
\Upsilon^{\E{i}}_{ij,\vec{k}}(\vec{x},\lambda) = \frac{\mathrm{e}^{i \Phi_{\vec
k}^{\E{i}}}}{\sqrt{N(\mathcal{G}^{\E{i}})}} \sum_{G_\alpha \in \mathcal{G}^{\E{i}}} (\mat{M}_{G_\alpha})_{ik}  e_{k\ell}(\unitvec{k},\lambda) (\transpose{\mat{M}}_{G_\alpha})_{\ell j} \mathrm{e}^{i\vec{k}\cdot G_\alpha\vec{x}}\,,
\end{equation}
where $N(\mathcal{G}^{\E{i}})$ is the number of elements in $\mathcal{G}^{\E{i}}$ and $\Phi_{\vec k}^{\E{i}}$ represents a constant phase that will be chosen to simplify the reality condition of the field, which varies for each topology.\footnote{Although this additional phase was not explicitly shown in \rcite{COMPACT:2023rkp}, it does not affect the resulting correlations.}
$\mathcal{G}^{\E{i}}$ includes one group element for each of the $O(3)$ matrices $\mat{M}^{(G_\alpha)}$ that appears when we explicitly write the action of the group elements,
\begin{equation}
    \label{eqn:actionofGalpha}
    G_\alpha:(\vec{x}-\vec{x}_0)\to \mat{M}^{(G_\alpha)} (\vec{x}-\vec{x}_0) + \vec{v}^{(G_\alpha)}\,.
\end{equation}
For the manifolds under consideration, we can always select a much smaller set of group elements
\begin{equation}
    \label{eqn:actionofgenerator}
    g_{a_j}^{\E{i}}: \vec{x} \to \mat{M}^{\E{i}}_a (\vec{x}-\vec{x}_0^{\E{i}}) + \vec{T}^{\E{i}}_{a_j} + \vec{x}^{\E{i}}_0 ,
\end{equation}
to act as generators of the group---3 for \E{1}--\E{6}, 2 for \E{11} and \E{12}, 1 for \slabh and \slabi.   
Here $a$ labels the distinct $O(3)$ matrices $\mat{M}^{\E{i}}_a$ in the choice of generators, while $j$ labels the distinct translations $\vec{T}^{\E{i}}_{a_j}$ for each $\mat{M}^{\E{i}}_a$. $\vec{x}_0^{\E{i}}$ is the origin of the coordinate system.
Each $G_\alpha$ can be written as a finite product of $g_{a_j}^{\E{i}}$ and their inverses.

Since we are considering only orientable manifolds in this paper, $\mat{M}^{(G_\alpha)}\in SO(3)$.
These $\mat{M}^{(G_\alpha)}$ are then just the matrices $\mat{M}^{\E{i}}_a$ that appear in the generators of every manifold, as described in \rcite{COMPACT:2023rkp}, plus all non-identical $SO(3)$ matrices that can be built from arbitrary products of those $\mat{M}^{\E{i}}_a$.
In the subsequent section, we will describe the $\mathcal{G}^{\E{i}}$ for each \E{i}.

Our previous paper on scalar eigenmodes of orientable Euclidean manifolds \cite{COMPACT:2023rkp} contains a  detailed discussion of the generators generally and for each manifold. 
We alert the reader to the discussion in \rcite{COMPACT:2023rkp} of the separation between ``topology parameters'' and ``observer parameters.''
Because the topological boundary conditions violate isotropy, and, for all but \E{1}, \E{11}, \slabh, and \E{18}, homogeneity, observables depend on both the position and orientation of the observer. 
When we present the generators of each topology below, we make specific simplifying choices for the orientation of the coordinate system that reduces the number of topology parameters---effectively by trading them for the Euler angles of the observer's orientation.
However, we do not do the same for the choice of origin.
In \rcite{COMPACT:2023rkp}, we described how the freedom to choose  the coordinate origin can be used to affect the representation of the generators, changing the components of the $\vec{T}^{\E{i}}_{a_j}$ that are in the plane of rotation of $\mat{M}^{\E{i}}_a$.
We pointed to two convenient choices---placing the origin on the axis of all rotations or placing the observer at the origin---but noted that the real observer (i.e., us) does not have the freedom to change their location, unlike the freedom to rotate their coordinate system.
Therefore, as in \rcite{COMPACT:2023rkp}, here we explicitly retain the dependence on the origin. 
We direct the reader to \rcite{COMPACT:2023rkp} for further discussion of this point, as well as to the discussion---in general and for each topology---of how to avoid over counting of topological parameters.

Equation \eqref{eqn:generaleigenmodeinvariance} must still be satisfied for every group element $G_\alpha \in \Gamma^{\E{i}}$. Among those group elements are a subgroup of pure translations, which are all the integer linear combinations of the $\vec{T}^{\E{i}}_j$. 
These are the translations of what we term the associated homogeneous space (\E{1}, \E{11}, or \slabh) of that manifold.
Considering the invariance of $\Upsilon^{\E{i}}_{ij,\vec{k}}(\vec{x},\lambda)$ under the translation by $\vec{T}^{\E{i}}_j$, and recognizing that 
$\mathcal{G}^{\E{i}}$ always includes the identity matrix, we learn that one must have

\begin{equation}
    \mat{e}(\unitvec{k},\lambda) \mathrm{e}^{i\vec{k}\cdot\left[(\vec{x}-\vec{x}_0)+\vec{T}^{E_i}_{j}\right]} = \mat{e}(\unitvec{k},\lambda) \mathrm{e}^{i\vec{k}\cdot(\vec{x}-\vec{x}_0)} 
\end{equation}

\noindent or more compactly
\begin{equation}
    \label{eqn:associatedhomogeneousdiscretization}
    \vec{k}\cdot\vec{T}^{\E{i}}_j = 2\pi n_j\,, \quad \mbox{for } n_j\in\integers.
\end{equation}
This is exactly the discretization condition that we get with an \E{1}, \E{11}, or \slabh.
In other words, the eigenmodes of the Laplacian on an \E{i} manifold are linear combinations of the Fourier modes that are eigenmodes of the associated homogeneous space.
For each \E{i} below, we present those discretization conditions.

Equation \eqref{eqn:generaleigenmodeformula} satisfies the invariance condition \eqref{eqn:generaleigenmodeinvariance} for all $\vec{k}$ allowed by 
\eqref{eqn:associatedhomogeneousdiscretization}, however in some cases the sum over $G_\alpha \in \mathcal{G}^{\E{i}}$ yields more than one identical term.
This occurs when $\mat{M}^{(G_\alpha)}\vec{k}=\vec{k}$ for certain $\vec{k}$ allowed by \eqref{eqn:associatedhomogeneousdiscretization}.
More specifically, for the manifolds in question, where $\mat{M}^{(G_\alpha)}$ are matrices $\mat{M}^{\E{i}}_a$  representing rotations by $2\pi p_a/q_a$ ($p_a\in\integers^{\neq 0}$, $q_a\in\integers^{> 0}$, $\vert p \vert$ and $q$ relatively prime) about one of the three coordinate axes, this occurs when $(\mat{M}^{\E{i}}_a)^{N_a}\vec{k} = \vec{k}$ has a solution for $N_a < q_a$.  We will consider those cases explicitly for each \E{i}.

Finally, we note that it will prove useful for many of the topologies to define
\begin{equation}
    \label{eqn:M0jdef}
    \mat{M}^{\E{i}}_{00} \equiv\mathbb{0}, \quad \mat{M}^{\E{i}}_{01} \equiv\identity, \quad
    \mbox{and} \quad \mat{M}^{\E{i}}_{0j} \equiv \sum_{r=0}^{j-1} (\mat{M}^{\E{i}}_B)^r \quad \mbox{for} \quad j>1.
\end{equation}

\subsection{\E{1}: 3-torus}

\noindent \textit{Properties}: Manifolds of this topology are compact, orientable, homogeneous, and anisotropic.
Further, all the compact topologies are roots of \E{1} (as described in detail in \rcite{COMPACT:2023rkp}). \\

\noindent
\textit{Generators}: The generators are given by
\begin{align} 
    \label{eqn:E1generalT}
    & \mat{M}^{\E{1}} = \identity, \quad \mbox{with} \nonumber \\
    & \vec{T}^{\E{1}}_1 = L_1 \begin{pmatrix} 1 \\ 0 \\ 0 \end{pmatrix}, \quad
    \vec{T}^{\E{1}}_2 = L_2 \begin{pmatrix} \cos\alpha \\ \sin\alpha, \\ 0 \end{pmatrix}, \quad
    \vec{T}^{\E{1}}_3 = L_3 \begin{pmatrix} \cos\beta\cos\gamma \\ \cos\beta\sin\gamma \\ \sin\beta \end{pmatrix} .
\end{align}

We will always choose the lengths to be positive (here meaning $0 < L_{i}$) with the orientation of the vector determined by the angles (here $\alpha$, $\beta$, and $\gamma$).   
In section $3.1$ of \rcite{COMPACT:2023rkp} we presented the set of conditions on the parameters of \E{1} ($L_i$, $\alpha$, $\beta$, and $\gamma$) that avoids double-counting identical manifolds with different choices of generators.\\

\noindent \textit{Volume}:  
\begin{equation}
    \label{eqn:E1volume}
    V_{\E{1}} 
        = \vert(\vec{T}^{\E{1}}_{{1}}\times\vec{T}^{\E{1}}_{{2}})\cdot\vec{T}^{\E{1}}_{{3}}\vert 
        = L_{{1}} L_{{2}} L_{{3}} \vert \sin{\alpha}\sin{\beta} \vert .
\end{equation}
    
\subsubsection{Eigenmodes and correlation matrices of \E{1}}
\label{secn:eigenmodesE1}

The 3-torus is the simplest of the compact Euclidean topologies and will serve as a model for determining the eigenspectrum and eigenmodes of all the Euclidean three-manifolds.
In this subsection, we determine which of the eigenvalues and eigenmodes of the tensor Laplacian acting on the covering space \E{18} are preserved by the isometries of the topology.
We then use that information to present the Fourier space and spherical-harmonic space correlation matrices of any fluctuations that are linearly related to independent Gaussian random fluctuations of the amplitudes of those eigenmodes.

While, in general, the eigenmodes of \E{i} can be linear combinations of the \E{18} eigenmodes, the \E{1} eigenmodes represent a subset of the \E{18} eigenmodes that adhere to the \E{1} symmetries. The eigenmode equation (\ref{eqn:generaleigenmodeinvariance}) for $E_1$ yields
\begin{equation}
    \label{eqn:E1eigenmodeinvariance}
\mat{e}(\unitvec{k},\lambda) \mathrm{e}^{i\vec{k}\cdot\left[(\vec{x}-\vec{x}_0)+\vec{T}^{E_1}_{j}\right]} = \mat{e}(\unitvec{k},\lambda) \mathrm{e}^{i\vec{k}\cdot(\vec{x}-\vec{x}_0)} 
\end{equation}
and more compactly 
\begin{equation}
\vec{k}\cdot\vec{T}^{E_1}_{j} = 2\pi n_j\,, \quad \mbox{for } n_j\in\integers.
\end{equation}

This is because all the group elements of $\Gamma^{\E{1}}$ are pure translations, i.e.,
$\mat{M}^{(G_\alpha)}=\identity$ for all $G_\alpha\in \mathcal{G}^{\E{1}}$, 
so $\mat{M}^{(G_\alpha)}\vec{k}=\vec{k}$ trivially.

As discussed above in general (cf.\ \eqref{eqn:associatedhomogeneousdiscretization}),
the symmetry condition \eqref{eqn:E1eigenmodeinvariance} leads to the discretization of the allowed $\vec{k}$ in \E{1}. Since the wavenumbers are now discretized, they are labeled by integers $n_i \in \integers$ and we denote this explicitly by writing the wavevector as $\vec{k}_{\vec{n}}$ for $\vec{n} = (n_1, n_2, n_3)$.
Here and below we will use either the $n_i$ or $(\vec{k}_{\vec{n}})_i$ labels as convenient for the situation.
Inverting these requirements, the components of the wavevectors are 

\begin{align}
    \label{eqn:E1_ki}
    (\vec{k}_{\vec{n}})_{x} &= \frac{2\pi n_1}{L_1} , \nonumber \\
    (\vec{k}_{\vec{n}})_{y} &= \frac{2\pi n_2}{L_2\sin\alpha} - \frac{2\pi n_1}{L_1}\frac{\cos\alpha}{\sin\alpha},\\
    (\vec{k}_{\vec{n}})_{z} &= \frac{2\pi n_3}{L_3\sin\beta}
        - 
        \frac{2\pi n_2}{L_2}
          \frac{\cos\beta\sin\gamma}{\sin\alpha\sin\beta} 
        - 
        \frac{2\pi n_1}{L_1}
          \frac{\cos\beta
            (\sin\alpha\cos\gamma-\cos\alpha\sin\gamma)
          }{\sin\alpha\sin\beta}.
          \nonumber    
\end{align}
Clearly the eigenvalues $-k_{\vec{n}}^2 = -\vert\vec{k}_{\vec{n}}\vert^2$ are a quadratic form in the $n_j$.
Thus
\begin{empheq}[box=\fbox]{equation}
   \Upsilon^{\E{1}}_{ij,\vec{k}_{\vec{n}}}(\vec{x},\lambda) = e_{ij}(\unitvec{k}_{\vec{n}},\lambda) \mathrm{e}^{i\vec{k}_{\vec{n}}\cdot(\vec{x}-\vec{x}_0)}, \quad \mbox{for } \vec{n}\in \setN^{\E{1}}
\end{empheq}
where
\begin{empheq}[box=\fbox]{equation}
    \setN^{\E{1}} \equiv \{(n_1,n_2,n_3) \vert n_i\in\integers\}\setminus (0,0,0)\,.
\end{empheq}

The analogue to \cref{eqn:CE18XY}, i.e., the helicity-dependent Fourier-mode-amplitude correlation matrix for \E{1} is

\begin{empheq}[box=\fbox]{align}
    \label{eqn:CE1XY}
    C^{\E{1};XY}_{iji'j', \vec{k}_{\vec{n}} \vec{k}_{\vecnp}, \lambda \lambda'} 
    &= V_{\E{1}}\frac{\pi^2}{2 k_{\vec{n}}^3} \Tpower(k_{\vec{n}}) 
    \Delta^X(k_{\vec{n}}) \DeltaYstar(k_{\vec{n}}) 
    \Kdelta_{\vec{k}_{\vec{n}}\vec{k}_{\vecnp}} \Kdelta_{\lambda \lambda'} \mathcal{E}_{iji'j'}(\unitvec{k}_{\vec{n}}, \unitvec{k}_{\vec{n}'},\lambda)\;.
\end{empheq}
In transitioning from the covering space \E{18} we have replaced $(2\pi)^3\Ddelta(\vec{k}-\vec{k}')$ with $V_{\E{1}}\Kdelta_{\vec{k}_{\vec{n}}\vec{k}_{\vecnp}}$, where the volume factor $V_{\E{1}}$ is given by \eqref{eqn:E1volume}.

As for \E{18} above, we can project a field $\field^X$ onto the sky by performing a radial integral with a suitable transfer function, giving
\begin{align}
    \label{eqn:almE1}
   a^{\E{1}; X}_{\ell m} = \frac{1}{V_{\E{1}}} \sum_{\lambda =\pm 2} \sum_{\vec{n} \in \setN^{\E{1}}}  \mathcal{D}(\vec{k}_{\vec{n}}, \lambda, 0) \xi^{\E{1};X,\unitvec{k}_{\vec{n}}}_{k_{\vec{n}},\ell m,\lambda} \ \Delta^X_{\ell} (k_{\vec{n}})
\end{align}
with
\begin{empheq}[box=\fbox]{equation}
 \label{eqn: E1coeff_T}
    \xi^{\E{1};T,\unitvec{k}_{\vec{n}}}_{k_{\vec{n}},\ell m,\lambda} \equiv i^{\ell} \sqrt{\frac{2\pi^2(\ell + 2)!}{(\ell - 2)!}} \ {}_{-\lambda}Y_{\ell m} (\unitvec{k}_{\vec{n}})  \mathrm{e}^{-i \vec{k}_{\vec{n}}\cdot \vec{x}_0},
\end{empheq}
\begin{empheq}[box=\fbox]{equation}
 \label{eqn: E1coeff_E}
    \xi^{\E{1};E,\unitvec{k}_{\vec{n}}}_{k_{\vec{n}},\ell m,\lambda} \equiv -i^{\ell} \ \sqrt{2 \pi^2} \ {}_{-\lambda}Y_{\ell m} (\unitvec{k}_{\vec{n}})\mathrm{e}^{-i \vec{k}_{\vec{n}}\cdot \vec{x}_0},
\end{empheq}
and
\begin{empheq}[box=\fbox]{equation}
\xi^{\E{1};B,\unitvec{k}_{\vec{n}}}_{k_{\vec{n}},\ell m,\lambda} = -\frac{\lambda}{2} \xi^{\E{1};E,\unitvec{k}_{\vec{n}}}_{k_{\vec{n}},\ell m,\lambda}.
\end{empheq}
Because $\setN^{\E{1}}$ labels only a discrete set of $\vec{k}_{\vec{n}}$, the integral over $\mathrm{d}^3 k$ in \cref{eqn:almE18} is replaced by a sum over $\vec{n}\in\setN^{\E{1}}$.

For the compact topologies \E{i} with $i \in \{1, \ldots, 6\}$, the spherical-harmonic covariance matrix has the general form\footnote{
    Note that, while in general $\Delta^Y_{ \ell}(\vec{k})$ can be complex, for the usual cases of CMB temperature and polarization $\Delta^Y_{ \ell}(\vec{k})$ is real; nevertheless, we retain the complex conjugate for generic $Y$.}
\begin{empheq}[box=\fbox]{align}
    \label{eqn:HarmonicCovariance}
    C^{\E{i}; XY}_{\ell m\ell'm'}  &\equiv \langle a^{\E{i}, X}_{\ell m} a^{\E{i}, Y*}_{\ell' m'} \rangle \nonumber\\ 
     &=\frac{\pi^2}{{2 V_{\E{i}}}} \sum_{\lambda = \pm 2}
    \sum_{\vec{n}\in \setN^{\E{i}}}
    \frac{\Tpower(k_n)}{k_{\vec{n}}^3 } 
    \Delta^{X}_{\ell}(k_{\vec{n}})
    \Delta^{Y*}_{\ell'}(k_{\vec{n}}) \ \xi^{\E{i};X,\unitvec{k}_{\vec{n}}}_{k_{\vec{n}},\ell m,\lambda}\ \xi^{\E{i};Y,\unitvec{k}_{\vec{n}}*}_{k_{\vec{n}},\ell' m',\lambda} .
\end{empheq}

\subsection{\E{2}: Half-turn space}
\label{secn:topologyE2}

\noindent \textit{Properties}: This manifold is compact, orientable, inhomogeneous, and anisotropic. \\

\noindent \textit{Generators}: In general the generators of \E{2} can be written as\footnote{Here and throughout all rotations will be treated as active.}
\begin{align}
    \label{eqn:E2generalT}
    & \mat{M}^{\E{2}}_A = \identity, \quad \mat{M}^{\E{2}}_B = \mat{R}_{\unitvec{z}}(\pi) = \diag(-1,-1,1), \quad \mbox{with} \nonumber \\
   & \vec{T}^{\E{2}}_{\A{1}} = L_{\A{1}} \begin{pmatrix} 1 \\ 0 \\ 0 \end{pmatrix}, \quad 
     \vec{T}^{\E{2}}_{\A{2}} = L_{\A{2}} \begin{pmatrix} \cos\alpha \\ \sin\alpha \\ 0 \end{pmatrix}, \quad
     \vec{T}^{\E{2}}_{B} = L_B \begin{pmatrix} \cos\beta\cos\gamma \\ \cos\beta\sin\gamma \\ \sin\beta \end{pmatrix}.
\end{align}
In section $3.2$ of \rcite{COMPACT:2023rkp}, we presented a set of conditions on the parameters of \E{2} ($L_{A_i}$, $L_B$, $\alpha$, $\beta$, and $\gamma$) that avoids double-counting identical manifolds with different generator choices.
\\
\\
\noindent\textit{Associated \E{1}:}
In addition to $\vec{T}^{\E{2}}_1\equiv\vec{T}^{\E{2}}_{\A{1}}$ and $\vec{T}^{\E{2}}_2\equiv\vec{T}^{\E{2}}_{\A{2}}$, a third independent translation is
\begin{equation}
    \label{eqn:E2assocE1}
    g^{\E{2}}_3 \equiv (g_B^{\E{2}})^2: \vec{x} \to \vec{x} + \vec{T}^{\E{2}}_3
\end{equation}
for
\begin{equation}
    \vec{T}^{\E{2}}_3 \equiv \begin{pmatrix} 0 \\ 0 \\ 2L_{Bz} \end{pmatrix} = 2 L_B \begin{pmatrix} 0 \\ 0 \\ \sin\beta \end{pmatrix}.
\end{equation}
The three vectors $\vec{T}^{\E{2}}_1$, $\vec{T}^{\E{2}}_2$, and $\vec{T}^{\E{2}}_3$ define the associated \E{1}. \\

\noindent \textit{Volume}:
\begin{equation}
    \label{eqn:VE2}
    V_{\E{2}} = \frac{1}{2}\vert(\vec{T}^{\E{2}}_{1}\times\vec{T}^{\E{2}}_{2})\cdot\vec{T}^{\E{2}}_{3}\vert 
      = \vert L_{1x} L_{2y} L_{Bz} \vert
      = L_{1} L_{2} L_B \vert \sin\alpha\sin\beta\vert.
\end{equation}

\subsubsection{Eigenmodes and correlation matrices of \E{2}}
\label{secn:eigenmodesE2}

The eigenspectrum and eigenmodes of the half-turn space can be derived using a method similar to that of the 3-torus. While one could start from the covering space, a more efficient approach is to recognize that \E{2} is essentially \E{1} with additional symmetries imposed. Consequently, the eigenspectrum of \E{2} becomes discretized, with wavevectors denoted by $\vec{k}_{\vec{n}}$, and its eigenfunctions, $\Upsilon^{\E{2}}_{ij,\vec{k}_{\vec{n}}}(\vec{x},\lambda)$, are expressed as linear combinations of the eigenfunctions $\Upsilon^{\E{1}}_{ij,\vec{k}_{\vec{n}}}(\vec{x},\lambda)$ from \E{1}. In the case of \E{2}, the discretization condition given in  \cref{eqn:associatedhomogeneousdiscretization}, derived from the translation vectors $\vec{T}^{\E{2}}_j$, determines the components of the allowed wavevectors,
\begin{align}
    \label{eqn:E2_kni}
    (\vec{k}_{\vec{n}})_{x} &= \frac{2\pi n_1}{L_1}, \nonumber \\
    (\vec{k}_{\vec{n}})_{y} &= \frac{2\pi n_2}{L_2\sin\alpha} - \frac{2\pi n_1}{L_1}\frac{\cos\alpha}{\sin\alpha},\\
    (\vec{k}_{\vec{n}})_{z} &= \frac{2\pi n_3}{2L_B\sin\beta}. \nonumber
\end{align}

Unlike in \E{1}, the eigenmodes of \E{2} can include a linear combination of two \E{1} eigenmodes. For \E{2}, the eigenmodes are determined by substituting its generators into \cref{eqn:generaleigenmodeformula} and identifying the minimum number of terms required to keep the eigenmodes invariant under these generators. This process can be simplified using the transformation properties of helicity tensors under a rotation: 
\begin{equation}
\label{eqn:helicityrotationtrans}
(\mat{M}^{E_i}_B)^N \mat{e}(\unitvec{k},\lambda)\left[(\mat{M}^{E_i}_B)^T\right]^N = \mat{e}\left(\left[(\mat{M}^{\E{i}}_B)^T\right]^N\unitvec{k},\lambda\right), \quad \mbox{for} \quad \unitvec{k}\neq \unitvec{z}.
\end{equation}
For the special case of $\unitvec{k} = \unitvec{z}$, the helicity tensors transform under rotation as
\begin{align}
    \label{eqn:helicityZroationtrans}
    (\mat{M}^{E_2}_B)_{ik} e_{k\ell}(\unitvec{z},\lambda)(\mat{M}^{E_2}_B)_{j\ell} &= \mathrm{e}^{-i\lambda \theta^{\E{i}}}e_{ij}(\unitvec{z},\lambda),
\end{align}
where $\theta^{\E{i}}$ is the rotation angle about $\unitvec{z}$, associated with the generator $\mat{M}_B^{\E{i}}$ for each topology.

The eigenmodes for \E{2} are then obtained by finding the smallest 
$N$ that satisfies $(\mat{M}^{\E{i}}_B)^N\vec{k}_{\vec{n}} = \vec{k}_{\vec{n}}$ depending on $\vec{k}_{\vec{n}}$, where, $N=1$ holds when $(\vec{k}_{\vec{n}})_{x} = (\vec{k}_{\vec{n}})_{y} = 0$, and $N=2$ otherwise. So, we have:
\begin{description}
    \item[\textbf{$N=1$ eigenmodes: }]  $\vec{k}_{\vec{n}}=\transpose{(0,0,(\vec{k}_{\vec{n}})_z)}$, 
        i.e., $\vec{n}=(0,0,n_3)$, 
        $n_3 \in 2\integers^{\neq 0}$, with
    \begin{empheq}[box=\fbox]{align}
       \Upsilon^{\E{2}}_{ij,(0,0,n_3)}(\vec{x},\lambda)& = e_{ij}(\unitvec{k}_{\vec{n}},\lambda) \mathrm{e}^{i\vec{k}_{\vec{n}} \cdot (\vec{x}-\vec{x}_0)} = \mathrm{e}^{i(\vec{k}_{\vec{n}})_{z} (z-z_0)}, 
    \end{empheq}
    \item[\textbf{$N=2$ eigenmodes: }] $((\vec{k}_{\vec{n}})_{x}, (\vec{k}_{\vec{n}})_{y}) \neq (0,0)$, 
        i.e., $(n_1,n_2) \neq (0,0)$, and per \cref{eqn:generaleigenmodeformula},
    \begin{empheq}[box=\fbox]{align}
        \Upsilon^{\E{2}}_{ij,\vec{n}}(\vec{x},\lambda) =&
            \ \frac{\mathrm{e}^{-i \vec{k}_{\vec{n}} \cdot \vec{T}^{\E{2}}_B/2}}{\sqrt{2}} \left( e_{ij}(\unitvec{k},\lambda) \mathrm{e}^{i\vec{k}_{\vec{n}}\cdot(\vec{x}-\vec{x}_0)}+ \right.\\ 
            & \left. \qquad \qquad \qquad \quad + e_{ij}\left((\mat{M}^{\E{2}}_B)^T\unitvec{k}_n,\lambda\right) \mathrm{e}^{i\vec{k}_{\vec{n}}\cdot(\mat{M}^{\E{2}}_B(\vec{x}-\vec{x}_0)+\vec{T}^{\E{2}}_B)} \right). \nonumber
    \end{empheq}
\end{description}
For the $N = 2$ modes, we choose $\Phi_{\vec{k}_{\vec{n}}}^{\E{2}} = - \vec{k}_{\vec{n}} \cdot \vec{T}^{\E{2}}_B/2$ in \cref{eqn:generaleigenmodeformula}---this will simplify the reality condition of the field.  
Moreover, note that $\transpose{\vec{k}}_{\vec{n}}\mat{M}^{\E{2}}_B = (-(\vec{k}_{\vec{n}})_x, -(\vec{k}_{\vec{n}})_y, (\vec{k}_{\vec{n}})_z)$, i.e., $\mat{M}^{\E{2}}_B$ maps $(n_1, n_2, n_3) \to (-n_1, -n_2, n_3)$.  
As a result, summing over $(n_1, n_2, n_3)$ cannot include all $n_1 \in \integers$ and $n_2 \in \integers$ simultaneously, as this would lead to double-counting eigenmodes.
Hence, we define two sets of allowed modes, one for $N=1$ and another for $N=2$:
\begin{empheq}[box=\fbox]{align}
    \setN^{\E{2}}_1 &= \{(0,0,n_3)|n_3\in 2\integers^{\neq0} \} , \nonumber\\
    \setN^{\E{2}}_2 &= \{(n_1,n_2,n_3)| n_1 \in \integers^{>0}, n_2 \in \integers, n_3 \in \integers\} 
                    \cup \{(0,n_2,n_3)| n_2 \in \integers^{>0}, n_3 \in \integers\} , \\
    \setN^{\E{2}} &= \setN^{\E{2}}_1 \cup \setN^{\E{2}}_2 . \nonumber
\end{empheq}
With these, the Fourier-mode correlation matrix can now be expressed as
 \begin{empheq}[box=\fbox]{align}
     \label{eqn:E2FourierCovarianceStandardConvention}
      C^{\E{2};XY}_{iji'j', \vec{k}_{\vec{n}} \vec{k}_{\vecnp}, \lambda \lambda'}
      &= {} V_{\E{2}} \frac{\pi^2}{2 k^3_{\vec{n}}}
         \Tpower(k_{\vec n})\Delta^X(k_{\vec{n}})\DeltaYstar(k_{\vec{n}})
         \mathrm{e}^{i(\vec{k}_{\vecnp}-\vec{k}_{\vec{n}})\cdot\vec{x}_0}
     \Kdelta_{\lambda \lambda'} \times\\
     & \times \mathcal{E}_{iji'j'}(\unitvec{k}_{\vec{n}}, \unitvec{k}_{\vec{n}'}, \lambda) \left[
             \sum_{\vec{\tilde{n}} \in \setN^{\E{2}}_1}
                  \Kdelta_{\vec{k}_{\vec{n}}\vec{k}_{\vec{\tilde{n}}}}
                  \Kdelta_{\vec{k}_{\vecnp}\vec{k}_{\vec{\tilde{n}}}} + \frac{1}{2}
                 \sum_{\vec{\tilde{n}}\in \setN^{\E{2}}_2}
                 \sum_{a=0}^1\sum_{b=0}^1 
         \right.
           \nonumber \\ 
            & \left. 
                \qquad \qquad \qquad  \quad \mathrm{e}^{i\vec{k}_{\vec{\tilde{n}}}\cdot(\vec{T}^{(a)}-\vec{T}^{(b)})}
                \Kdelta_{\vec{k}_{\vec{n}}([(\mat{M}^{\E{2}}_B){}^T]^a\vec{k}_{\vec{\tilde{n}}})}
                 \Kdelta_{\vec{k}_{\vecnp}([(\mat{M}^{\E{2}}_B){}^T]^b\vec{k}_{\vec{\tilde{n}}})} 
         \right], 
         \nonumber
 \end{empheq}
where $V_{\E{2}}$ is given in \eqref{eqn:VE2}, $\vec{T}^{(0)}\equiv\vec{0}$, and $\vec{T}^{(1)}\equiv\vec{T}^{\E{2}}_B$.\footnote{
    In  \cref{eqn:E2FourierCovarianceStandardConvention}, $\vec{k}_{\vec{n}}$ and $\vec{k}_{\vecnp}$ are wavevectors of the associated \E{1}, as specified above in \eqref{eqn:E2_kni}.
    $C^{\E{2};XY}_{iji'j', \vec{k}_{\vec{n}} \vec{k}_{\vecnp}, \lambda \lambda'}$ describes correlations between amplitudes of the plane gravitational waves that comprise the tensor eigenmodes of a specific manifold---i.e., of a specific topology, with specific values of its parameters.
    It is this object that would be used, for example, in creating realizations of initial conditions of the gravitational wave field in three dimensions.
    If one was, instead, constructing a likelihood function to compare data with expectations from \E{2} manifolds, one would need to convolve $C^{\E{2};XY}_{iji'j', \vec{k}_{\vec{n}} \vec{k}_{\vecnp}, \lambda \lambda'}$ with a kernel characterizing the Fourier structure of the survey of interest.  However, there is not likely to be interest in 3-dimensional realizations of the tensor field in the foreseeable future.
}
Note that $C^{\E{2};XY}_{iji'j', \vec{k}_{\vec{n}} \vec{k}_{\vecnp}, \lambda \lambda'}=0$ for $\vert \vec{k}_{\vecnp}\vert\neq\vert \vec{k}_{\vec{n}}\vert$, so $\Tpower(k_{\vec n})$, $\Delta^X(k_{\vec{n}})$, and $\DeltaYstar(k_{\vec{n}})$ are each a function only of $k_{\vec{n}}$.

Another consequence of the rotation in $\mat{M}^{\E{2}}_B$ appears when expressing the eigenmodes in the harmonic basis.  
In this case, we combine modes that share the same eigenvalue $-k^2_{\vec{n}}$ and orientations $\unitvec{k}_{\vec{n}}$ and $(\mat{M}^{\E{2}}_B)^T \unitvec{k}_{\vec{n}}$.  
Since one of the generators of the half-turn space (see \cref{eqn:E2generalT}) includes a rotation by $\pi$ around the $z$-axis, we can leverage the rotation properties of spin-weighted spherical harmonics to simplify the resulting expressions.
In particular
\begin{equation}
    {}_{\lambda}Y_{\ell m} ((\mat{M}^{\E{2}}_B){}^T \unitvec{k}_{\vec{n}}) = \mathrm{e}^{i m \pi} {}_{\lambda}Y_{\ell m}  (\unitvec{k}_{\vec{n}}) = (-1)^m {}_{\lambda}Y_{\ell m} .
\end{equation}
This gives for the eigenmodes in the harmonic basis
\begin{empheq}[box=\fbox]{align}
    \xi^{\E{2};T,\unitvec{k}_{\vec{n}}}_{k_{\vec{n}},\ell m,\lambda} &\equiv 
           i^{\ell} \sqrt{\frac{2\pi^2(\ell + 2)!}{(\ell - 2)!}} \ {}_{-\lambda}Y_{\ell m} (\unitvec{k}_{\vec{n}}) \  \mathrm{e}^{-i \vec{k}_{\vec{n}} \cdot \vec{x}_0} , \quad \mbox{for } \vec{n}\in \setN^{\E{2}}_1\\
    \xi^{\E{2};T,\unitvec{k}_{\vec{n}}}_{k_{\vec{n}},\ell m,\lambda}& \equiv 
          \frac{i^{\ell}}{\sqrt{2}} \sqrt{\frac{2\pi^2(\ell + 2)!}{(\ell - 2)!}} \ {}_{-\lambda}Y_{\ell m} (\unitvec{k}_{\vec{n}}) \times \nonumber\\
         & \quad \times \mathrm{e}^{-i \vec{k}_{\vec{n}} \cdot \vec{T}^{\E{2}}_B/2} \left( \mathrm{e}^{-i \vec{k}_n\cdot \vec{x}_0} +  (-1)^m \mathrm{e}^{-i \vec{k}_n\cdot \left[ M^{\E{2}}_B \vec{x}_0- \vec{T}^{\E{2}}_B \right]} \right)  \quad \mbox{for } \vec{n} \in \setN^{\E{2}}_2 \nonumber
\end{empheq}

\begin{empheq}[box=\fbox]{align}
    \xi^{\E{2};E,\unitvec{k}_{\vec{n}}}_{k_{\vec{n}},\ell m,\lambda} &\equiv 
       -i^{\ell}\ \sqrt{2 \pi^2} \ {}_{-\lambda}Y_{\ell m} (\unitvec{k}_{\vec{n}}) \  \mathrm{e}^{-i \vec{k}_{\vec{n}} \cdot \vec{x}_0} , \quad \mbox{for } \vec{n}\in \setN^{\E{2}}_1\\
    \xi^{\E{2};E,\unitvec{k}_{\vec{n}}}_{k_{\vec{n}},\ell m,\lambda}& \equiv 
       - \frac{i^{\ell}}{\sqrt{2}}\ \sqrt{2 \pi^2} \ {}_{-\lambda}Y_{\ell m} (\unitvec{k}_{\vec{n}}) \times \nonumber\\
         & \quad \times \mathrm{e}^{-i \vec{k}_{\vec{n}} \cdot \vec{T}^{\E{2}}_B/2} \left( \mathrm{e}^{-i \vec{k}_n\cdot \vec{x}_0} +  (-1)^m \mathrm{e}^{-i \vec{k}_n\cdot \left[ M^{\E{2}}_B \vec{x}_0- \vec{T}^{\E{2}}_B \right]} \right), \quad \mbox{for } \vec{n} \in \setN^{\E{2}}_2 \nonumber
\end{empheq}
and
\begin{empheq}[box=\fbox]{equation}
\xi^{\E{2};B,\unitvec{k}_{\vec{n}}}_{k_{\vec{n}},\ell m,\lambda} = -\frac{\lambda}{2} \xi^{\E{2};E,\unitvec{k}_{\vec{n}}}_{k_{\vec{n}},\ell m,\lambda}.
\end{empheq}
In simplifying the previous expressions we have made use of \cref{eqn:SHzaxis}. Finally, the spherical-harmonic space covariance matrix has the form \eqref{eqn:HarmonicCovariance}.

\subsection{\E{3}: Quarter-turn space}
\label{secn:topologyE3}

\textit{Properties}: This manifold is compact, orientable, inhomogeneous, and anisotropic. \\

\noindent \textit{Generators}:  In general the generators of \E{3} can be written as
\begin{align}
    \label{eqn:E3generalT}
    & \mat{M}^{\E{3}}_A = \identity, \quad \mat{M}^{\E{3}}_B = \mat{R}_{\unitvec{z}}(\pi/2) =
        \begin{pmatrix}
        0 & -1 & 0\\
        1 &  \hphantom{-}0 & 0\\
        0 &  \hphantom{-}0 & 1
        \end{pmatrix} ,
        \quad \mbox{with}
        \nonumber \\
   &  \vec{T}^{\E{3}}_{\A{1}} = L_{A} \begin{pmatrix} 1 \\ 0 \\ 0 \end{pmatrix}, \quad
    \vec{T}^{\E{3}}_{\A{2}} = L_{A} \begin{pmatrix} 0 \\ 1 \\ 0 \end{pmatrix}, \quad
    \vec{T}^{\E{3}}_{B} = L_B \begin{pmatrix} \cos\beta\cos\gamma \\ \cos\beta\sin\gamma \\ \sin\beta \end{pmatrix} .
\end{align}
In section $3.3$ of \rcite{COMPACT:2023rkp}, we presented a set of conditions on the parameters of \E{3} ($L_{A}$, $L_B$, $\beta$, and $\gamma$) that avoids double-counting identical manifolds with different generator choices.
\\
\\
\noindent \textit{Associated \E{1}}: In addition to $\vec{T}^{\E{3}}_1\equiv\vec{T}^{\E{3}}_{\A{1}}$ and $\vec{T}^{\E{3}}_2\equiv\vec{T}^{\E{3}}_{\A{2}}$, a third independent translation follows from $(\mat{M}^{\E{3}}_B)^4=\identity$:
\begin{equation}
    g^{\E{3}}_3 \equiv (g^{\E{3}}_B )^4:\vec{x}\to \vec{x} + \vec{T}^{\E{3}}_3,
\end{equation}
for
\begin{equation}
    \label{eqn:E3assocE1}
    \vec{T}^{\E{3}}_3 \equiv \begin{pmatrix} 0 \\ 0 \\ 4 L_{Bz} \end{pmatrix}
      = 4 L_B \begin{pmatrix} 0 \\ 0 \\ \sin\beta \end{pmatrix}.
\end{equation}
\\
\noindent \textit{Volume}:
\begin{equation}
    \label{eqn:VE3}
    V_{\E{3}} = \frac{1}{4}
        \vert(\vec{T}^{\E{3}}_{{1}}\times\vec{T}^{\E{3}}_{{2}})\cdot\vec{T}^{\E{3}}_{3}\vert 
      = L_A^2 L_B |\sin\beta|.
\end{equation}

\subsubsection{Eigenmodes and correlation matrices of \E{3}}

The eigenspectrum and eigenmodes of the quarter-turn space can be derived similarly to the case of \E{2}.  
For \E{3}, the discretization condition \eqref{eqn:associatedhomogeneousdiscretization}, which follows from the translation vectors $\vec{T}^{\E{3}}_j$, yields the components of the allowed wavevectors,
\begin{equation}
    (\vec{k}_{\vec{n}})_{x} = \frac{2\pi n_1}{L_A }, \quad
    (\vec{k}_{\vec{n}})_{y} = \frac{2\pi n_2}{L_A},  \quad
     (\vec{k}_{\vec{n}})_{z} = \frac{2\pi n_3}{4 L_B \sin\beta}.
\end{equation}

As in the case of \E{2}, the eigenmodes of \E{3} include linear combinations of \E{1} eigenmodes, as $(\mat{M}^{\E{3}}_B)^N \vec{k}_{\vec{n}} = \vec{k}_{\vec{n}}$ admits multiple solutions for the minimum positive $N$, depending on $\vec{k}_{\vec{n}}$.  Here we have:
\begin{description}
    \item[\textbf{$N=1$ eigenmodes: }]  $\vec{k}_{\vec{n}}=\transpose{(0,0,(\vec{k}_{\vec{n}})_z)}$, 
        i.e., $\vec{n}=(0,0,n_3)$, 
        $n_3 = 4 m + 2$, where $ m \in \integers$, with
    \begin{empheq}[box=\fbox]{align}
        &\Upsilon^{\E{3}}_{ij,(0,0,n_3)}(\vec{x},\lambda) = e_{ij}(\unitvec{k}_{\vec{n}},\lambda) \mathrm{e}^{i\vec{k}_{\vec{n}} \cdot (\vec{x}-\vec{x}_0)}\,,
    \end{empheq}
    \item[\textbf{$N=4$ eigenmodes: }] $((\vec{k}_{\vec{n}})_{x}, (\vec{k}_{\vec{n}})_{y}) \neq (0,0)$, 
        i.e., $(n_1,n_2) \neq (0,0)$, with
    \begin{empheq}[box=\fbox]{align}
        \Upsilon^{\E{3}}_{ij,\vec{n}}(\vec{x},\lambda) =& 
            \frac{\mathrm{e}^{-i\vec{k}_{\vec{n}}\cdot \mat{M}^{\E{3}}_{02}\vec{T}^{\E{3}}_B/2}}{\sqrt{4}} \sum_{a=0}^3 e_{ij}\left(\left[(\mat{M}^{\E{3}}_B)^T\right]^a\unitvec{k}_n,\lambda\right) \times\\  
            &\qquad \qquad \qquad \qquad \qquad \qquad
             \quad \times \mathrm{e}^{i\vec{k}_{\vec{n}}\cdot \left((\mat{M}^{\E{3}}_B)^a(\vec{x}-\vec{x}_0)+\mat{M}^{\E{3}}_{0a}\vec{T}^{\E{3}}_B\right)}, \nonumber
    \end{empheq}
\end{description}
and $\mat{M}^{\E{3}}_{0a}$ defined in \eqref{eqn:M0jdef}. For the $N = 4$ modes, we choose $\Phi_{\vec{k}_{\vec{n}}}^{\E{3}} = -\vec{k}_{\vec{n}}\cdot \mat{M}^{\E{3}}_{02}\vec{T}^{\E{3}}_B/2$ in \eqref{eqn:generaleigenmodeformula}, this will simplify the reality condition of the field.  
The repeated action of $\mat{M}^{\E{3}}_B$ cyclically maps $(n_1, n_2) \to (n_2, -n_1) \to (-n_1, -n_2) \to (-n_2, n_1)$. 
 Consequently, as with \E{2}, the cyclic properties of $\mat{M}^{\E{3}}_B$ lead to overcounting of eigenmodes if all $n_1 \in \integers$ and $n_2 \in \integers$ are included.
To avoid this, we define two sets of allowed modes, now $N = 1$ and $N = 4$,
\begin{empheq}[box=\fbox]{align}
    \setN^{\E{3}}_1 &= \{(0,0,n_3 = 4m+2)|m \in \integers \} , \nonumber \\
    \setN^{\E{3}}_4 &= \{(n_1,n_2,n_3)| n_1 \in \integers^{\geq 0}, n_2 \in \integers^{>0}, n_3 \in \integers\} , \\
    \setN^{\E{3}} &= \setN^{\E{3}}_1 \cup \setN^{\E{3}}_4 . \nonumber
\end{empheq}
We note that $\setN^{\E{3}}_1$ differs from the set of allowed wavenumbers in the scalar case. This difference stems from the fact that not only does the complex exponential give an overall phase when transformed by an element of the isometry group, but also the helicity tensor can give a complex phase that needs to be taken care of by appropriate discretization of the wavevectors. Using these, the helicity-dependent Fourier-mode correlation matrix can be written as
\begin{empheq}[box=\fbox]{align}
\label{eqn:E3FourierCovarianceStandardConvention}
     C^{\E{3};XY}_{iji'j', \vec{k}_{\vec{n}} \vec{k}_{\vecnp}, \lambda \lambda'}&
     = {} V_{\E{3}} \frac{\pi^2}{2k^3_{\vec{n}}}
        \Tpower(k_{\vec n})\Delta^X(k_{\vec{n}})\DeltaYstar(k_{\vec{n}})
        \mathrm{e}^{i(\vec{k}_{\vecnp}-\vec{k}_{\vec{n}})\cdot\vec{x}_0}  \Kdelta_{\lambda \lambda'} \ \times \\
        & \times \mathcal{E}_{iji'j'}(\unitvec{k}_{\vec{n}}, \unitvec{k}_{\vec{n}'}, \lambda)\left[
            \sum_{\vec{\tilde{n}} \in \setN^{\E{3}}_1}
                 \Kdelta_{\vec{k}_{\vec{n}}\vec{k}_{\vec{\tilde{n}}}}
                 \Kdelta_{\vec{k}_{\vecnp}\vec{k}_{\vec{\tilde{n}}}} 
                + \frac{1}{4}
                \sum_{\vec{\tilde{n}}\in \setN^{\E{3}}_4}
                \sum_{a=0}^3\sum_{b=0}^3 
                \right. \nonumber\\
        & \left. \qquad \qquad \qquad \quad
            \mathrm{e}^{i\vec{k}_{\vec{\tilde{n}}}\cdot(\vec{T}^{(a)}-\vec{T}^{(b)})}
             \Kdelta_{\vec{k}_{\vec{n}}([(\mat{M}^{\E{3}}_B){}^T]^a\vec{k}_{\vec{\tilde{n}}})}
             \Kdelta_{\vec{k}_{\vecnp}([(\mat{M}^{\E{3}}_B){}^T]^b\vec{k}_{\vec{\tilde{n}}})} 
         \right] ,
         \nonumber
\end{empheq}
where $V_{\E{3}}$ is given in \eqref{eqn:VE3}, $\vec{T}^{(0)} \equiv \vec{0}$, and $\vec{T}^{(a)} \equiv \mat{M}^{\E{3}}_{0a} \vec{T}^{\E{3}}_B$ for $a \in \{ 1,2,3 \}$.

Similarly to \E{2}, we can apply the rotation properties of spin-weighted spherical harmonics and utilize the fact that $\mat{M}^{\E{3}}_B$ represents a rotation by $\pi/2$ around the $z$-axis, which leads to

\begin{equation}
    {}_{\lambda}Y_{\ell -m} \left(\left[(\mat{M}^{\E{3}}_B){}^T \right]^j \unitvec{k}_{\vec{n}}\right) = \mathrm{e}^{-i m j\pi/2} \ {}_{\lambda}Y_{\ell -m}  (\unitvec{k}_{\vec{n}}) = (i)^{j m} \ {}_{\lambda}Y_{\ell -m} (\unitvec{k}) .
\end{equation}
This gives for the eigenmodes in the harmonic basis
\begin{empheq}[box=\fbox]{align}
    \xi^{\E{3};T,\unitvec{k}_{\vec{n}}}_{k_{\vec{n}},\ell m,\lambda} 
        & \equiv i^{\ell} \sqrt{\frac{2\pi^2(\ell + 2)!}{(\ell - 2)!}} \ {}_{-\lambda}Y_{\ell m} (\unitvec{k}_{\vec{n}})  \  \mathrm{e}^{-i \vec{k}_{\vec{n}} \cdot \vec{x}_0} , \quad \mbox{for } \vec{n}\in \setN^{\E{3}}_1\\
    \xi^{\E{3};T,\unitvec{k}_{\vec{n}}}_{k_{\vec{n}},\ell m,\lambda}
        &\equiv \frac{i^{\ell}}{\sqrt{4}} \sqrt{\frac{2\pi^2(\ell + 2)!}{(\ell - 2)!}} \ {}_{-\lambda}Y_{\ell m} (\unitvec{k}_{\vec{n}}) \ \mathrm{e}^{-i\vec{k}_{\vec{n}}\cdot \mat{M}^{\E{3}}_{02}\vec{T}^{\E{3}}_B/2} \times \nonumber\\
        & \qquad \qquad \qquad \quad \times \sum_{j=0}^{3} \mathrm{e}^{-i m j \pi/2} \ \mathrm{e}^{-i \vec{k}_n\cdot \left[ (\mat{M}^{\E{3}}_B)^j \vec{x}_0- \mat{M}^{\E{3}}_{0j} \vec{T}^{\E{3}}_B \right]}, \quad \mbox{for } \vec{n} \in \setN^{\E{3}}_4 \nonumber
\end{empheq}

\begin{empheq}[box=\fbox]{align}
    \xi^{\E{3};E,\unitvec{k}_{\vec{n}}}_{k_{\vec{n}},\ell m,\lambda}
        &\equiv -i^{\ell} \sqrt{2 \pi^2} \ {}_{-\lambda}Y_{\ell m} (\unitvec{k}_{\vec{n}}) \  \mathrm{e}^{-i \vec{k}_{\vec{n}} \cdot \vec{x}_0} , \quad \mbox{for } \vec{n}\in \setN^{\E{3}}_1\\
    \xi^{\E{3};E,\unitvec{k}_{\vec{n}}}_{k_{\vec{n}},\ell m,\lambda}
        & \equiv -\frac{i^{\ell}}{\sqrt{4}} \ \sqrt{2 \pi^2} \ {}_{-\lambda}Y_{\ell m} (\unitvec{k}_{\vec{n}}) \ \mathrm{e}^{-i\vec{k}_{\vec{n}}\cdot \mat{M}^{\E{3}}_{02}\vec{T}^{\E{3}}_B/2} \times \nonumber\\
        & \qquad \qquad \qquad \quad \times  \sum_{j=0}^{3} \mathrm{e}^{-i m j \pi/2} \mathrm{e}^{-i \vec{k}_n\cdot \left[ (\mat{M}^{\E{3}}_B)^j \vec{x}_0- \mat{M}^{\E{3}}_{0j} \vec{T}^{\E{3}}_B \right]}, \quad \mbox{for } \vec{n} \in \setN^{\E{3}}_4 \nonumber
\end{empheq}
and
\begin{empheq}[box=\fbox]{equation}
\xi^{\E{3};B,\unitvec{k}_{\vec{n}}}_{k_{\vec{n}},\ell m,\lambda} = - \frac{\lambda}{2} \xi^{\E{3};E,\unitvec{k}_{\vec{n}}}_{k_{\vec{n}},\ell m,\lambda}\,.
\end{empheq}
The spherical-harmonic space covariance matrix has the form \eqref{eqn:HarmonicCovariance}.

\subsection{\E{4}: Third-turn space}
\label{secn:topologyE4}

\textit{Properties}: This manifold is compact, orientable, inhomogeneous, and anisotropic. \\

\noindent \textit{Generators}: In general the generators of \E{4} can be written as
\begin{align}
    \label{eqn:E4generalT}
    & \mat{M}^{\E{4}}_A = \identity, \quad \mat{M}^{\E{4}}_B = \mat{R}_{\unitvec{z}}(2\pi/3) =
        \begin{pmatrix}
        -1/2 & -\sqrt{3}/2 & 0\\
        \sqrt{3}/2 & -1/2 & 0\\
        0 & \hphantom{-}0 & 1
        \end{pmatrix} ,
        \quad \mbox{with}
        \nonumber \\
    &  \vec{T}^{\E{4}}_{\A{1}} = L_{A} \begin{pmatrix} 1 \\ 0 \\ 0 \end{pmatrix}, \quad
    \vec{T}^{\E{4}}_{\A{2}} = L_{A} \begin{pmatrix} -1/2 \\ \sqrt{3}/2 \\ 0 \end{pmatrix}, \quad
    \vec{T}^{\E{4}}_{B} = L_B \begin{pmatrix} \cos\beta\cos\gamma \\ \cos\beta\sin\gamma \\ \sin\beta \end{pmatrix} .
\end{align}
In section $3.4$ of \rcite{COMPACT:2023rkp}, we presented a set of conditions on the parameters of \E{4} ($L_{A}$, $L_B$, $\beta$, and $\gamma$) that avoids double-counting identical manifolds with different generator choices.
\\
\\
\noindent \textit{Associated \E{1}}: In addition to $\vec{T}^{\E{4}}_1\equiv\vec{T}^{\E{4}}_{\A{1}}$ and $\vec{T}^{\E{4}}_2\equiv\vec{T}^{\E{4}}_{\A{2}}$, a third independent translation follows from $(\mat{M}^{\E{4}}_B)^3=\identity$:
\begin{equation}
    g^{\E{4}}_3 \equiv (g^{\E{4}}_B )^3:\vec{x}\to \vec{x} + \vec{T}^{\E{4}}_3,
\end{equation}
for
\begin{equation}
    \label{eqn:E4assocE1}
    \vec{T}^{\E{4}}_3 \equiv \begin{pmatrix} 0 \\ 0 \\ 3 L_{Bz} \end{pmatrix}
      = 3 L_B \begin{pmatrix} 0 \\ 0 \\ \sin\beta \end{pmatrix}.
\end{equation}
\\
\noindent \textit{Volume}:
\begin{equation}
    \label{eqn:VE4}
    V_{\E{4}}= \frac{1}{3}
    \vert(\vec{T}^{\E{4}}_{{1}}\times\vec{T}^{\E{4}}_{{2}})\cdot\vec{T}^{\E{4}}_{3}\vert 
    = \frac{\sqrt{3}}{2} L_A^2 L_B |\sin\beta|.
\end{equation}

\subsubsection{Eigenmodes and correlation matrices of \E{4}}
\label{secn:eigenmodesE4}
The eigenspectrum and eigenmodes of the third-turn space can be derived similarly to those from above.
For \E{4}, the discretization condition \eqref{eqn:associatedhomogeneousdiscretization}, which follows from the translation vectors $\vec{T}^{\E{4}}_j$, yields the components of the allowed wavevectors,

\begin{equation}
    (\vec{k}_{\vec{n}})_{x} = \frac{2\pi n_1}{L_A}, \quad
    (\vec{k}_{\vec{n}})_{y} = \frac{2\pi }{\sqrt{3} L_A}(n_1 + 2n_2), \quad
    (\vec{k}_{\vec{n}})_{z} = \frac{2\pi n_3}{3 L_B \sin\beta}.
\end{equation}

As in the case of \E{2}, the eigenmodes of \E{4} include linear combinations of \E{1} eigenmodes, as $(\mat{M}^{\E{4}}_B)^N \vec{k}_{\vec{n}} = \vec{k}_{\vec{n}}$ admits multiple solutions for the minimum positive $N$, depending on $\vec{k}_{\vec{n}}$.  
Here we have:
\begin{description}
    \item[\textbf{$N=1$ eigenmodes: }]  $\vec{k}_{\vec{n}}=\transpose{(0,0,(\vec{k}_{\vec{n}})_z)}$, 
      i.e., $\vec{n}=(0,0,n_3)$, 
        $n_3 = 3m+\lambda$, where $ m \in \integers$, with
    \begin{empheq}[box=\fbox]{align}
     \Upsilon^{\E{4}}_{ij,(0,0,n_3)}(\vec{x},\lambda) = \mathrm{e}^{i\vec{k}_{\vec{n}}\cdot (\vec{x}-\vec{x}_0))} e_{ij}(\unitvec{k}_{\vec{n}},\lambda), 
    \end{empheq}
    \item[\textbf{$N=3$ eigenmodes: }] $((\vec{k}_{\vec{n}})_{x}, (\vec{k}_{\vec{n}})_{y}) \neq (0,0)$, 
        i.e., $(n_1,n_2) \neq (0,0)$, with
    \begin{empheq}[box=\fbox]{align}
       \Upsilon^{\E{4}}_{ij,\vec{n}}(\vec{x},\lambda) =
           \frac{1}{\sqrt{3}} \sum_{a=0}^2 e_{ij}\left(\left[(\mat{M}^{\E{4}}_B)^T\right]^a\unitvec{k}_n,\lambda\right)
            \mathrm{e}^{i\vec{k}_{\vec{n}}\cdot \left( (\mat{M}^{\E{4}}_B)^a(\vec{x}-\vec{x}_0)+\mat{M}^{\E{4}}_{0a}\vec{T}^{\E{4}}_B \right)} , 
    \end{empheq}
\end{description}
and $\mat{M}^{\E{4}}_{0a}$ defined in \eqref{eqn:M0jdef}. For the $N = 1$ and $N = 3$, we have $\Phi^{\E{4}}_{\vec{k}_{\vec{n}}} = 1$, as no additional phase factors arise to simplify the correlations further.
As in \E{2}, the cyclic properties of $\mat{M}^{\E{4}}_B$ would lead to repeated counting of eigenmodes if all $n_1\in\integers$ and $n_2\in\integers$ were included.
To avoid this, we define two sets of allowed modes, now for $N=1$ and $N=3$,
\begin{empheq}[box=\fbox]{align}
    {}_\lambda\setN^{\E{4}}_1 &= \{(0,0, 3m + \lambda)| m \in \integers\}, \mbox{ for } \lambda \in \{-2, 2\} \nonumber \\
    \setN^{\E{4}}_3 &= \{(n_1,n_2,n_3)| n_1 \in \integers^{\neq 0}, n_2 \in \integers, n_1 n_2 \geq 0,  n_3 \in \integers\},  \\
    \setN^{\E{4}} &= {}_\lambda\setN^{\E{4}}_1 \cup \setN^{\E{4}}_3 . \nonumber
\end{empheq}
In this case, not only is ${}_\lambda\setN^{\E{4}}_1$ different from its scalar counterpart, but as it is apparent from the employed notation, the set of allowed wavevectors is different for each of the helicities. Using these, the helicity-dependent Fourier-mode correlation matrix can be written as
\begin{empheq}[box=\fbox]{align}
\label{eqn:E3FourierCovarianceStandardConvention}
     C^{\E{4};XY}_{iji'j', \vec{k}_{\vec{n}} \vec{k}_{\vecnp}, \lambda \lambda'}&
     = {} V_{\E{4}} \frac{\pi^2}{2k^3_{\vec{n}}}
        \Tpower(k_{\vec n})\Delta^X(k_{\vec{n}})\DeltaYstar(k_{\vec{n}})
        \mathrm{e}^{i(\vec{k}_{\vecnp}-\vec{k}_{\vec{n}})\cdot\vec{x}_0}  \Kdelta_{\lambda \lambda'}\ \times \\
        & \times \mathcal{E}_{iji'j'}(\unitvec{k}_{\vec{n}}, \unitvec{k}_{\vec{n}'}, \lambda) \left[
            \sum_{\vec{\tilde{n}} \in {}_\lambda\setN^{\E{4}}_1}
                 \Kdelta_{\vec{k}_{\vec{n}}\vec{k}_{\vec{\tilde{n}}}}
                 \Kdelta_{\vec{k}_{\vecnp}\vec{k}_{\vec{\tilde{n}}}}  + \frac{1}{3}
                \sum_{\vec{\tilde{n}}\in \setN^{\E{4}}_3}
                \sum_{a=0}^2\sum_{b=0}^2 
                 \right. \nonumber\\
        & \left. \qquad \qquad \qquad \quad 
        \mathrm{e}^{i\vec{k}_{\vec{\tilde{n}}}\cdot(\vec{T}^{(a)}-\vec{T}^{(b)})}
             \Kdelta_{\vec{k}_{\vec{n}}([(\mat{M}^{\E{4}}_B){}^T]^a\vec{k}_{\vec{\tilde{n}}})}
             \Kdelta_{\vec{k}_{\vecnp}([(\mat{M}^{\E{4}}_B){}^T]^b\vec{k}_{\vec{\tilde{n}}})} 
         \right] ,
         \nonumber
\end{empheq}
where $V_{\E{4}}$ is given in \eqref{eqn:VE4}, $\vec{T}^{(0)} \equiv \vec{0}$, and $\vec{T}^{(a)} \equiv \mat{M}^{\E{4}}_{0a} \vec{T}^{\E{4}}_B$ for $a \in \{ 1,2 \}$.

Also as in \E{2}, we can use the rotation properties of the spin-weighted spherical harmonics along with the fact that $\mat{M}^{\E{4}}_B$ is a rotation around the $z$-axis by $2\pi/3$ to note that
\begin{equation}
    {}_{\lambda}Y_{\ell -m}\! \left(\left[(\mat{M}^{\E{4}}_B){}^T \right]^j \unitvec{k}_{\vec{n}}\right) = \mathrm{e}^{-i 2 m j \pi/3} \ {}_{\lambda}Y_{\ell -m}  (\unitvec{k}_{\vec{n}})\,.
\end{equation}
This gives for the eigenmodes in the harmonic basis
\begin{empheq}[box=\fbox]{align}
    \xi^{\E{4};T,\unitvec{k}_{\vec{n}}}_{k_{\vec{n}},\ell m,\lambda} 
        &\equiv i^{\ell} \sqrt{\frac{2\pi^2(\ell + 2)!}{(\ell - 2)!}} \ {}_{-\lambda}Y_{\ell m} (\unitvec{k}_{\vec{n}}) \  \mathrm{e}^{-i \vec{k}_{\vec{n}} \cdot \vec{x}_0} , \quad \mbox{for } \vec{n}\in {}_\lambda\setN^{\E{4}}_1\\
    \xi^{\E{4};T,\unitvec{k}_{\vec{n}}}_{k_{\vec{n}},\ell m,\lambda}
        & \equiv \frac{i^{\ell}}{\sqrt{3}} \sqrt{\frac{2\pi^2(\ell + 2)!}{(\ell - 2)!}} \ {}_{-\lambda}Y_{\ell m} (\unitvec{k}_{\vec{n}}) \times \nonumber \\
        & \qquad\qquad\qquad\times \sum_{j=0}^{2} \mathrm{e}^{-i 2 m j \pi/3} \mathrm{e}^{-i \vec{k}_n\cdot \left[ (\mat{M}^{\E{4}}_B)^j \vec{x}_0- \mat{M}^{\E{4}}_{0j} \vec{T}^{\E{4}}_B \right]}, \quad \mbox{for } \vec{n} \in \setN^{\E{4}}_3 \nonumber
\end{empheq}

\begin{empheq}[box=\fbox]{align}
    \xi^{\E{4};E,\unitvec{k}_{\vec{n}}}_{k_{\vec{n}},\ell m,\lambda} 
        &\equiv -i^{\ell} \sqrt{2 \pi^2} \ {}_{-\lambda}Y_{\ell m} (\unitvec{k}_{\vec{n}}) \  \mathrm{e}^{-i \vec{k}_{\vec{n}} \cdot \vec{x}_0} , \quad \mbox{for } \vec{n}\in {}_\lambda\setN^{\E{4}}_1\\
    \xi^{\E{4};E,\unitvec{k}_{\vec{n}}}_{k_{\vec{n}},\ell m,\lambda}
        & \equiv -\frac{i^{\ell}}{\sqrt{3}} \ \sqrt{2 \pi^2} \ {}_{-\lambda}Y_{\ell m} (\unitvec{k}_{\vec{n}})\times \nonumber \\
        & \qquad\qquad\qquad \times  \sum_{j=0}^{2}  \mathrm{e}^{-i 2 m j \pi/3} \mathrm{e}^{-i \vec{k}_n\cdot \left[ (\mat{M}^{\E{4}}_B)^j \vec{x}_0- \mat{M}^{\E{4}}_{0j} \vec{T}^{\E{4}}_B \right]}, \quad \mbox{for } \vec{n} \in \setN^{\E{4}}_3 \nonumber
\end{empheq}
and
\begin{empheq}[box=\fbox]{equation}
\xi^{\E{4};B,\unitvec{k}_{\vec{n}}}_{k_{\vec{n}},\ell m,\lambda} = -\frac{\lambda}{2} \xi^{\E{4};E,\unitvec{k}_{\vec{n}}}_{k_{\vec{n}},\ell m,\lambda}\,.
\end{empheq}
The spherical-harmonic space covariance matrix has the form \eqref{eqn:HarmonicCovariance}.

\subsection{\E{5}: Sixth-turn space}
\label{secn:topologyE5}

\textit{Properties}: This manifold is compact, orientable, inhomogeneous, and anisotropic. \\

\noindent \textit{Generators}: In general the generators of \E{5} can be written as
\begin{align}
    \label{eqn:E5generalT}
    & \mat{M}^{\E{5}}_A = \identity, \quad \mat{M}^{\E{5}}_B = \mat{R}_{\unitvec{z}}(\pi/3) =
        \begin{pmatrix}
        1/2 & -\sqrt{3}/2 & 0\\
        \sqrt{3}/2 & 1/2 & 0\\
        0 & \hphantom{-}0 & 1
        \end{pmatrix} ,
        \quad \mbox{with}
        \nonumber \\
    &  \vec{T}^{\E{5}}_{\A{1}} = L_{A} \begin{pmatrix} 1 \\ 0 \\ 0 \end{pmatrix}, \quad
    \vec{T}^{\E{5}}_{\A{2}} = L_{A} \begin{pmatrix} -1/2 \\ \sqrt{3}/2 \\ 0 \end{pmatrix}, \quad
    \vec{T}^{\E{5}}_{B} = L_B \begin{pmatrix} \cos\beta\cos\gamma \\ \cos\beta\sin\gamma \\ \sin\beta \end{pmatrix} .
\end{align}
In section $3.5$ of \rcite{COMPACT:2023rkp}, we presented a set of conditions on the parameters of \E{5} ($L_{A}$, $L_B$, $\beta$, and $\gamma$) that avoids double-counting identical manifolds with different generator choices.
\\
\\
\noindent \textit{Associated \E{1}}: In addition to $\vec{T}^{\E{5}}_1\equiv\vec{T}^{\E{5}}_{\A{1}}$ and $\vec{T}^{\E{5}}_2\equiv\vec{T}^{\E{5}}_{\A{2}}$, a third independent translation follows from $(\mat{M}^{\E{5}}_B)^6=\identity$:
\begin{equation}
    g^{\E{5}}_3 \equiv (g^{\E{5}}_B )^6:\vec{x}\to \vec{x} + \vec{T}^{\E{5}}_3,
\end{equation}
for
\begin{equation}
    \label{eqn:E5assocE1}
    \vec{T}^{\E{5}}_3 \equiv \begin{pmatrix} 0 \\ 0 \\ 6 L_{Bz} \end{pmatrix}
      = 6 L_B \begin{pmatrix} 0 \\ 0 \\ \sin\beta \end{pmatrix}.
\end{equation}
\\
\noindent \textit{Volume}:
\begin{equation}
    \label{eqn:VE5}
    V_{\E{5}}= \frac{1}{6}
    \vert(\vec{T}^{\E{5}}_{{1}}\times\vec{T}^{\E{5}}_{{2}})\cdot\vec{T}^{\E{5}}_{3}\vert 
    = \frac{\sqrt{3}}{2} L_A^2 L_B |\sin\beta|.
\end{equation}

\subsubsection{Eigenmodes and correlation matrices of \E{5}}

The eigenspectrum and eigenmodes of the third-turn space can be derived similarly to those from above, in particular, it is very similar to \E{4}.
For \E{5}, the discretization condition \eqref{eqn:associatedhomogeneousdiscretization} from the translation vectors $\vec{T}^{\E{5}}_j$ yields the components of the allowed wavevectors,
\begin{equation}
    (\vec{k}_{\vec{n}})_{x} = \frac{2\pi n_1}{L_A}, \quad
    (\vec{k}_{\vec{n}})_{y} = \frac{2\pi}{\sqrt{3} L_A}(n_1 + 2n_2), \quad
    (\vec{k}_{\vec{n}})_{z} = \frac{2\pi n_3}{6 L_B \sin\beta}.
\end{equation}

As in \E{2}, the eigenmodes of \E{5} can include linear combinations of \E{1} eigenmodes since $(\mat{M}^{\E{5}}_B)^N \vec{k}_{\vec{n}} = \vec{k}_{\vec{n}}$ has more than one solution for the minimum positive $N$, depending on $\vec{k}_{\vec{n}}$.
Here we have:
\begin{description}
   \item[\textbf{$N=1$ eigenmodes: }]  $\vec{k}_{\vec{n}}=\transpose{(0,0,(\vec{k}_{\vec{n}})_z)}$, 
      i.e., $\vec{n}=(0,0,n_3)$, 
        $n_3 = 6m+\lambda$, where $ m \in \integers$, with
    \begin{empheq}[box=\fbox]{align}
    \Upsilon^{\E{5}}_{ij,(0,0,n_3)}(\vec{x},\lambda) = \mathrm{e}^{i\vec{k}_{\vec{n}}\cdot (\vec{x}-\vec{x}_0))} e_{ij}(\unitvec{z},\lambda), 
    \end{empheq}
    \item[\textbf{$N=6$ eigenmodes: }] $((\vec{k}_{\vec{n}})_{x}, (\vec{k}_{\vec{n}})_{y}) \neq (0,0)$, 
        i.e., $(n_1,n_2) \neq (0,0)$, with
    \begin{empheq}[box=\fbox]{align}
        \Upsilon^{\E{5}}_{ij,\vec{n}}(\vec{x},\lambda) =&
            \frac{\mathrm{e}^{-i\vec{k}_{\vec{n}}\cdot \mat{M}^{\E{5}}_{03}\vec{T}^{\E{5}}_B/2}}{\sqrt{6}} \sum_{a=0}^5 e_{ij}\left(\left[(\mat{M}^{\E{5}}_B)^T\right]^a\unitvec{k}_n,\lambda\right)\\
            & \qquad \qquad \qquad \qquad \qquad \qquad \qquad
            \mathrm{e}^{i\vec{k}_{\vec{n}}\cdot \left((\mat{M}^{\E{5}}_B)^a(\vec{x}-\vec{x}_0)+\mat{M}^{\E{5}}_{0a}\vec{T}^{\E{5}}_B \right)}, \nonumber
    \end{empheq}
\end{description}
and $\mat{M}^{\E{5}}_{0a}$ defined in \eqref{eqn:M0jdef}. For the $N = 6$ modes, we choose $\Phi_{\vec{k}_{\vec{n}}}^{\E{5}} = -\vec{k}_{\vec{n}}\cdot \mat{M}^{\E{5}}_{03}\vec{T}^{\E{5}}_B/2$ in \eqref{eqn:generaleigenmodeformula}---this will simplify the reality condition of the field. 
As in \E{2}, the cyclic properties of $\mat{M}^{\E{5}}_B$ would lead to overcounting of eigenmodes if all $n_1\in\integers$ and $n_2\in\integers$ were included.
To avoid this, we define two sets of allowed modes, now for $N=1$ and $N=6$,
\begin{empheq}[box=\fbox]{align}
    {}_\lambda\setN^{\E{5}}_1 &= \{(0,0,n_3 = 6m + \lambda)| m \in \integers\} \mbox{ for } \lambda \in \{2, -2\}  , \nonumber \\
    \setN^{\E{5}}_6 &= \{(n_1,n_2,n_3)| n_1 \in \integers^{> 0}, n_2 \in \integers^{\geq 0}, n_3 \in \integers\}
    , \\
    \setN^{\E{5}} &= {}_\lambda\setN^{\E{5}}_1 \cup \setN^{\E{5}}_6 . \nonumber
\end{empheq}
Using these, the helicity-dependent Fourier-mode correlation matrix can be written as
\begin{empheq}[box=\fbox]{align}
\label{eqn:E5FourierCovarianceStandardConvention}
     C^{\E{5};XY}_{iji'j', \vec{k}_{\vec{n}} \vec{k}_{\vecnp}, \lambda \lambda'}&
     = {} V_{\E{5}} \frac{\pi^2}{2 k^3_{\vec{n}}}
        \Tpower(k_{\vec n})\Delta^X(k_{\vec{n}})\DeltaYstar(k_{\vec{n}})
        \mathrm{e}^{i(\vec{k}_{\vecnp}-\vec{k}_{\vec{n}})\cdot\vec{x}_0}  \Kdelta_{\lambda \lambda'} \times\\
        & \times \mathcal{E}_{iji'j'}(\unitvec{k}_{\vec{n}}, \unitvec{k}_{\vec{n}'}, \lambda) \left[
            \sum_{\vec{\tilde{n}} \in {}_\lambda\setN^{\E{5}}_1}
                 \Kdelta_{\vec{k}_{\vec{n}}\vec{k}_{\vec{\tilde{n}}}}
                 \Kdelta_{\vec{k}_{\vecnp}\vec{k}_{\vec{\tilde{n}}}}  + \frac{1}{6}
                \sum_{\vec{\tilde{n}}\in \setN^{\E{5}}_6}
                \sum_{a=0}^5\sum_{b=0}^5 
                 \right. \nonumber\\
        & \left. \qquad \qquad \qquad \quad 
        \mathrm{e}^{i\vec{k}_{\vec{\tilde{n}}}\cdot(\vec{T}^{(a)}-\vec{T}^{(b)})}
             \Kdelta_{\vec{k}_{\vec{n}}([(\mat{M}^{\E{5}}_B){}^T]^a\vec{k}_{\vec{\tilde{n}}})}
             \Kdelta_{\vec{k}_{\vecnp}([(\mat{M}^{\E{5}}_B){}^T]^b\vec{k}_{\vec{\tilde{n}}})} 
         \right] ,
         \nonumber
\end{empheq}
where $V_{\E{5}}$ is given in \eqref{eqn:VE5}, $\vec{T}^{(0)} \equiv \vec{0}$, and $\vec{T}^{(a)} \equiv \mat{M}^{\E{5}}_{0a} \vec{T}^{\E{5}}_B$ for $a \in \{ 1,\ldots, 5 \}$.

Also as in \E{2}, we can use the rotation properties of the spherical harmonics along with the fact that $\mat{M}^{\E{5}}_B$ is a rotation around the $z$-axis by $\pi/3$ to note that
\begin{equation}
    {}_{\lambda}Y_{\ell -m}\!\left(\left[(\mat{M}^{\E{5}}_B){}^T \right]^j \unitvec{k}_{\vec{n}}\right) = \mathrm{e}^{-i m j \pi/3} \ {}_{\lambda}Y_{\ell -m}  (\unitvec{k}_{\vec{n}}).
\end{equation}
This gives for the eigenmodes in the harmonic basis
\begin{empheq}[box=\fbox]{align}
    \xi^{\E{5};T,\unitvec{k}_{\vec{n}}}_{k_{\vec{n}},\ell m,\lambda} 
        &\equiv i^{\ell} \sqrt{\frac{2\pi^2(\ell + 2)!}{(\ell 
           - 2)!}} \ {}_{-\lambda}Y_{\ell m} (\unitvec{k}_{\vec{n}}) \  \mathrm{e}^{-i \vec{k}_{\vec{n}} \cdot \vec{x}_0} , \quad \mbox{for } \vec{n}\in {}_{\lambda}\setN^{\E{5}}_1 \\
    \xi^{\E{5};T,\unitvec{k}_{\vec{n}}}_{k_{\vec{n}},\ell m,\lambda}
        & \equiv \frac{i^{\ell}}{\sqrt{6}} \sqrt{\frac{2\pi^2(\ell + 2)!}{(\ell - 2)!}} \ {}_{-\lambda}Y_{\ell m} (\unitvec{k}_{\vec{n}}) \ \mathrm{e}^{-i\vec{k}_{\vec{n}}\cdot \mat{M}^{\E{5}}_{03}\vec{T}^{\E{5}}_B/2} \times \nonumber \\
        & \qquad\qquad\qquad\times\sum_{j=0}^{5} \mathrm{e}^{-i m j \pi/3} \mathrm{e}^{-i \vec{k}_n\cdot \left[ (\mat{M}^{\E{5}}_B)^j \vec{x}_0- \mat{M}^{\E{5}}_{0j} \vec{T}^{\E{5}}_B \right]}, \quad \mbox{for } \vec{n} \in \setN^{\E{5}}_6 \nonumber
\end{empheq}

\begin{empheq}[box=\fbox]{align}
    \xi^{\E{5};E,\unitvec{k}_{\vec{n}}}_{k_{\vec{n}},\ell m,\lambda} 
        &\equiv -i^{\ell} \ \sqrt{2 \pi^2} \ {}_{-\lambda}Y_{\ell m} (\unitvec{k}_{\vec{n}}) \  \mathrm{e}^{-i \vec{k}_{\vec{n}} \cdot \vec{x}_0} , \quad \mbox{for } \vec{n}\in {}_{\lambda}\setN^{\E{5}}_1\\
    \xi^{\E{5};E,\unitvec{k}_{\vec{n}}}_{k_{\vec{n}},\ell m,\lambda}
        & \equiv-\frac{i^{\ell}}{\sqrt{6}} \ \sqrt{2 \pi^2} \ {}_{-\lambda}Y_{\ell m} (\unitvec{k}_{\vec{n}}) \ \mathrm{e}^{-i\vec{k}_{\vec{n}}\cdot \mat{M}^{\E{5}}_{03}\vec{T}^{\E{5}}_B/2} \times \nonumber \\
        & \qquad\qquad\qquad\times \sum_{j=0}^{5}  \mathrm{e}^{-i  m j \pi/3} \mathrm{e}^{-i \vec{k}_n\cdot \left[ (\mat{M}^{\E{5}}_B)^j \vec{x}_0- \mat{M}^{\E{5}}_{0j} \vec{T}^{\E{5}}_B \right]}, \quad \mbox{for } \vec{n} \in \setN^{\E{5}}_6 \nonumber
\end{empheq}
and
\begin{empheq}[box=\fbox]{equation}
\xi^{\E{5};B,\unitvec{k}_{\vec{n}}}_{k_{\vec{n}},\ell m,\lambda} = -\frac{\lambda}{2} \xi^{\E{5};E,\unitvec{k}_{\vec{n}}}_{k_{\vec{n}},\ell m,\lambda}\,.
\end{empheq}
The spherical-harmonic space covariance matrix has the form \eqref{eqn:HarmonicCovariance}.

\subsection{\E{6}: Hantzsche-Wendt space}
\label{secn:topologyE6}

\textit{Properties}: This manifold is compact, orientable, inhomogeneous, and anisotropic (for more information, see \rcite{Aurich:2014sea}). \\

\noindent \textit{Generators}: In general the generators of \E{6} can be written as
\begin{align}
    \label{eqn:E6generalT}
    & \mat{M}^{\E{6}}_{A} = 
    \begin{pmatrix}
        1 & \hphantom{-}0 & \hphantom{-}0\\
        0 & -1 & \hphantom{-}0\\
        0 & \hphantom{-}0 & -1
    \end{pmatrix}, \quad
        \mat{M}^{\E{6}}_{B} = 
    \begin{pmatrix}
        -1 & \hphantom{-}0 & \hphantom{-}0\\
        \hphantom{-}0 & \hphantom{-}1 & \hphantom{-}0\\
        \hphantom{-}0 & \hphantom{-}0 & -1
    \end{pmatrix} , \quad
        \mat{M}^{\E{6}}_{C} = 
    \begin{pmatrix}
        -1 & \hphantom{-}0 & \hphantom{-}0\\
        \hphantom{-}0 & -1 & \hphantom{-}0\\
        \hphantom{-}0 & \hphantom{-}0 & \hphantom{-}1
    \end{pmatrix}, \quad \mbox{with} \nonumber \\
    & \vec{T}^{\E{6}}_{A} = \begin{pmatrix} L_{Ax} \\ (r_y + \frac{1}{2})L_{By} \\ (r_z- \frac{1}{2})L_{Cz} \end{pmatrix} , \quad
    \vec{T}^{\E{6}}_{B} = \begin{pmatrix} (r_x- \frac{1}{2})L_{Ax} \\ L_{By} \\ (r_z + \frac{1}{2})L_{Cz} \end{pmatrix} , \quad
    \vec{T}^{\E{6}}_{C} = \begin{pmatrix} (r_x + \frac{1}{2})L_{Ax} \\ (r_y- \frac{1}{2})L_{By} \\ L_{Cz} \end{pmatrix} .
\end{align}
Here, we introduce an alternative representation by replacing the orientation angles with three additional parameters, $\{r_x, r_y, r_z\} \in (-1/2, 1/2]$. The standard (special origin, i.e., ``untilted'') form corresponds to setting $r_x = r_y = r_z = \frac{1}{2}$. In section $3.6$ of \rcite{COMPACT:2023rkp}, we presented a set of conditions on the parameters of \E{6} (i.e., $L_{A_x}$, $L_{B_y}$, and $L_{C_z}$) to avoid double-counting identical manifolds with different generator choices.
\\

\noindent \textit{Associated \E{1}}: Since $(\mat{M}^{\E{6}}_A)^2 = (\mat{M}^{\E{6}}_B)^2 = (\mat{M}^{\E{6}}_C)^2 = \identity$,
$(g^{\E{6}}_A)^2$, $(g^{\E{6}}_B)^2$, and $(g^{\E{6}}_C)^2$ are (independent) pure translations: 
\begin{align}
    \label{eqn:E6assocE1}
    g^{\E{6}}_1 & \equiv (g^{\E{6}}_A)^2: \vec{x} \to \vec{x} + \vec{T}^{\E{6}}_1, \nonumber \\
    g^{\E{6}}_2 & \equiv (g^{\E{6}}_B)^2: \vec{x} \to \vec{x} + \vec{T}^{\E{6}}_2, \\
    g^{\E{6}}_3 & \equiv (g^{\E{6}}_C)^2: \vec{x} \to \vec{x} + \vec{T}^{\E{6}}_3, \nonumber
\end{align}
for
\begin{equation}
    \vec{T}^{\E{6}}_1 \equiv \begin{pmatrix} 2 L_{Ax} \\ 0 \\ 0 \end{pmatrix}, \quad
    \vec{T}^{\E{6}}_2 \equiv \begin{pmatrix} 0 \\ 2 L_{By} \\ 0 \end{pmatrix}, \quad
    \vec{T}^{\E{6}}_3 \equiv \begin{pmatrix} 0 \\ 0 \\ 2 L_{Cz} \end{pmatrix}.
\end{equation}
\\
\noindent \textit{Volume}:
\begin{equation}
    \label{eqn:VE6}
    V_{\E{6}}= \frac{1}{4}
    \vert(\vec{T}^{\E{6}}_{{1}}\times\vec{T}^{\E{6}}_{{2}})\cdot\vec{T}^{\E{6}}_{3}\vert 
    = 2 \vert L_{Ax} L_{By} L_{Cz} \vert .
\end{equation}

\subsubsection{Eigenmodes and correlation matrices of \E{6}}

\label{secn:eigenmodesE6}

The eigenspectrum and eigenmodes of the Hantzsche-Wendt space can be determined in a manner analogous to those from above.
Complications arise from the fact that \E{6} contains rotations around multiple axes so a more careful discussion is warranted.
The discretization condition \eqref{eqn:associatedhomogeneousdiscretization} from the translation vectors $\vec{T}^{\E{6}}_j$ is still straightforward and leads to the components of the allowed wavevectors,
\begin{equation}
    (\vec{k}_{\vec{n}})_{x} = \frac{2\pi n_1}{2 L_{Ax}}, \quad
    (\vec{k}_{\vec{n}})_{y} = \frac{2\pi n_2}{2 L_{By}}, \quad
    (\vec{k}_{\vec{n}})_{z} = \frac{2\pi n_3}{2 L_{Cz}}.
\end{equation}

As in \E{2}, the eigenmodes of \E{6} can include linear combinations of \E{1} eigenmodes since $(\mat{M}^{\E{6}}_a)^{N_a} \vec{k}_{\vec{n}} = \vec{k}_{\vec{n}}$ has more than one solution for the minimum positive $N_a$, depending on $\vec{k}_{\vec{n}}$, for $a\in\{A, B, C\}$.
Since all the rotations are half turns, i.e., all $(\mat{M}^{\E{6}}_a)^2 = \identity$, there are linear combinations with the $N_a=1$ and $N_a=2$.
At first glance \cref{eqn:generaleigenmodeformula} seems to suggest that the eigenmodes with $N_a=1$ will be a linear combination of four eigenmodes of \E{1}.
However, since $\mat{M}^{\E{6}}_A \mat{M}^{\E{6}}_B = \mat{M}^{\E{6}}_C$ (and all permutations of $\{A, B, C\}$) along with the invariance of the eigenmodes under the group action \eqref{eqn:generaleigenmodeinvariance}, a linear combination of only two eigenmodes of \E{1} is required in this case.
The results of tensor eigenmodes are quite similar to the scalar eigenmodes, with an additional contribution of the tensor part. It is useful to examine how the helicity tensors transform under these generators, as demonstrated in \cref{eqn:helicityrotationtrans,eqn:helicityZroationtrans}:
\begin{align}
    \label{equ:E6 tensorTransformations}
    \mat{M}^{\E{6}}_A \cdot \mat{e}(\unitvec{x},\lambda)\cdot (\mat{M}^{\E{6}}_A)^T &= \mat{e}(\unitvec{x},\lambda), \nonumber\\
    \mat{M}^{\E{6}}_A \cdot \mat{e}(\unitvec{y},\lambda)\cdot (\mat{M}^{\E{6}}_A)^T &= \mat{e}(-\unitvec{y},\lambda), \nonumber\\
    \mat{M}^{\E{6}}_A \cdot \mat{e}(\unitvec{z},\lambda) \cdot(\mat{M}^{\E{6}}_A)^T &= \mat{e}(-\unitvec{z},\lambda), \nonumber\\
    \mat{M}^{\E{6}}_B \cdot \mat{e}(\unitvec{x},\lambda) \cdot(\mat{M}^{\E{6}}_B)^T &= \mat{e}(-\unitvec{x},\lambda), \nonumber\\
    \mat{M}^{\E{6}}_B \cdot \mat{e}(\unitvec{y},\lambda)\cdot (\mat{M}^{\E{6}}_B)^T &= \mat{e}(\unitvec{y},\lambda), \\
    \mat{M}^{\E{6}}_B \cdot \mat{e}(\unitvec{z},\lambda)\cdot (\mat{M}^{\E{6}}_B)^T &= \mat{e}(-\unitvec{z},\lambda), \nonumber\\
    \mat{M}^{\E{6}}_C \cdot \mat{e}(\unitvec{x},\lambda) \cdot(\mat{M}^{\E{6}}_C)^T &= \mat{e}(-\unitvec{x},\lambda), \nonumber\\
    \mat{M}^{\E{6}}_C \cdot \mat{e}(\unitvec{y},\lambda) \cdot(\mat{M}^{\E{6}}_C)^T &= \mat{e}(-\unitvec{y},\lambda), \nonumber\\
    \mat{M}^{\E{6}}_C \cdot \mat{e}(\unitvec{z},\lambda) \cdot(\mat{M}^{\E{6}}_C)^T &= \mat{e}(\unitvec{z},\lambda) . \nonumber
\end{align}
Based on this we can choose:
\begin{description}
    \item[\textbf{$N_A=1$ eigenmodes: }]  $\vec{k}_{\vec{n}}=\transpose{((\vec{k}_{\vec{n}})_x, 0, 0)}$, 
        i.e., $\vec{n}=(n_1,0,0)$, 
        $n_1 \in 2\integers^{> 0} $, with\footnote{Here, we replace $\unitvec{k}_{\vec{n}}$ with $\unitvec{x}$ and $-\unitvec{x}$ in the argument of $e_{ij}$. Furthermore, since $\vec{k}_{\vec{n}}$ has only $x$ components, the application of $M_B$ and $M_C$ results in the transformation $\vec{k}_{\vec{n}} \rightarrow -\vec{k}_{\vec{n}}$.}
    \begin{empheq}[box=\fbox]{align}
            \Upsilon^{\E{6}}_{ij,(n_1,0,0)}(\vec{x},\lambda) =& 
           \ \frac{\mathrm{e}^{-i \vec{k}_{\vec{n}} \cdot \vec{T}^{\E{6}}_B/2}}{\sqrt{2}} 
            \left[e_{ij}(\unitvec{x},\lambda) \mathrm{e}^{i\vec{k}_{\vec{n}} \cdot (\vec{x}-\vec{x}_0)} \ + \right.\nonumber\\
            & \left. \qquad \qquad \qquad \qquad
              + \ e_{ij}(-\unitvec{x},\lambda) \mathrm{e}^{-i\vec{k}_{\vec{n}} \cdot (\vec{x}-\vec{x}_0)} \mathrm{e}^{i\vec{k}_{\vec{n}}\cdot\vec{T}^{\E{6}}_B}  \right], 
    \end{empheq}
    \item[\textbf{$N_B=1$ eigenmodes: }]  $\vec{k}_{\vec{n}}=\transpose{(0,(\vec{k}_{\vec{n}})_y, 0)}$, 
        i.e., $\vec{n}=(0,n_2,0)$, 
        $n_2 \in  2\integers^{> 0} $, with \footnote{Here, we replace $\unitvec{k}_{\vec{n}}$ with $\unitvec{y}$ and $-\unitvec{y}$ in the argument of $e_{ij}$. Furthermore, since $\vec{k}_{\vec{n}}$ has only $y$ components, the application of $M_A$ and $M_C$ results in the transformation $\vec{k}_{\vec{n}} \rightarrow -\vec{k}_{\vec{n}}$.}
    \begin{empheq}[box=\fbox]{align}
          \Upsilon^{\E{6}}_{ij,(0,n_2,0)}(\vec{x},\lambda) = &
               \ \frac{\mathrm{e}^{-i \vec{k}_{\vec{n}} \cdot \vec{T}^{\E{6}}_C/2}}{\sqrt{2}} \left[e_{ij}(\unitvec{y},\lambda) \mathrm{e}^{i\vec{k}_{\vec{n}} \cdot (\vec{x}-\vec{x}_0)} \ + \right.\nonumber\\
            & \left. \qquad \qquad \qquad \qquad
              + \ e_{ij}(-\unitvec{y},\lambda) \mathrm{e}^{-i\vec{k}_{\vec{n}} \cdot (\vec{x}-\vec{x}_0)} \mathrm{e}^{i\vec{k}_{\vec{n}}\cdot\vec{T}^{\E{6}}_C}  \right], 
    \end{empheq}
    \item[\textbf{$N_C=1$ eigenmodes: }]  $\vec{k}_{\vec{n}}=\transpose{((0,0,(\vec{k}_{\vec{n}})_z)}$, 
        i.e., $\vec{n}=(0,0,n_3)$, 
        $n_3 \in  2\integers^{> 0} $, with\footnote{Here, we replace $\unitvec{k}_{\vec{n}}$ with $\unitvec{z}$ and $-\unitvec{z}$ in the argument of $e_{ij}$. Furthermore, since $\vec{k}_{\vec{n}}$ has only $z$ components, the application of $M_A$ and $M_B$ results in the transformation $\vec{k}_{\vec{n}} \rightarrow -\vec{k}_{\vec{n}}$.}
    \begin{empheq}[box=\fbox]{align}
        \Upsilon^{\E{6}}_{ij,(0,0,n_3)}(\vec{x},\lambda) = &
             \ \frac{\mathrm{e}^{-i \vec{k}_{\vec{n}} \cdot \vec{T}^{\E{6}}_A/2}}{\sqrt{2}} \left[e_{ij}(\unitvec{z},\lambda) \mathrm{e}^{i\vec{k}_{\vec{n}} \cdot (\vec{x}-\vec{x}_0)}  \ + \right.\nonumber\\
            & \left. \qquad \qquad \qquad \qquad
              + \ e_{ij}(-\unitvec{z},\lambda) \mathrm{e}^{-i\vec{k}_{\vec{n}} \cdot (\vec{x}-\vec{x}_0)} \mathrm{e}^{i\vec{k}_{\vec{n}}\cdot\vec{T}^{\E{6}}_A}  \right], 
    \end{empheq}
    \item[\textbf{$N_a=2$ eigenmodes: }] at most one component of $\vec{k}_{\vec{n}}$ zero, i.e., at most one of the $n_i$ equal to zero, with 
    \begin{empheq}[box=\fbox]{align}
        \Upsilon^{\E{6}}_{ij,\vec{n}}(\vec{x},\lambda) = &
             \ \frac{1}{\sqrt{4}} \left[ e_{ij}(\unitvec{k}_{\vec{n}},\lambda)
            \mathrm{e}^{i\vec{k}_{\vec{n}} \cdot (\vec{x}-\vec{x}_0)} \ + \right.\nonumber\\
            & \left. \quad + \
            \sum_{a \in\{A, B, C\}} e_{ij}\left((\mat{M}^{\E{6}}_a)^T\unitvec{k}_{\vec{n}},\lambda\right) \ \mathrm{e}^{i\vec{k}_{\vec{n}} \cdot \mat{M}^{\E{6}}_a (\vec{x}-\vec{x}_0)} \mathrm{e}^{i \vec{k}_{\vec{n}} \cdot \vec{T}^{\E{6}}_a} \right].
    \end{empheq}
\end{description}
For the $N_{a} = 1$ modes, we choose $\Phi_{\vec{k}_{\vec{n}}}^{\E{6}} = -\vec{k}_{\vec{n}} \cdot \vec{T}^{\E{2}}_a/2$ for $a \in \{A, B, C\}$ in \eqref{eqn:generaleigenmodeformula}---this will simplify the reality condition of the field.
The cyclic properties of the $\mat{M}^{\E{6}}_a$ would lead to repeated counting of eigenmodes.
To avoid this, we define the sets of allowed modes as
\begin{empheq}[box=\fbox]{align}
    \setN^{\E{6}}_{1A} &= \{(n_1,0,0)| n_1\in 2\integers^{> 0} \} , \nonumber \\
    \setN^{\E{6}}_{1B} &= \{(0,n_2,0)| n_2\in  2\integers^{> 0}  \} , \nonumber \\
    \setN^{\E{6}}_{1C} &= \{(0,0,n_3)| n_3\in  2\integers^{> 0}  \} , \\
    \setN^{\E{6}}_2 &= \{(n_1,n_2,n_3)| n_1 \in \integers^{> 0}, n_2 \in \integers^{> 0}, n_3 \in \integers\} \nonumber \\
    & {} \qquad \cup \{(0, n_2, n_3)| n_2 \in \integers^{>0}, n_3 \in \integers^{>0} \} \nonumber \\
    & {} \qquad \cup \{(n_1, 0, n_3)| n_1 \in \integers^{>0}, n_3 \in \integers^{>0} \}, \nonumber \\
    \setN^{\E{6}} &= \setN^{\E{6}}_{1A} \cup \setN^{\E{6}}_{1B} \cup \setN^{\E{6}}_{1C} \cup \setN^{\E{6}}_2 . \nonumber
\end{empheq}
Using these, the helicity-dependent Fourier-mode correlation matrix can be written as
\begin{empheq}[box=\fbox]{align}
\label{eqn:E6FourierCovarianceStandardConvention}
      & C^{\E{6};XY}_{iji'j', \vec{k}_{\vec{n}} \vec{k}_{\vecnp}, \lambda \lambda'}
     = {} V_{\E{6}} \frac{\pi^2}{2 k^3_{\vec{n}}}
        \Tpower(k_{\vec n})\Delta^X(k_{\vec{n}})\DeltaYstar(k_{\vec{n}}) \times \nonumber \\
        & \qquad \qquad \qquad \qquad \qquad \qquad \qquad \qquad \quad
            \times \mathrm{e}^{i(\vec{k}_{\vecnp}-\vec{k}_{\vec{n}})\cdot\vec{x}_0}  \Kdelta_{\lambda \lambda'} \mathcal{E}_{iji'j'}(\unitvec{k}_{\vec{n}}, \unitvec{k}_{\vec{n}'}, \lambda) \times \\
        & \qquad \quad \times \left[
            \frac{1}{2} \sum_{\vec{\tilde{n}} \in \setN^{\E{6}}_{1A}}
                \sum_{a,b\in\{0, B\}} \mathrm{e}^{i\vec{k}_{\vec{n}} \cdot (\vec{T}^{(a)} - \vec{T}^{(b)})}
                 \Kdelta_{\vec{k}_{\vec{n}}((\mat{M}^{\E{6}}_a){}^T \vec{k}_{\tilde{\vec{n}}})}
                 \Kdelta_{\vec{k}_{\vecnp}((\mat{M}^{\E{6}}_b){}^T \vec{k}_{\tilde{\vec{n}}})} + {}
            \right. \nonumber \\
        & \qquad \qquad \quad  + \frac{1}{2} \sum_{\vec{\tilde{n}} \in \setN^{\E{6}}_{1B}}
                \sum_{a,b\in\{0, C\}} \mathrm{e}^{i\vec{k}_{\vec{n}} \cdot (\vec{T}^{(a)} - \vec{T}^{(b)})}
                 \Kdelta_{\vec{k}_{\vec{n}}((\mat{M}^{\E{6}}_a){}^T \vec{k}_{\tilde{\vec{n}}})}
                 \Kdelta_{\vec{k}_{\vecnp}((\mat{M}^{\E{6}}_b){}^T \vec{k}_{\tilde{\vec{n}}})} + {} \nonumber \\
        &\qquad  \qquad \quad+ \frac{1}{2} \sum_{\vec{\tilde{n}} \in \setN^{\E{6}}_{1C}}
                \sum_{a,b\in\{0, A\}} \mathrm{e}^{i\vec{k}_{\vec{n}} \cdot (\vec{T}^{(a)} - \vec{T}^{(b)})}
                 \Kdelta_{\vec{k}_{\vec{n}}((\mat{M}^{\E{6}}_a){}^T \vec{k}_{\tilde{\vec{n}}})}
                 \Kdelta_{\vec{k}_{\vecnp}((\mat{M}^{\E{6}}_b){}^T \vec{k}_{\tilde{\vec{n}}})} + {} \nonumber \\
        & \qquad \qquad \quad \left.
                 + \frac{1}{4}
                \sum_{\vec{\tilde{n}}\in \setN^{\E{6}}_2}
                \sum_{a,b\in\{0, A, B, C\}} \mathrm{e}^{i\vec{k}_{\vec{n}} \cdot (\vec{T}^{(a)} - \vec{T}^{(b)})}
                 \Kdelta_{\vec{k}_{\vec{n}}((\mat{M}^{\E{6}}_a){}^T \vec{k}_{\tilde{\vec{n}}})}
                 \Kdelta_{\vec{k}_{\vecnp}((\mat{M}^{\E{6}}_b){}^T \vec{k}_{\tilde{\vec{n}}})} 
         \right],
         \nonumber
\end{empheq}
where $V_{\E{6}}$ is given in \eqref{eqn:VE6}, $\vec{T}^{(0)} \equiv \vec{0}$, and $\vec{T}^{(a)} \equiv \vec{T}^{\E{6}}_a$ for $a \in \{ A, B, C \}$.

The rotation properties of the spin-weighted spherical harmonics can again be used to simplify the eigenmodes of \E{6} in the harmonic basis.
Though there are multiple axes of rotation, the fact that they are half turns allows direct computation to show that
\begin{align}
    {}_{-\lambda}Y_{\ell m}\! \left((\mat{M}^{\E{6}}_A){}^T \unitvec{k}_{\vec{n}}\right) &= (-1)^{\ell + m} \ {}_{\lambda}Y^*_{\ell m}    (\unitvec{k}_{\vec{n}}), \nonumber \\
    {}_{-\lambda}Y_{\ell m}\! \left((\mat{M}^{\E{6}}_B){}^T \unitvec{k}_{\vec{n}}\right) &= (-1)^{\ell} \ 
      {}_{\lambda}Y^*_{\ell m}  (\unitvec{k}_{\vec{n}}),\\
    {}_{-\lambda}Y_{\ell m}\! \left((\mat{M}^{\E{6}}_C){}^T \unitvec{k}_{\vec{n}}\right) &= (-1)^m \ {}_{-\lambda}Y_{\ell m}    (\unitvec{k}_{\vec{n}})\,. \nonumber
\end{align}
With these, the eigenmodes in the harmonic basis can be written in a number of useful forms.
We have general expressions patterned after the expressions given for \E{1}--\E{5},
\begin{empheq}[box=\fbox]{align}
 \xi^{\E{6};T,\unitvec{k}_{\vec{n}}}_{k_{\vec{n}},\ell m,\lambda} &= \frac{i^{\ell}}{\sqrt{2}} 
                    \sqrt{\frac{2\pi^2(\ell + 2)!}{(\ell - 2)!}} \ 
                    {}_{-\lambda}Y_{\ell m} (\unitvec{k}_{n_x}) \ \mathrm{e}^{-i \vec{k}_{\vec{n}} \cdot \vec{T}^{\E{6}}_B/2} \times \\ & \qquad \qquad \qquad \qquad \quad \times \left[ \mathrm{e}^{-i \vec{k}_n\cdot \vec{x}_0} 
                    + (-1)^{m} \mathrm{e}^{i\vec{k}_n\cdot \vec{x}_0} \mathrm{e}^{i \vec{k}_n \cdot \vec{T}^{\E{6}}_B}\right],
                    \ \mbox{ for } \vec{n}\in \setN^{\E{6}}_A \nonumber \\
 \nonumber \xi^{\E{6};T,\unitvec{k}_{\vec{n}}}_{k_{\vec{n}},\ell m,\lambda} 
         &= \frac{i^{\ell}}{\sqrt{2}} \sqrt{\frac{2\pi^2(\ell + 2)!}{(\ell - 2)!}} \ 
                    {}_{-\lambda}Y_{\ell m} (\unitvec{k}_{n_y}) \ \mathrm{e}^{-i \vec{k}_{\vec{n}} \cdot \vec{T}^{\E{6}}_C/2} \times \\ & \qquad \qquad \qquad \qquad \quad \times \left[ \mathrm{e}^{-i \vec{k}_n\cdot \vec{x}_0} 
                    +(-1)^{m} \mathrm{e}^{i \vec{k}_n\cdot \vec{x}_0} \mathrm{e}^{i \vec{k}_n \cdot \vec{T}^{\E{6}}_C}\right],
                    \mbox{ for } \vec{n}\in \setN^{\E{6}}_B \ \nonumber \\
 \nonumber \xi^{\E{6};T,\unitvec{k}_{\vec{n}}}_{k_{\vec{n}},\ell m,\lambda} &= \frac{i^{\ell}}{\sqrt{2}}\sqrt{\frac{2\pi^2(\ell + 2)!}{(\ell - 2)!}} \ \mathrm{e}^{-i \vec{k}_{\vec{n}} \cdot \vec{T}^{\E{6}}_A/2}
                    \left[ {}_{-\lambda}Y_{\ell m} (\unitvec{k}_{n_z}) \mathrm{e}^{-i \vec{k}_n\cdot \vec{x}_0} + \right. \nonumber \\
                    & \left. \qquad \qquad \qquad \qquad \qquad +{}_{\lambda}Y_{\ell m} (\unitvec{k}_{n_z})  \mathrm{e}^{i \vec{k}_n\cdot \vec{x}_0} 
                    \mathrm{e}^{i \vec{k}_n \cdot \vec{T}^{\E{6}}_A}\right], \qquad  \mbox{ for } \vec{n}\in \setN^{\E{6}}_C \nonumber \\
\nonumber \xi^{\E{6};T,\unitvec{k}_{\vec{n}}}_{k_{\vec{n}},\ell m,\lambda} &= \frac{i^{\ell}}{\sqrt{4}} \sqrt{\frac{2\pi^2(\ell + 2)!}{(\ell - 2)!}} 
                    \left[  {}_{-\lambda}Y_{\ell m} (\unitvec{k}_{\vec{n}}) \right. \left( \mathrm{e}^{-i\vec{k}_{\vec{n}} \cdot \vec{x}_0 } + (-1)^m  \mathrm{e}^{-i\vec{k}_{\vec{n}} \cdot (\mat{M}^{\E{6}}_C \vec{x}_0 - \vec{T}^{\E{6}}_C)}\right) \nonumber \\
                 & \qquad \qquad \qquad \qquad \quad {} + (-1)^\ell  {}_{\lambda}Y^*_{\ell m} (\unitvec{k}_{\vec{n}}) 
                  \left( (-1)^m \mathrm{e}^{-i\vec{k}_{\vec{n}} \cdot (\mat{M}^{\E{6}}_A \vec{x}_0 - \vec{T}^{\E{6}}_A)} + \right. \nonumber \\
                  &\left. \left. \qquad \qquad \qquad \qquad \qquad \quad \ + \mathrm{e}^{-i\vec{k}_{\vec{n}} \cdot (\mat{M}^{\E{6}}_B \vec{x}_0 - \vec{T}^{\E{6}}_B)} \right)\right], \qquad \ \mbox{ for } \vec{n}\in \setN^{\E{6}}_2 \nonumber
\end{empheq}

\begin{empheq}[box=\fbox]{align}\xi^{\E{6};E,\unitvec{k}_{\vec{n}}}_{k_{\vec{n}},\ell m,\lambda} &= -\frac{i^{\ell}}{\sqrt{2}} 
                    \ \sqrt{2 \pi^2} \ 
                    {}_{-\lambda}Y_{\ell m} (\unitvec{k}_{n_x}) \ \mathrm{e}^{-i \vec{k}_{\vec{n}} \cdot \vec{T}^{\E{6}}_B/2} \times \\ & \qquad \qquad \qquad \qquad \times \left[ \mathrm{e}^{-i \vec{k}_n\cdot \vec{x}_0} 
                    + (-1)^{m} \mathrm{e}^{i\vec{k}_n\cdot \vec{x}_0} \mathrm{e}^{i \vec{k}_n \cdot \vec{T}^{\E{6}}_B}\right],
                    \ \mbox{ for } \vec{n}\in \setN^{\E{6}}_A \nonumber \\
 \nonumber \xi^{\E{6};E,\unitvec{k}_{\vec{n}}}_{k_{\vec{n}},\ell m,\lambda} &= -\frac{i^{\ell}}{\sqrt{2}} \ \sqrt{2 \pi^2} \ 
                    {}_{-\lambda}Y_{\ell m} (\unitvec{k}_{n_y}) \ \mathrm{e}^{-i \vec{k}_{\vec{n}} \cdot \vec{T}^{\E{6}}_C/2} \times \\ & \qquad \qquad \qquad \qquad \times \left[ \mathrm{e}^{-i \vec{k}_n\cdot \vec{x}_0} 
                    +(-1)^{m} \mathrm{e}^{i \vec{k}_n\cdot \vec{x}_0} \mathrm{e}^{i \vec{k}_n \cdot \vec{T}^{\E{6}}_C}\right],
                    \mbox{ for } \vec{n}\in \setN^{\E{6}}_B \nonumber \\
 \nonumber \xi^{\E{6};E,\unitvec{k}_{\vec{n}}}_{k_{\vec{n}},\ell m,\lambda} &= -\frac{i^{\ell}}{\sqrt{2}} \sqrt{2 \pi}
                     \ \mathrm{e}^{-i \vec{k}_{\vec{n}} \cdot \vec{T}^{\E{6}}_A/2} \left[ {}_{-\lambda}Y_{\ell m} (\unitvec{k}_{n_z}) \mathrm{e}^{-i \vec{k}_n\cdot \vec{x}_0} + \right. \nonumber \\
                    & \left. \qquad \qquad \qquad \qquad \quad \ +{}_{\lambda}Y_{\ell m} (\unitvec{k}_{n_z})  \mathrm{e}^{i \vec{k}_n\cdot \vec{x}_0} 
                    \mathrm{e}^{i \vec{k}_n \cdot \vec{T}^{\E{6}}_A}\right], \qquad  \mbox{ for } \vec{n}\in \setN^{\E{6}}_C \nonumber \\
        \xi^{\E{6};E,\unitvec{k}_{\vec{n}}}_{k_{\vec{n}},\ell m,\lambda} &= -\frac{i^{\ell}}{\sqrt{4}}\ \sqrt{2 \pi^2}
                    \left[  {}_{-\lambda}Y_{\ell m} (\unitvec{k}_{\vec{n}}) \right. \left( \mathrm{e}^{-i\vec{k}_{\vec{n}} \cdot \vec{x}_0 } + (-1)^m  \mathrm{e}^{-i\vec{k}_{\vec{n}} \cdot (\mat{M}^{\E{6}}_C \vec{x}_0 - \vec{T}^{\E{6}}_C)}\right) \nonumber \\
                 & \qquad \qquad \qquad \qquad \quad {} + (-1)^\ell  {}_{\lambda}Y^*_{\ell m} (\unitvec{k}_{\vec{n}}) 
                  \left( (-1)^m \mathrm{e}^{-i\vec{k}_{\vec{n}} \cdot (\mat{M}^{\E{6}}_A \vec{x}_0 - \vec{T}^{\E{6}}_A)} + \right. \nonumber \\
                  &\left. \left. \qquad \qquad \qquad \qquad \qquad \quad \ + \mathrm{e}^{-i\vec{k}_{\vec{n}} \cdot (\mat{M}^{\E{6}}_B \vec{x}_0 - \vec{T}^{\E{6}}_B)} \right)\right], \qquad \ \mbox{ for } \vec{n}\in \setN^{\E{6}}_2 \nonumber 
\end{empheq}
and
\begin{empheq}[box=\fbox]{equation}
    \xi^{\E{6};B,\unitvec{k}_{\vec{n}}}_{k_{\vec{n}},\ell m,\lambda} = -\frac{\lambda}{2} \xi^{\E{6};E,\unitvec{k}_{\vec{n}}}_{k_{\vec{n}},\ell m,\lambda}\,.
\end{empheq}
The spherical-harmonic space covariance matrix has the form \eqref{eqn:HarmonicCovariance}.

\subsection{\E{11}: Chimney space}
\label{secn:topologyE11}

The chimney space, \E{11}, serves as the basis for all Euclidean manifolds with two compact dimensions (i.e., compact cross-sections), much like \E{1} does for all compact manifolds. It can be regarded as \E{1} with one non-compact dimension. Additionally, \E{12} through \E{15} all originate from \E{11}. \\

\noindent \textit{Properties}:
This manifold has compact cross-sections and is orientable, homogeneous, and anisotropic. \\

\noindent \textit{Generators}: Since \E{11} has only two compact dimensions, it is described by two generators.
In general (see \rcite{COMPACT:2023rkp}) the $z$ direction is chosen to be non-compact and the generators of \E{11} are given by
\begin{align} 
    \label{eqn:E11generalT}
    & \mat{M}^{\E{11}}_A = \identity, \quad\mbox{with} \nonumber \\
    & \vec{T}^{\E{11}}_{\A{1}} = L_{\A{1}} \begin{pmatrix} 1 \\ 0 \\ 0 \end{pmatrix}, \quad
    \vec{T}^{\E{11}}_{\A{2}} = L_{\A{2}} \begin{pmatrix} \cos\alpha \\ \sin\alpha \\ 0 \end{pmatrix}.
\end{align}
Similarly to \E{1}, it is often convenient to simplify notation by dropping the $A$ label and instead use
\begin{align}
    L_{iw} &\equiv L_{\A{i}w}, \quad \mbox{for } i\in \{1,2\}, \, w\in\{x,y,z\}; \nonumber \\
    L_{i} &\equiv L_{\A{i}}, \quad \mbox{for } i\in \{1,2\} .
\end{align}
In section $3.7$ of \rcite{COMPACT:2023rkp}, we presented a set of conditions on the parameters of \E{11} ($L_i$ and $\alpha$) that avoids double-counting identical manifolds with different generator choices. \\
\\
\noindent\textit{Cross-sectional area}: 
Since the chimney spaces have two compact dimensions, their volumes are infinite, but their cross-sections perpendicular to the non-compact direction are finite:
\begin{equation}
    \label{eqn:AE11}
    A_{\E{11}} = \vert \vec{T}^{\E{11}}_1 \times \vec{T}^{\E{11}}_2 \vert = L_1 L_2 \vert \sin\alpha \vert.
\end{equation}

\subsubsection{Eigenmodes and correlation matrices of \E{11}}
\label{secn:eigenmodesE11}

The chimney space resembles the 3-torus but has only two compact dimensions, resulting in a finite cross-sectional area \eqref{eqn:AE11}. Similar to \E{1}, we start with the eigenmodes of the covering space (\E{18}) \eqref{eqn:EuclideanFourierBasis} and select those that preserve the symmetries of \E{11}.

\begin{equation}
    \label{eqn:E11eigenmodeinvariance}
  G_\alpha (\Upsilon^{\E{11}}_{ij,\vec{k}}(\vec{x},\lambda)) = \Upsilon^{\E{11}}_{ij,\vec{k}}(\vec{x},\lambda)\,.
\end{equation}
Similar to the scalar case of \E{11}, since all elements of $\Gamma^{\E{11}}$ are pure translations, as in \E{1}, only one \E{18} eigenmode contributes for each allowed $\vec{k}$. The two generators of \E{11} \eqref{eqn:E11generalT} impose two discretization conditions--rather than three, as in \E{1}--while the helicity tensor does not introduce any additional constraints, following \eqref{eqn:associatedhomogeneousdiscretization},
\begin{align}
    \label{eqn:E11_ki}
    (\vec{k}_{\vec{n}})_{x} &= \frac{2\pi n_1}{L_1}, \\
    (\vec{k}_{\vec{n}})_{y} &= \frac{2\pi n_2}{L_2\sin\alpha} - \frac{2\pi n_1}{L_1}\frac{\cos\alpha}{\sin\alpha} , \nonumber
\end{align}
while $(\vec{k}_{\vec{n}})_{z}\equiv k_z$ is unconstrained.
We will write $\vec{k}_{\vec{n}}$ as a shorthand for the wavevector parametrized by $(n_1,n_2; k_z)$, i.e., by the integers $n_1$ and $n_2$ and the real variable $k_z$.
The usual (untilted) results are recovered for $\alpha=\pi/2$.
Thus
\begin{empheq}[box=\fbox]{equation}
    \Upsilon^{\E{11}}_{ij, \vec{n}}(\vec{x},\lambda) = e_{ij}(\unitvec{k}_{\vec{n}}, \lambda) \mathrm{e}^{i\vec{k}_{\vec{n}}\cdot(\vec{x}-\vec{x}_0)}\,, \quad \mbox{for } \vec{n}\in \setN^{\E{11}}
\end{empheq}
where
\begin{empheq}[box=\fbox]{equation}
    \setN^{\E{11}} \equiv \{(n_1,n_2) \vert n_i\in\integers\}\,.
\end{empheq}
Following \eqref{eqn:CE18XY} the Fourier-mode correlation matrix for \E{11} is
\begin{empheq}[box=\fbox]{align}
    \label{eqn:CE11XY}
    C^{\E{11};XY}_{iji'j', \vec{k}_{\vec{n}} \vec{k}_{\vecnp}, \lambda \lambda'} 
    &= 2\pi A_{\E{11}}\frac{\pi^2}{2 k_{\vec{n}}^3}\Tpower(k_{\vec{n}}) 
    \Delta^X(k_{\vec{n}}) \DeltaYstar(k_{\vec{n}}) 
    \Kdelta_{(\vec{k}_{\vec{n}})_x (\vec{k}_{\vecnp})_x}
    \Kdelta_{(\vec{k}_{\vec{n}})_y (\vec{k}_{\vecnp})_y} \times {}\nonumber \\
    & \qquad \qquad \qquad \qquad \qquad 
    {}\times \Ddelta(k_z - k'_z) \Kdelta_{\lambda\lambda'}\mathcal{E}_{iji'j'}(\unitvec{k}_{\vec{n}}, \unitvec{k}_{\vec{n}'},\lambda)
    \,.
\end{empheq}
In transitioning from \E{1} (compare \cref{eqn:CE11XY} to \cref{eqn:CE1XY}) we have replaced $V_{\E{1}}$ with $2\pi A_{\E{11}}$ (the cross-sectional area given by \eqref{eqn:AE11}) and  $\Kdelta_{\vec{k}_{\vec{n}}\vec{k}_{\vecnp}}$ with a Kronecker delta for the $x$ and $y$ components of $\vec{k}_{\vec{n}}$ and a Dirac delta function for the $z$ component.

We can project the field $\field^X$ onto the sky by performing a radial integral with a suitable weight function and transfer function, giving
\begin{align}
    \label{eqn:almE11}
    a^{\E{11};X}_{\ell m} 
    &= \frac{1}{2 \pi A_{\E{1}}} 
    \sum_{(n_1,n_2)\in\setN^{\E{11}} }  \sum_{\lambda =\pm 2} \int_{-\infty}^{\infty} \mathrm{d} k_z \,
     \mathcal{D}(\vec{k}_{\vec{n}}, \lambda, 0) \xi^{\E{11};X,\unitvec{k}_{\vec{n}}}_{k_{\vec{n}},\ell m,\lambda} \ \Delta^X_{\ell} (k_{\vec{n}}),
\end{align}
with
\begin{empheq}[box=\fbox]{equation}
  \xi^{\E{11};T,\unitvec{k}_{\vec{n}}}_{k_{\vec{n}},\ell m,\lambda} \equiv i^{\ell} \sqrt{\frac{2\pi^2(\ell + 2)!}{(\ell - 2)!}} \ {}_{-\lambda}Y_{\ell m} (\unitvec{k}_{\vec{n}})  \mathrm{e}^{-i \vec{k}_{\vec{n}}\cdot \vec{x}_0},
\end{empheq}

\begin{empheq}[box=\fbox]{equation}
 \label{eqn: E11coeff_E}
    \xi^{\E{11};E,\unitvec{k}_{\vec{n}}}_{k_{\vec{n}},\ell m,\lambda} \equiv -i^{\ell}  \ \sqrt{2 \pi^2} \ {}_{-\lambda}Y_{\ell m} (\unitvec{k}_{\vec{n}})\mathrm{e}^{-i \vec{k}_{\vec{n}}\cdot \vec{x}_0},
\end{empheq}
and
\begin{empheq}[box=\fbox]{equation}
\xi^{\E{11};B,\unitvec{k}_{\vec{n}}}_{k_{\vec{n}},\ell m,\lambda} = -\frac{\lambda}{2} \xi^{\E{11};E,\unitvec{k}_{\vec{n}}}_{k_{\vec{n}},\ell m,\lambda}\,,
\end{empheq}
where a similar transition to that above was performed in starting from \cref{eqn:almE1} and replacing the sum over the $(\vec{k}_{\vec{n}})_z$ with the integral over $k_z$.

For the chimney spaces \E{i} with $i=11$ or $i=12$, the spherical-harmonic space covariance matrix now has the form
\begin{empheq}[box=\fbox]{equation}
    \label{eqn:HarmonicCovarianceE11}
    C^{\E{i};XY}_{\ell m\ell'm'}  =
     \frac{\pi}{{4 A_{\E{i}}}}
        \sum_{\lambda = \pm 2}\sum_{(n_1,n_2)\in \setN^{\E{i}}} \int_{-\infty}^{\infty} \mathrm{d} k_z \,
        \frac{\Tpower(k_{\vec{n}})}{k_{\vec{n}}^3 } 
        \Delta^{X}_{\ell}(k_{\vec{n}})
        \Delta^{Y*}_{\ell'}(k_{\vec{n}}) \ 
        \xi^{\E{i};X,\unitvec{k}_{\vec{n}}}_{k_{\vec{n}},\ell m,\lambda}\ 
        \xi^{\E{i};Y,\unitvec{k}_{\vec{n}}*}_{k_{\vec{n}},\ell' m',\lambda} .
\end{empheq}

\subsection{\E{12}: Chimney space with half turn}
\label{secn:topologyE12}

The chimney space with half turn is a root of \E{11} and can be thought of as \E{2} with one non-compact dimension. \\

\noindent \textit{Properties}:
This manifold has compact cross-sections and is orientable, inhomogeneous, and anisotropic. \\

\noindent \textit{Generators}:
Similar to \E{11}, there are two generators, and similar to \E{2}, one of the matrices is a rotation by $\pi$.
Conventionally this rotation is chosen to be around the $y$-axis (cf.\ \rcite{Riazuelo2004:prd}).
Here we instead choose the rotation to be around the $z$-axis, as is done in \E{2}.
This makes it clear that \E{12} is the limit of \E{2} with $ \vert\vec{T}^{\E{2}}_{\A{2}} \vert \to \infty$.
In general (see \rcite{COMPACT:2023rkp}), the generators of \E{12} can be written as
\begin{align} 
    \label{eqn:E12generalT}
    & \mat{M}^{\E{12}}_A = \identity, \quad \mat{M}^{\E{12}}_B = \mat{R}_{\unitvec{z}}(\pi) = \diag(-1, -1, 1), \quad \mbox{with} \nonumber \\
    & \vec{T}^{\E{12}}_A \equiv \vec{T}^{\E{12}}_{1} = L_{A} \begin{pmatrix} 1 \\ 0 \\ 0 \end{pmatrix}, \quad \quad
    \vec{T}^{\E{12}}_{B} = L_{B} \begin{pmatrix} \cos\beta\cos\gamma \\ \cos\beta\sin\gamma \\ \sin\beta \end{pmatrix}.
\end{align}

In section $3.8$ of \rcite{COMPACT:2023rkp}, we presented a set of conditions on the parameters of \E{12} ($L_A$, $L_B$, $\gamma$ and $\beta$) that avoids double-counting identical manifolds with different generator choices. \\

\noindent \textit{Associated \E{11}}: In addition to $\vec{T}^{\E{12}}_1$ defined above, a second independent translation is
\begin{equation}
    ( g^{\E{12}}_B )^2: \vec{x} \to \vec{x} + \vec{T}^{\E{12}}_2,
\end{equation}
for
\begin{equation}
    \label{eqn:E12assocE11}
    \vec{T}^{\E{12}}_2 \equiv \begin{pmatrix} 0 \\ 0 \\ 2 L_{Bz} \end{pmatrix}
        = 2 L_B \begin{pmatrix} 0 \\ 0 \\ \sin\beta \end{pmatrix}\,.
\end{equation}
\\
\noindent\textit{Cross-sectional area}:
\begin{equation}
    \label{eqn:AE12}
    A_{\E{12}} = \frac{1}{2} \vert \vec{T}^{\E{12}}_1 \times \vec{T}^{\E{12}}_2 \vert = L_A L_B \vert \sin\beta \vert.
\end{equation}

\subsubsection{Eigenmodes and correlation matrices of \E{12}}
\label{secn:eigenmodesE12}

The eigenspectrum and eigenmodes of the chimney space with half turn  can be derived using a method similar to that of the chimney space in much the same way that these quantities for $\E{2}$ were determined from \E{1}.
For \E{12}, the discretization condition \eqref{eqn:associatedhomogeneousdiscretization} leads to the components of the allowed wavevectors,
\begin{equation}
    (\vec{k}_{\vec{n}})_{x} = \frac{2\pi n_1}{L_{Ax}} , \quad (\vec{k}_{\vec{n}})_{z} = \frac{\pi n_2}{L_{Bz}} ,
\end{equation}
with $k_y$ again unconstrained.
As in \E{11}, we will write $\vec{k}_{\vec{n}}$ as a shorthand for the wavevector parametrized by the integer array $\vec{n}=(n_1,n_2)$ and the real variable $k_y$.

Unlike in \E{11}, the eigenmodes of \E{12} can include a linear combination of two \E{11} eigenmodes.
Here $(\mat{M}^{\E{12}}_B)^N \vec{k}_{\vec{n}} =\vec{k}_{\vec{n}}$ has two solutions, $N=1$ and $N=2$.
Written explicitly:
\begin{description}
    \item[\textbf{$N=1$ eigenmodes: }]  $\vec{k}_{\vec{n}}=\transpose{(0,0,(\vec{k}_{\vec{n}})_z)}$, 
        i.e., $\vec{n}=(0,n_2)$, $n_2 \in 2\integers^{\neq 0}$, $k_y = 0$, with
    \begin{empheq}[box=\fbox]{equation}
         \Upsilon^{\E{12}}_{ij, \vec{n}}(\vec{x},\lambda) = e_{ij}( \unitvec{k}_{\vec{n}}, \lambda) \mathrm{e}^{i\vec{k}_{\vec{n}}\cdot(\vec{x}-\vec{x}_0)}
        =  e_{ij}(\unitvec{k}_{\vec{n}}, \lambda) \mathrm{e}^{i \pi n_2 (z - z_0) / L_{Bz}},
    \end{empheq}
    \item[\textbf{$N=2$ eigenmodes: }] $((\vec{k}_{\vec{n}})_{x}, (\vec{k}_{\vec{n}})_{z}) \neq (0,0)$, 
        i.e., $(n_1,n_2) \neq (0,0)$, with
    \begin{empheq}[box=\fbox]{align}
         \Upsilon^{\E{12}}_{ij,\vec{n}}(\vec{x},\lambda) =&
            \frac{\mathrm{e}^{-i \vec{k}_{\vec{n}} \cdot \vec{T}^{\E{12}}_B/2}}{\sqrt{2}} \left( e_{ij}(\unitvec{k}_{\vec{n}},\lambda) \mathrm{e}^{i\vec{k}_{\vec{n}}\cdot(\vec{x}-\vec{x}_0)} + \right. \nonumber\\
            & \left. \qquad \qquad \qquad \qquad + e_{ij}\left((\mat{M}^{\E{12}}_B)^T\unitvec{k}_{\vec{n}},\lambda\right) \mathrm{e}^{i\vec{k}_{\vec{n}}\cdot(\mat{M}^{\E{12}}_B(\vec{x}-\vec{x}_0)+\vec{T}^{\E{12}}_B)} \right).
    \hspace{3mm}
    \end{empheq}
\end{description}
For the $N = 2$ modes, we choose $\Phi_{\vec{k}_{\vec{n}}}^{\E{12}} = -\vec{k}_{\vec{n}}\cdot \vec{T}^{\E{12}}_B/2$ in \eqref{eqn:generaleigenmodeformula}---this will simplify the reality condition of the field. Here the two sets of allowed modes are defined by
\begin{empheq}[box=\fbox]{align}
     \setN^{\E{12}}_1 &\equiv \{(0,n_2) \vert n_2\in 2\integers^{\neq0}\}\,, \nonumber\\
    \setN^{\E{12}}_2 &\equiv \{(n_1,n_2) \vert n_1 \in \integers^{>0}, n_2 \in \integers\} , \\
    \setN^{\E{12}} &\equiv \setN^{\E{12}}_1 \cup \setN^{\E{12}}_2 \,.\nonumber
\end{empheq}
With these the Fourier-mode correlation matrix can now be expressed as 
\begin{empheq}[box=\fbox]{align}
\label{eqn:E12FourierCovarianceStandardConvention}
     C^{\E{12};XY}_{iji'j', \vec{k}_{\vec{n}} \vec{k}_{\vecnp}, \lambda \lambda'} 
     = {} & 2\pi A_{\E{12}} \frac{\pi^2}{2k^3_{\vec{n}}}
        \Tpower(k_{\vec n})\Delta^X(k_{\vec{n}})\DeltaYstar(k_{\vec{n}})
         \mathrm{e}^{i(\vec{k}_{\vecnp}-\vec{k}_{\vec{n}})\cdot\vec{x}_0} \Kdelta_{\lambda\lambda'} \times \nonumber
        \\
        & \quad \times \mathcal{E}_{iji'j'}(\unitvec{k}_{\vec{n}}, \unitvec{k}_{\vec{n}'}, \lambda)
            \frac{1}{2}
                \sum_{(\tilde{n}_1, \tilde{n}_2)\in \setN^{\E{12}}_2}
                \int_{-\infty}^{\infty} \mathrm{d}\tilde{k}_y \, \sum_{a=0}^1\sum_{b=0}^1
                  \nonumber \\
        & \qquad \ \mathrm{e}^{i\vec{k}_{\vec{\tilde{n}}}\cdot(\vec{T}^{(a)}-\vec{T}^{(b)})} \Kdelta_{(\vec{k}_{\vec{n}})_x (\vec{k}^{(a)}_{\vec{\tilde{n}}})_x}
                \Kdelta_{(\vec{k}_{\vecnp})_x (\vec{k}^{(b)}_{\vec{\tilde{n}}})_x} \times
                \\
        & \qquad \ {} \times
             \Kdelta_{(\vec{k}_{\vec{n}})_z (\vec{k}^{(a)}_{\vec{\tilde{n}}})_z}
             \Kdelta_{(\vec{k}_{\vecnp})_z (\vec{k}^{(b)}_{\vec{\tilde{n}}})_z}
             \Ddelta(k_y - \tilde{k}^{(a)}_y)
             \Ddelta(k'_y - \tilde{k}^{(b)}_y)
         , 
         \nonumber
\end{empheq}
where the terms with $\vec{n} \in \setN^{\E{12}}_1$ are of measure zero and have been dropped, $A_{\E{12}}$ is given in \eqref{eqn:AE12}, $\vec{T}^{(0)}\equiv\vec{0}$, $\vec{T}^{(1)}\equiv\vec{T}^{\E{12}}_B$, $\vec{k}^{(a)}_{\tilde{\vec{n}}} \equiv [(\mat{M}^{\E{12}}_B){}^T]^a \vec{k}_{\vec{\tilde{n}}}$, and $\tilde{k}^{(a)}_y \equiv (\vec{k}^{(a)}_{\tilde{\vec{n}}})_y$.

In the harmonic basis we have:
\begin{empheq}[box=\fbox]{align}
   \xi^{\E{12};T,\unitvec{k}_{\vec{n}}}_{k_{\vec{n}},\ell m,\lambda} &\equiv 
           i^{\ell} \sqrt{\frac{2\pi^2(\ell + 2)!}{(\ell 
           - 2)!}} \ {}_{-\lambda}Y_{\ell m} (\unitvec{k}_{\vec{n}}) \  \mathrm{e}^{-i \vec{k}_{\vec{n}} \cdot \vec{x}_0} , \quad \mbox{for } \vec{n}\in \setN^{\E{12}}_1\\
    \xi^{\E{12};T,\unitvec{k}_{\vec{n}}}_{k_{\vec{n}},\ell m,\lambda}& \equiv 
        \frac{i^{\ell}}{\sqrt{2}} \sqrt{\frac{2\pi^2(\ell + 2)!}{(\ell - 2)!}} \ {}_{-\lambda}Y_{\ell m} (\unitvec{k}_{\vec{n}}) \ \mathrm{e}^{-i \vec{k}_{\vec{n}} \cdot \vec{T}^{\E{12}}_B/2} \times \nonumber \\ 
        & \qquad\qquad\quad \times \left( \mathrm{e}^{-i \vec{k}_n\cdot \vec{x}_0} +  (-1)^m \mathrm{e}^{-i \vec{k}_n\cdot \left[ \mat{M}^{\E{12}}_B \vec{x}_0- \vec{T}^{\E{12}}_B \right]} \right), \mbox{ for } \vec{n} \in \setN^{\E{12}}_2 \nonumber  
\end{empheq}

\begin{empheq}[box=\fbox]{align}
   \xi^{\E{12};E,\unitvec{k}_{\vec{n}}}_{k_{\vec{n}},\ell m,\lambda} &\equiv 
        -i^{\ell} \sqrt{2 \pi^2}
       \ {}_{-\lambda}Y_{\ell m} (\unitvec{k}_{\vec{n}}) \  \mathrm{e}^{-i \vec{k}_{\vec{n}} \cdot \vec{x}_0}, \quad \mbox{for } \vec{n}\in \setN^{\E{12}}_1\\
    \xi^{\E{12};E,\unitvec{k}_{\vec{n}}}_{k_{\vec{n}},\ell m,\lambda}&
        \equiv -\frac{i^{\ell}}{\sqrt{2}} \ \sqrt{2 \pi^2} \ {}_{-\lambda}Y_{\ell m} (\unitvec{k}_{\vec{n}}) \ \mathrm{e}^{-i \vec{k}_{\vec{n}} \cdot \vec{T}^{\E{12}}_B/2} \times \nonumber\\
        & \qquad \qquad \quad \times \left( \mathrm{e}^{-i \vec{k}_n\cdot \vec{x}_0} + (-1)^m \mathrm{e}^{-i \vec{k}_n\cdot \left[ \mat{M}^{\E{12}}_B \vec{x}_0- \vec{T}^{\E{12}}_B \right]} \right), \ \mbox{for } \vec{n} \in \setN^{\E{12}}_2 \nonumber
\end{empheq}
and
\begin{empheq}[box=\fbox]{equation}
\xi^{\E{12};B,\unitvec{k}_{\vec{n}}}_{k_{\vec{n}},\ell m,\lambda} = -\frac{\lambda}{2} \xi^{\E{12};E,\unitvec{k}_{\vec{n}}}_{k_{\vec{n}},\ell m,\lambda}\,.
\end{empheq}
The spherical-harmonic space covariance matrix has the form \eqref{eqn:HarmonicCovarianceE11}.

\subsection{\E{16}: Slab space including rotation}
\label{secn:topologyE16}

The slab space \E{16} serves as the basis for all Euclidean three-manifolds with one compact dimension (i.e., compact lengths), much like \E{1} and \E{11} do for fully compact and two-compact-dimensional spaces, respectively.  
The possibility of a corkscrew transformation (as opposed to a pure translation) in the slab space was largely overlooked until recently \cite{COMPACT:2023rkp}. While topologically, the corkscrew deformation is continuously reducible to the unrotated slab space, it produces a physically distinct pattern of clones.  To account for this distinction, we categorize \E{16} into two cases: \slabh\ and \slabi. \\

\subsubsection{\slabh: Conventional unrotated slab space}

The conventional definition of \E{16} only includes a translation.
Here we call this choice \slabh, because the space is homogeneous.\\

\noindent \textit{Properties:} This manifold has a compact length and is orientable, homogeneous, and anisotropic. \\

\noindent \textit{Generators}: In general, since \E{16} has one compact dimension it is described by one generator, which we may take to be a translation in the $z$ direction (see \rcite{COMPACT:2023rkp}), so the generator of \slabh\ is
\begin{equation} 
    \label{eqn:E16hgeneralT}
    \mat{M}^{\slabh}_A = \identity, \quad \mbox{with} \quad \vec{T}^{\slabh}_A = L\begin{pmatrix} 0 \\ 0 \\ 1 \end{pmatrix}.
\end{equation}
Even though there is only one generator, since this generator is a pure translation, we follow the convention of using $A$ to label it. As in \E{1} and \E{11}, $L$ is always positive.\\

\noindent \textit{Length}: Since the slab spaces have only one compact dimension, their volumes and cross-sectional areas are infinite.
The shortest path length around the manifold at any point is $L$.

\subsubsection{\slabi: General rotated slab space}

The orientable slab space also allows a corkscrew motion, which is physically distinct and must be considered separately. For the eigenmodes of the Laplacian to form a complete basis for general smooth functions on the manifold, they must not have azimuthal symmetry around the corkscrew axis. This requirement is met only if the rotation angle is a rational multiple of $2 \pi$,
\\

\noindent \textit{Properties}: Due to the corkscrew motion this differs from \slabh\ in that it is inhomogeneous.
This manifold has a compact length and is orientable, inhomogeneous, and anisotropic. \\

\noindent \textit{Generators}: Similarly to \slabh, there is one generator.
In general the generator of \slabi\ can be written as
\begin{align} 
    \label{eqn:E16igeneralT}
    &\mat{M}^{\slabi}_B = \mat{R}_{\unitvec{z}}(2\pi p/q)
     = \begin{pmatrix} \cos(2\pi p/q) & -\sin(2\pi p/q) & 0 \\
       \sin(2\pi p/q) & \hphantom{-}\cos(2\pi p/q) & 0 \\
       0 & \hphantom{-}0 & 1 \end{pmatrix} ,
       \quad \mbox{with}  \nonumber \\
     &\vec{T}^{\slabi}_B = \begin{pmatrix} L_{x} \\ 0 \\ L_{z} \end{pmatrix} = L\begin{pmatrix} \cos\beta \\ 0 \\ \sin\beta \end{pmatrix} ,
\end{align}
where $p\in \integers^{\neq 0}$, $q\in \integers^{>1}$, and $|p|$ and $q$ are relatively prime.
As in \slabh, here since the generator is a rotation, we use $B$ to label it.
\\

\noindent \textit{Associated \slabh}: A pure translation can be defined for \slabi\ as
\begin{equation}
    \label{eqn:E16iassocE16h}
    g^{\slabi}_1 \equiv (g^{\slabi}_B)^q: \vec{x} \to \vec{x} + \vec{T}^{\slabi}_1, \quad \mbox{for }
    \vec{T}^{\slabi}_1 \equiv \begin{pmatrix} 0 \\ 0 \\ q L_{z} \end{pmatrix} = q L \sin\beta \begin{pmatrix} 0 \\ 0 \\ 1 \end{pmatrix} .
\end{equation}
\\
\noindent \textit{Length}: The length of \slabi\ is 
\begin{equation}
    \label{eqn:LE16i}
    L_{\slabi} = L \vert \sin\beta \vert.
\end{equation}

\subsubsection{Eigenmodes and correlation matrices of \slabh}

Similarly to \E{1} and \E{11}, \slabh\ is homogeneous.
It is compact in only one dimension so there is only one generator of the topology.
The one discretization condition following from \eqref{eqn:associatedhomogeneousdiscretization} leads to
\begin{equation}
     \label{eqn:E16h_ki}
    (\vec{k}_{\vec{n}})_{z} = \frac{2\pi n}{L}\,,
\end{equation}
while $(\vec{k}_{\vec{n}})_x \equiv k_x$ and $(\vec{k}_{\vec{n}})_y \equiv k_y$ are unconstrained.
As in the chimney spaces, we will again write $\vec{k}_{\vec{n}}$ as a shorthand for the wavevector characterized by the integer $n$ and the real variables $k_x$ and $k_y$.

Thus
 \begin{empheq}[box=\fbox]{align}
    \Upsilon^{\slabh}_{ij, \vec{n}}(\vec{x}, \lambda) = 
       e_{ij}(\unitvec{k}_{\vec{n}}, \lambda) \mathrm{e}^{i\vec{k}_{\vec{n}}\cdot(\vec{x}-\vec{x}_0)}, \quad \mbox{for } n\in \setN^{\slabh}
\end{empheq}
where
\begin{empheq}[box=\fbox]{equation}
    \setN^{\slabh} = \{n\in \integers \} .
\end{empheq}
Following \eqref{eqn:CE18XY} the Fourier-mode correlation matrix for \slabh\ is
\begin{empheq}[box=\fbox]{align}
     \label{eqn:CE16hXY}
     C^{\slabh;XY}_{iji'j', \vec{k}_{\vec{n}} \vec{k}_{\vecnp}, \lambda \lambda'}
        &= (2\pi)^2 L \frac{\pi^2}{2 k^3}
        \Tpower(k_{\vec{n}}) \Delta^X(k_{\vec{n}}) \DeltaYstar(k_{\vec{n}}) \ \Kdelta_{\lambda\lambda'} \ \mathcal{E}_{iji'j'}(\unitvec{k}_{\vec{n}}, \unitvec{k}_{\vec{n}'}, \lambda) \times\\
        &\qquad \qquad \qquad \qquad {} \times
        \Ddelta(k_x - k'_x) \Ddelta(k_y - k'_y)
        \Kdelta_{(\vec{k}_{\vec{n}})_z (\vec{k}_{\vecnp})_z}.
        \nonumber
\end{empheq}
In transitioning from \E{1} (compare \cref{eqn:CE16hXY} to \cref{eqn:CE1XY}) we have replaced $V_{\E{1}}$ with $(2\pi)^2 L$ and $\Kdelta_{\vec{k}_{\vec{n}}\vec{k}_{\vecnp}}$ with a Kronecker delta for the $z$ component of $\vec{k}_{\vec{n}}$ and Dirac delta functions for the $x$ and $y$ components.

We can project the field $\field^X$ onto the sky by performing a radial integral with a suitable weight function and a transfer function, giving
\begin{equation}
    \label{eqn:almE16h}
    a^{\slabh;X}_{\ell m} 
     = \frac{1}{(2\pi)^2 L}
        \sum_{n\in \setN^{\slabh}} \sum_{\lambda =\pm 2}
        \int_{-\infty}^\infty \mathrm{d} k_x\, \int_{-\infty}^\infty \mathrm{d} k_y \,
       \mathcal{D}(\vec{k}_{\vec{n}}, \lambda, 0) \xi^{\slabh;X,\unitvec{k}_{\vec{n}}}_{k_{\vec{n}},\ell m,\lambda} \ \Delta^X_{\ell} (k_{\vec{n}}),
\end{equation}
with
\begin{empheq}[box=\fbox]{equation}
  \xi^{\slabh;T,\unitvec{k}_{\vec{n}}}_{k_{\vec{n}},\ell m,\lambda} \equiv i^{\ell} \sqrt{\frac{2\pi^2(\ell + 2)!}{(\ell - 2)!}} \ {}_{-\lambda}Y_{\ell m} (\unitvec{k}_{\vec{n}})  \mathrm{e}^{-i \vec{k}_{\vec{n}}\cdot \vec{x}_0},
\end{empheq}

\begin{empheq}[box=\fbox]{equation}
 \label{eqn: E16coeff_E}
    \xi^{\slabh;E,\unitvec{k}_{\vec{n}}}_{k_{\vec{n}},\ell m,\lambda} \equiv - i^{\ell}\ \sqrt{2 \pi^2} \ {}_{-\lambda}Y_{\ell m} (\unitvec{k}_{\vec{n}})\mathrm{e}^{-i \vec{k}_{\vec{n}}\cdot \vec{x}_0},
\end{empheq}
and
\begin{empheq}[box=\fbox]{equation}
\xi^{\slabh;B,\unitvec{k}_{\vec{n}}}_{k_{\vec{n}},\ell m,\lambda} = -\frac{\lambda}{2} \xi^{\slabh;E,\unitvec{k}_{\vec{n}}}_{k_{\vec{n}},\ell m,\lambda}\,,
\end{empheq}
where a transition similar to that above was performed in starting from \cref{eqn:almE1}.

For the slab spaces the spherical-harmonic space covariance matrix now has the form
\begin{empheq}[box=\fbox]{align}\label{eqn:HarmonicCovarianceE16h}
    C^{\E{16}^{(a)};XY}_{\ell m\ell'm'} 
    = \frac{1}{8 L q}
         \sum_{\lambda = \pm 2} \sum_{n\in \setN^{\E{16}^{(a)}}}
         &\int_{-\infty}^\infty \mathrm{d} k_x\, \int_{-\infty}^\infty \mathrm{d} k_y \, 
       \frac{\Tpower(k_n)}{k_{\vec{n}}^3 } \times  {} \\ 
      &{}\times \Delta^{X}_{\ell}(k_{\vec{n}})
       \Delta^{Y*}_{\ell'}(k_{\vec{n}}) \ 
           \xi^{\E{16}^{(a)};X,\unitvec{k}_{\vec{n}}}_{k_{\vec{n}},\ell m,\lambda}\ 
           \xi^{\E{16}^{(a)};Y,\unitvec{k}_{\vec{n}}*}_{k_{\vec{n}},\ell' m',\lambda}\,, \nonumber
\end{empheq}
where $a \in \{\mathrm{h},\mathrm{i}\}$ labels the homogeneous (\slabh) and inhomogeneous (\slabi) slab spaces, with $q = 1$ for the \slabh\ space.

\subsubsection{Eigenmodes and correlation matrices of \slabi}

The eigenspectrum and eigenmodes of \slabi\ can be derived from \slabh\ like that for \E{12} from \E{11} since $\mat{M}^{\slabi}_B$ is a rotation around the $z$-axis.
For \slabi, the discretization condition \eqref{eqn:associatedhomogeneousdiscretization} leads to the component of the allowed wavevectors,
\begin{equation}
    \label{eqn:kzinE16i}
    (\vec{k}_{\vec{n}})_{z} = \frac{2\pi n}{q L\sin\beta} ,
\end{equation}
while $(\vec{k}_{\vec{n}})_x \equiv k_x$ and $(\vec{k}_{\vec{n}})_y \equiv k_y$ are again unconstrained.
As in \slabh, we will again write $\vec{k}_{\vec{n}}$ as a shorthand for the wavevector characterized by the integer $n$ and the real variables $k_x$ and $k_y$.

Unlike in \slabh, the eigenmodes of \slabi\ can include linear combinations of \slabh\ eigenmodes.
Here $(\mat{M}^{\slabi}_B)^N \vec{k}_{\vec{n}} = \vec{k}_{\vec{n}}$ has two solutions, $N=1$ and $N=q$. Written explicitly:
\begin{description}
    \item[\textbf{$N=1$ eigenmodes: }]  $\vec{k}_{\vec{n}}=\transpose{(0, 0, (\vec{k}_{\vec{n}})_z)}$, 
        i.e., $n = qm+\lambda$, $m \in \integers$, $k_x = k_y = 0$, with
    \begin{empheq}[box=\fbox]{equation}
        \Upsilon^{\slabi}_{ij, \vec{k}_{\vec{n}}}(\vec{x}, \lambda) = e_{ij}(\unitvec{k}_{\vec{n}}, \lambda) \mathrm{e}^{i\vec{k}_{\vec{n}} \cdot (\vec{x}-\vec{x}_0)},
    \end{empheq}
    \item[\textbf{$N=q$ eigenmodes: }] $(k_x, k_y) \neq (0,0)$, $n \in \integers$, with
    \begin{empheq}[box=\fbox]{equation}
        \Upsilon^{\slabi}_{ij, \vec{k}_{\vec{n}}}(\vec{x},\lambda) = \frac{1}{\sqrt{q}} \sum_{a=0}^{q-1} e_{ij}\left(\left[\left(\mat{M}_B^{\slabi}\right)^T\right]^a\vec{k}_{\vec{n}},\lambda\right) \mathrm{e}^{i\vec{k}_{\vec{n}}\cdot(\mat{M}^{\slabi}_B)^a(\vec{x}-\vec{x}_0)} \mathrm{e}^{i \vec{k}_{\vec{n}} \cdot \mat{M}^{\slabi}_{0a} \vec{T}^{\slabi}_B} ,
    \end{empheq}
\end{description}
and $\mat{M}^{\slabi}_{00} \equiv 0$ and $\mat{M}^{\slabi}_{0a}$ defined in \eqref{eqn:M0jdef}.
Here the two sets of allowed modes are defined by

\begin{empheq}[box=\fbox]{align}
    {}_\lambda\setN^{\slabi}_1 &= \{n = q m +\lambda \mid m \in \integers\} , \nonumber \\
    \setN^{\slabi}_q &= \{n\in \integers \} , \\
    \setN^{\slabi} &= {}_\lambda\setN^{\slabi}_1 \cup \setN^{\slabi}_q . \nonumber
\end{empheq}

Note that for $\setN^{\slabi}_2$ we require $(k_x, k_y)$ to lie in a wedge in the $xy$-plane with opening angle $2\pi/q$. With these the Fourier-mode correlation matrix can now be expressed as
\begin{empheq}[box=\fbox]{align}
     \label{eqn:E16iFourierCovarianceStandardConvention}
     C^{\slabi;XY}_{iji'j', \vec{k}_{\vec{n}} \vec{k}_{\vecnp}, \lambda \lambda'}
     &= (2\pi)^2 L_{\slabi}
        \frac{\pi^2}{2 k_{\vec{n}}^3} \Tpower(k_{\vec{n}}) \Delta^X(k_{\vec{n}}) \DeltaYstar(k_{\vec{n}})
            \mathrm{e}^{i(\vec{k}_{\vecnp}-\vec{k}_{\vec{n}})\cdot\vec{x}_0} \Kdelta_{\lambda \lambda'} \times \\
        & \quad \times \mathcal{E}_{iji'j'}(\unitvec{k}_{\vec{n}}, \unitvec{k}_{\vec{n}'}, \lambda) 
      \ \frac{1}{q^2} 
     \sum_{\tilde{n} \in \setN^{\slabi}_q} \int_{-\infty}^{\infty} \mathrm{d} k_x\, \int_{-\infty}^{\infty} k_y\, \sum_{a=0}^{q-1}\sum_{b=0}^{q-1} \nonumber \\
     & \qquad 
        \mathrm{e}^{i\vec{k}_{\tilde{\vec{n}}} \cdot (\vec{T}^{(a)} - \vec{T}^{(b)})} 
        \Kdelta_{(\vec{k}_{\vec{n}})_z (\vec{k}_{\tilde{\vec{n}}}^{(a)})_z} \Kdelta_{(\vec{k}_{\vecnp})_z (\vec{k}_{\tilde{\vec{n}}}^{(b)})_z} \times
    \nonumber \\
     & \quad {} \times
        \Ddelta(k_x - \tilde{k}^{(a)}_{x}) \Ddelta(k'_x - \tilde{k}^{(b)}_{x})
        \Ddelta(k_y - \tilde{k}^{(a)}_{y}) \Ddelta(k'_y - \tilde{k}^{(b)}_{y}) , \nonumber
\end{empheq}
where the terms with $n\in{}_\lambda\setN^{\slabi}_1$ are of measure zero and have been dropped, $L_{\slabi}$ is given in \eqref{eqn:LE16i}, $\vec{T}^{(0)} \equiv \vec{0}$, $\vec{T}^{(a)} \equiv \mat{M}^{\slabi}_{0a} \vec{T}^{\slabi}_B$ for $a \in \{ 1,\ldots, q-1 \}$, $\vec{k}^{(a)}_{\tilde{\vec{n}}} \equiv [(\mat{M}^{\slabi}_B){}^T]^a \vec{k}_{\vec{\tilde{n}}}$, and $\tilde{k}^{(a)}_w \equiv (\vec{k}^{(a)}_{\tilde{\vec{n}}})_w$ for $w \in \{x, y\}$.

In the harmonic basis we have
\begin{empheq}[box=\fbox]{align}
    \xi^{\slabi;T,\unitvec{k}_{\vec{n}}}_{k_{\vec{n}},\ell m,\lambda} &\equiv i^{\ell} \sqrt{\frac{2\pi^2(\ell + 2)!}{(\ell 
           - 2)!}} \ {}_{-\lambda}Y_{\ell m} (\unitvec{k}_{\vec{n}}) \  \mathrm{e}^{-i \vec{k}_{\vec{n}} \cdot \vec{x}_0}, \quad \mbox{for } \vec{n}\in \setN^{\slabi}_1\\
    \xi^{\slabi;T,\unitvec{k}_{\vec{n}}}_{k_{\vec{n}},\ell m,\lambda}& \equiv 
                \frac{i^{\ell}}{\sqrt{q}} \sqrt{\frac{2\pi^2(\ell +           2)!}{(\ell 
             - 2)!}} \ {}_{-\lambda}Y_{\ell m} (\unitvec{k}_{\vec{n}}) \times \nonumber \\
             & \qquad \qquad \times \sum_{j=0}^{q-1} \mathrm{e}^{-i m j p/q} \ \mathrm{e}^{-i \vec{k}_n\cdot \left[ (\mat{M}^{\slabi}_B)^j \vec{x}_0- \mat{M}^{\slabi}_{0j} \vec{T}^{\slabi}_B \right]}, \ \mbox{for } \vec{n} \in \setN^{\slabi}_2 , \nonumber
\end{empheq}

\begin{empheq}[box=\fbox]{align}
   \xi^{\slabi;E,\unitvec{k}_{\vec{n}}}_{k_{\vec{n}},\ell m,\lambda} &\equiv  -i^{\ell} \sqrt{2 \pi^2}
       \ {}_{-\lambda}Y_{\ell m} (\unitvec{k}_{\vec{n}}) \  \mathrm{e}^{-i \vec{k}_{\vec{n}} \cdot \vec{x}_0}, \quad \mbox{for } \vec{n}\in \setN^{\slabi}_1\\
    \xi^{\slabi;E,\unitvec{k}_{\vec{n}}}_{k_{\vec{n}},\ell m,\lambda}& \equiv 
        -\frac{i^{\ell}}{\sqrt{q}} \ \sqrt{2 \pi^2} \   {}_{-\lambda}Y_{\ell m} (\unitvec{k}_{\vec{n}})  \times \nonumber\\ 
        & \qquad \qquad \times \sum_{j=0}^{q-1}  \mathrm{e}^{-i  m j p/q} \ \mathrm{e}^{-i\vec{k}_n\cdot \left[ (\mat{M}^{\slabi}_B)^j \vec{x}_0- \mat{M}^{\slabi}_{0j} \vec{T}^{\slabi}_B \right]},                    \ \mbox{for } \vec{n} \in \setN^{\slabi}_2 , \nonumber
\end{empheq}
and
\begin{empheq}[box=\fbox]{equation}
\xi^{\slabi;B,\unitvec{k}_{\vec{n}}}_{k_{\vec{n}},\ell m,\lambda} = -\frac{\lambda}{2} \xi^{\slabi;E,\unitvec{k}_{\vec{n}}}_{k_{\vec{n}},\ell m,\lambda}\,,
\end{empheq}
The spherical-harmonic space covariance matrix has the form \eqref{eqn:HarmonicCovarianceE16h}.

\section{Numerical analysis}
\label{secn:numerical_results}

In the preceding section, we have presented the covariance matrices of CMB temperature and $E$-mode and $B$-mode polarizations for each of the orientable topologies of $E^3$ as a function of the topology parameters characterizing manifolds of those topologies and of the location of the observer. The CMB temperature and polarization anisotropies are regarded as well-characterized as Gaussian fields on the sphere of the sky, with no significant deviations detected in {\it Planck} data, though the variance of the temperature is low ($p\leq 2.5\%$ for all values of $N_\mathrm{side}$), and there is even similar evidence of low skewness at large scales ($N_\mathrm{side}\leq64$) \cite{Planck:2019kim,Planck:2013lks,Planck:2015igc,Planck:2019evm}. 
The information in a set of (correlated) Gaussian fields on the sky is fully encapsulated in the expectation values and the covariance matrices of their spherical harmonic coefficients, i.e., $\langle a_{\ell m}^{\E{i};X} \rangle$ and $C_{\ell m\ell'm'}^{\E{i};XY} \equiv \langle a_{\ell m}^{\E{i};X}  a_{\ell' m'}^{\E{i};Y*} \rangle$ \cite{Hu2002}. The expected values of these harmonic coefficients are zero in both trivial and non-trivial topologies, but the covariance matrix can be significantly different. 
In the trivial topology, this matrix has non-zero terms only along the diagonals of $\ell$ and $m$ (i.e., $\ell=\ell'$, $m=m'$), and additionally is only non-zero for $XY \in \{TT, EE, BB, TE\}$. 
Non-trivial topologies can break various symmetries, specifically isotropy and parity invariance, inducing otherwise-forbidden non-zero off-diagonal elements in the covariance matrix in all cross- and auto-correlations.

In \rcite{COMPACT:2024cud}, we presented normalized polarization covariance matrices for \E{1}--\E{3}, highlighting the violation of different symmetries in these topologies and their impact on the polarization covariance matrices. Building on this work, we now extend the analysis to the full $TEB$ covariance matrix for the compact, orientable Euclidean topologies \E{1}--\E{6}.\footnote{
    The non-compact cases \E{11}, \E{12}, and \slabh\ can be treated as limiting cases of \E{1} and \E{2}, while a detailed study of \slabi\ is reserved for future work.}
We perform numerical computations of the covariance matrix elements for representative manifolds of each topology, exploring the patterns of non-zero elements and their connection to the symmetries broken in these spaces. 
Additionally, we investigate the distinguishability of the resulting covariance matrices from the covering space by calculating their KL divergence, to be explained in the next section.

\subsection{KL divergence}

A key question in studying non-trivial topologies of the Universe is whether the associated probability distributions of observables, such as CMB fluctuations, are sufficiently distinct from those in the trivial topology (the covering space) to enable the discovery of the topology. We denote the probability distributions of the $a_{\ell m}$ of the CMB under the hypothesis of non-trivial and trivial topology by $p(\{a_{\ell m}\})$ and $q(\{a_{\ell m}\})$, respectively.
One effective way to quantify the detectability of the difference between $p$ and $q$  is using the KL divergence \cite{kullback1951, kullback1959information}, also known as the relative entropy.
      
The KL divergence quantifies the information lost when assuming an  (``incorrect'') model $q$ instead of the (``correct'') model $p$ and is defined as
\begin{equation}
    D_{\mathrm{KL}}(p || q) = \int \mathrm{d}\{a_{\ell m}\} \,\, p(\{a_{\ell m}\}) \ln \! \left[\frac{p(\{a_{\ell m}\})}{q(\{a_{\ell m}\})} \right]\;.
\end{equation}
For the CMB the $a^X_{\ell m}$ coefficients follow zero-mean Gaussian distributions allowing the KL divergence to be simplified to
\begin{equation}
    D_{\mathrm{KL}}(p || q) = \frac12 \sum_j \left(\ln |\lambda_j|+\lambda_j^{-1} - 1 \right)\,,\label{eq:DKLpq}
\end{equation}
where the $\{\lambda_i\}$ are the eigenvalues of the matrix 
\begin{equation*}
 C_{\ell m\ell' m'}^{XY,\,p} \, (C_{\ell m\ell' m'}^{XY,\, q})^{-1}\,.
\end{equation*}
We can also reverse the question: ``How much information is lost if we assume the model $q$ is represented by the model $p$?'' In our case, this quantifies whether CMB observations coming from the trivial topology (model $q$) are expected to be distinguishable from those coming from non-trivial topologies (model $p$). Since the eigenvalues of the inverse of the matrix $C_{\ell m\ell' m'}^{XY,\,p} \, (C_{\ell m\ell' m'}^{XY,\, q})^{-1}$ are simply \( 1/\lambda_j \), we find
\begin{equation}
    D_{\mathrm{KL}}(q || p) = \frac{1}{2} \sum_i \left(-\ln |\lambda_j|+\lambda_j - 1 \right).\label{eq:DKLqp}
\end{equation}
In our case, both questions yield similar answers.
We will focus primarily on \( D_{\mathrm{KL}}(p || q) \), studying what happens when observations coming from a non-trivial topology are interpreted as coming from $\Lambda$CDM.

\subsection{Evaluation of CMB covariance matrices}
\label{subsecn:numerical_covariances}
In the covering space \E{18}, the covariance matrix elements are efficiently computed numerically as a one-dimensional integral over the magnitude of $|\vec{k}| \in (0, \infty)$.
For non-trivial topologies, the discretization of the wavevectors transform integrals into sums whose dimensionality depends on whether the discretization is partial or complete, i.e., on how many of the three dimensions are compactified.
For the fully compact, orientable Euclidean topologies, this summation can be expressed in the form of Eq.~\eqref{eqn:HarmonicCovariance}, which we rewrite here for the auto- and cross-correlations of the CMB anisotropies $XY \in \{TT, EE, BB, TE, EB, TB\}$:

\begin{align}
\label{eqn:general_covariance_matrix}
C^{\E{i}; XY}_{\ell m\ell'm'}  = 
     \frac{\pi^2}{{2 V_{\E{i}}}} \sum_{\lambda = \pm 2}
    \sum_{\vec{n}\in \setN^{\E{i}}}
    \frac{\Tpower(k_{\vec{n}})}{k_{\vec{n}}^3 } 
    \Delta^{X}_{\ell}(k_{\vec{n}})
    \Delta^{Y*}_{\ell'}(k_{\vec{n}}) \ 
    \xi^{\E{i};X,\unitvec{k}_{\vec{n}}}_{k_{\vec{n}},\ell m,\lambda}\ 
    \xi^{\E{i};Y,\unitvec{k}_{\vec{n}}*}_{k_{\vec{n}},\ell' m',\lambda} .
\end{align}
This summation is over an infinite set of wavevectors, making it computationally infeasible to evaluate directly. In our previous work on scalar eigenmodes \cite{COMPACT:2023rkp}, we developed a method to compute this sum for reasonably large topological scales, particularly for Dirichlet domains that fully contain the LSS. This method consists in truncating the summation at a maximum wavevector magnitude \(|\vec{k}_{\mathrm{max}}(\ell)|\), allowing for practical computation while maintaining high accuracy.

This requires a precise definition for our multipole-dependent cutoff, \( |\vec{k}_{\textrm{max}}(\ell)| \). To address this, we employ two distinct methods: one for parity-even correlations ($TT$, $EE$, $BB$, $TE$) and another for parity-odd correlations ($EB$, $BT$), which are exactly zero in the covering space.

For \(XY \in \{TT, EE, BB, TE\}\), we define the ratio function
\begin{equation}
    \label{eqn:cut_off_definition}
    R_\ell(|\vec{k}|) \equiv \frac{C^{|\vec{k}|, XY}_{\ell}}{C^{\Lambda\textrm{CDM}, XY}_{\ell}},
\end{equation}
where
\begin{align}
    C^{|\vec{k}|, TT}_\ell &= \frac{\pi}{4} \frac{(\ell + 2)!}{(\ell - 2)!} 
    \int^{|\vec{k}|}_0 \frac{\mathrm{d}k'}{k'} \, \Tpower(k') \Delta^{T}_\ell(k')^2, \\
    C^{|\vec{k}|, EE}_\ell &= \frac{\pi}{4} 
    \int^{|\vec{k}|}_0 \frac{\mathrm{d}k'}{k'} \, \Tpower(k') \Delta^E_\ell(k')^2, \\
    C^{|\vec{k}|, BB}_\ell &= \frac{\pi}{4} 
    \int^{|\vec{k}|}_0 \frac{\mathrm{d}k'}{k'} \, \Tpower(k') \Delta^B_\ell(k')^2, \\
    C^{|\vec{k}|, TE}_\ell &= \frac{\pi}{4} \sqrt{\frac{(\ell + 2)!}{(\ell - 2)!}} 
    \int^{|\vec{k}|}_0 \frac{\mathrm{d}k'}{k'} \, \Tpower(k') \Delta^{T}_\ell(k') \Delta^E_\ell(k'),
\end{align}
\noindent and \(C^{\Lambda\textrm{CDM}, XY}_\ell\) is the standard \(\Lambda\)CDM angular power spectrum generated by \texttt{CAMB} \cite{Lewis:1999bs,2011ascl.soft02026L} for \E{18}. 

We define \( |\vec{k}_{\textrm{max}, XY}(\ell)| \) 
as the smallest $|\vec{k}|$ (among allowed $\vec{k}$)
satisfying
$R_\ell(|\vec{k}_{\textrm{max}, XY}(\ell)|) \ge 0.99$. 
In the limit \(L \to \infty\), this approach yields a power spectrum that is 1\% smaller than the actual power spectrum. For the computation of off-diagonal elements where \(\ell \neq \ell'\), we perform the summation in \cref{eqn:general_covariance_matrix} up to 
\( 
    |\vec{k}_{\textrm{max}}(\max(\ell, \ell'))| 
    \equiv
    \max(|\vec{k}_{\textrm{max}}(\ell)|,|\vec{k}_{\textrm{max}}(\ell')|) 
\). 
We find that increasing the precision beyond 99\% has a negligible impact on the KL divergence.

For $EB$ and $TB$ correlations, which are strictly zero in the standard \(\Lambda\)CDM framework, we adopt a different strategy. Initially, we select the maximum $|\vec{k}_{\textrm{max}, XY}(\ell)|$ for each \(\ell\) among the $TT$, $EE$, $BB$, and $TE$ correlations. We then increase this value slightly to check whether it leads to any significant changes in the $EB$ and $TB$ correlations. Since no changes are observed, we retain the same maximum $|\vec{k}_{\textrm{max}, XY}(\ell)|$ for these correlations as well.
For the computation of off-diagonal elements where \(\ell \neq \ell'\), we similarly perform the summation in \cref{eqn:general_covariance_matrix} up to \( |\vec{k}_{\textrm{max}, XY}(\max(\ell, \ell'))| \). As before, increasing the precision beyond 99\% has a negligible impact on the KL divergence.

We solve these equations using a specialized Python code developed for this purpose.  
To compute the transfer function, $\Delta^{X}_\ell(k)$, and the primordial tensor power spectrum, $\Tpower(k)$, we use \texttt{CAMB} with the {\it Planck} 2018 best-fit \(\Lambda\textrm{CDM}\) cosmological parameters \cite{Planck:2018vyg} and the tensor-to-scalar ratio $n_T=-0.0128$.  (Note that the results are independent of $r$, for $r\neq0$, since we are considering pure tensor perturbations in the absence of noise.)

\begin{figure}[h!]
    \centering
    \textbf{\E{1}}\par\medskip
    \begin{subfigure}[b]{\linewidth}
        \centering
        \includegraphics[width=\linewidth]{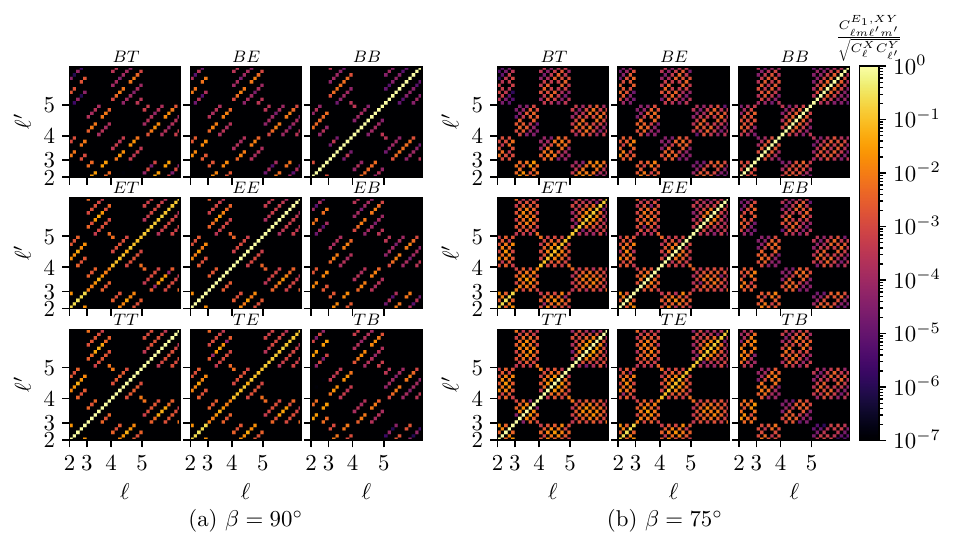}
    \end{subfigure}

    \begin{subfigure}[b]{\linewidth}
        \centering
        \includegraphics[width=\linewidth]{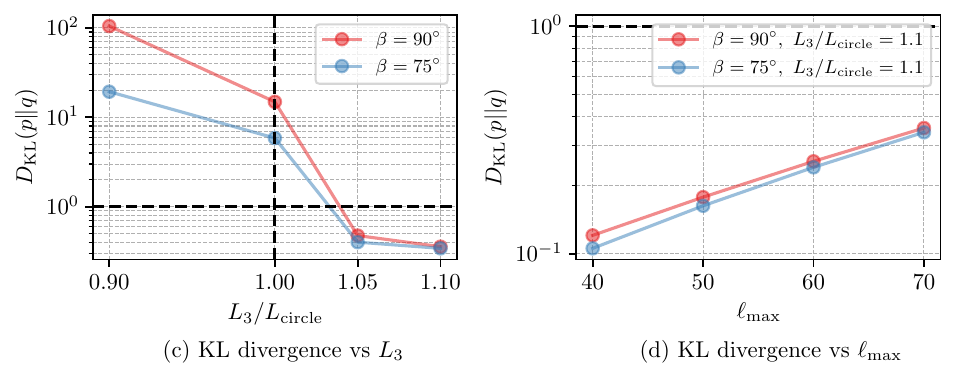}
    \end{subfigure}
    \caption{Upper panels: absolute values of the rescaled full \emph{TEB} CMB covariance matrix, $\Xi^{\E{1}; \mathrm{XY}}_{\ell m \ell' m'}$, for the \E{1} topology at low multipoles ($\ell_{\rm max} = 5$). The topology parameters are set to $L_1 = L_2 = 1.4 L_{\mathrm{LSS}}$ and $L_3 = L_{\mathrm{LSS}}$, with the left panel corresponding to the untilted case ($\beta = 90^\circ$) and the right panel to the tilted case ($\beta = 75^\circ$).
    Lower panels: KL divergence for the \E{1} topology. The left panel illustrates the KL divergence as a function of $L_3 / L_{\mathrm{circle}}$ for both untilted ($\beta = 90^\circ$) and tilted ($\beta = 75^\circ$) configurations, calculated up to $\ell_{\mathrm{max}} = 70$. Circles denote computed data points, while solid lines connect these points for visual continuity. Here, $L_{\mathrm{circle}} = L_{\mathrm{LSS}}$ represents the smallest value of $L_3$ at which no pairs of identical circles appear in the CMB. The right panel presents the behavior of the KL divergence as a function of $\ell_{\mathrm{max}}$ for $L_3 = 1.1 L_{\mathrm{LSS}}$, comparing the tilted and untilted configurations.}
    \label{fig:cov_matrix_E1}
\end{figure}

Before presenting the covariance matrices for all compact, orientable topologies, it is important to note that the choice of the orientation of the coordinate system affects the appearance and symmetries of the covariance matrix. To facilitate comparison across topologies, for our figures we always 
consider cases where the translation vectors $\vec{T}_i$ of the associated \E{1} of the manifold are orthogonal to one another and parallel to the axes of our coordinate system (except as described for \E{1} in \cref{fig:cov_matrix_E1}).
Moreover, in order to compare with the trivial topology, we plot the \emph{rescaled} covariance matrix, 
\begin{equation}
    \Xi^{\E{i}; XY}_{\ell m\ell'm'} \equiv \frac{C^{\E{i}; XY}_{\ell m\ell'm'}} {\sqrt{C^{\Lambda \mathrm{CDM}; XX}_{\ell}C^{\Lambda \mathrm{CDM}; YY}_{\ell'}}}.
\end{equation}

\begin{figure}[h!]
     \centering
    \textbf{\E{2}}\par\medskip
    \begin{subfigure}[b]{\linewidth}
        \centering
        \includegraphics[width=\linewidth]{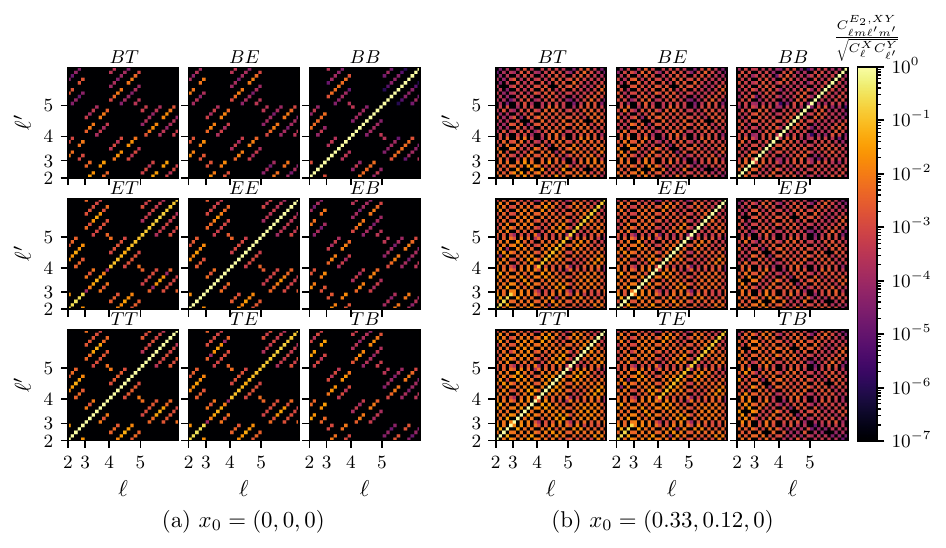}
    \end{subfigure}

    \begin{subfigure}[b]{\linewidth}
        \centering
        \includegraphics[width=\linewidth]{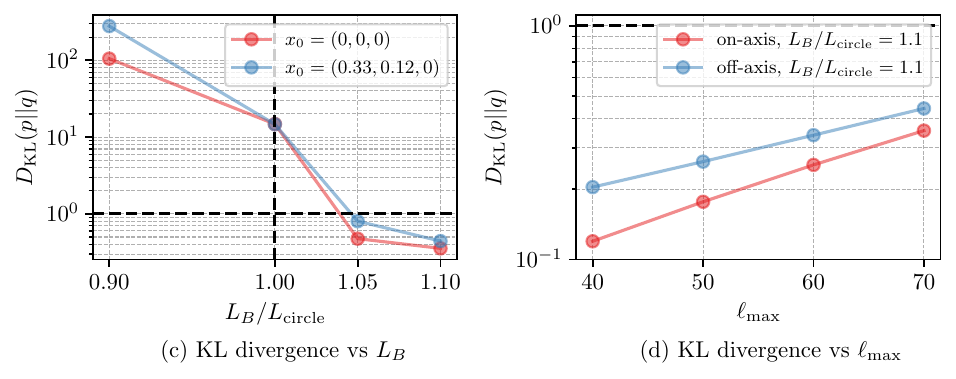}
    \end{subfigure}
    
    \caption{Upper panels: absolute values of the rescaled full \emph{TEB} CMB covariance matrix, $\Xi^{\E{2}; \mathrm{XY}}_{\ell m \ell' m'}$, for low multipoles ($\ell_{\rm max} = 5$) in the \E{2} topology. The results are shown for an untilted \E{2} topology with $L_{A_1} = L_{A_2} = 1.4 \LSS$ and $L_B = L_{\mathrm{circle}}$. The left panel corresponds to an on-axis observer, while the right panel represents an off-axis observer located at $\vec{x}_0 = (0.33, 0.12,0)L_\mathrm{LSS}$.
    Lower panels: KL divergence for the covariance matrices above. The left panel shows the KL divergence as a function of $L_B / L_{\mathrm{circle}}$, calculated up to $\ell_{\mathrm{max}} = 70$. For the on-axis observer, $L_B = L_{\mathrm{circle}} = L_{\mathrm{LSS}}$, while for the off-axis observer, $L_B = L_{\mathrm{circle}} \approx 0.71 L_{\mathrm{LSS}}$. The right panel shows the KL divergence as a function of $\ell_{\mathrm{max}}$ for $L_B = 1.1 L_{\mathrm{LSS}}$, highlighting differences between on-axis and off-axis observers. Circles denote computed data points, while solid lines connect these points for visual continuity.}
    \label{fig:cov_matrix_E2}
\end{figure}

\begin{figure}[h!]
     \centering
    \textbf{\E{3}}\par\medskip
    \begin{subfigure}[b]{\linewidth}
        \centering
        \includegraphics[width=\linewidth]{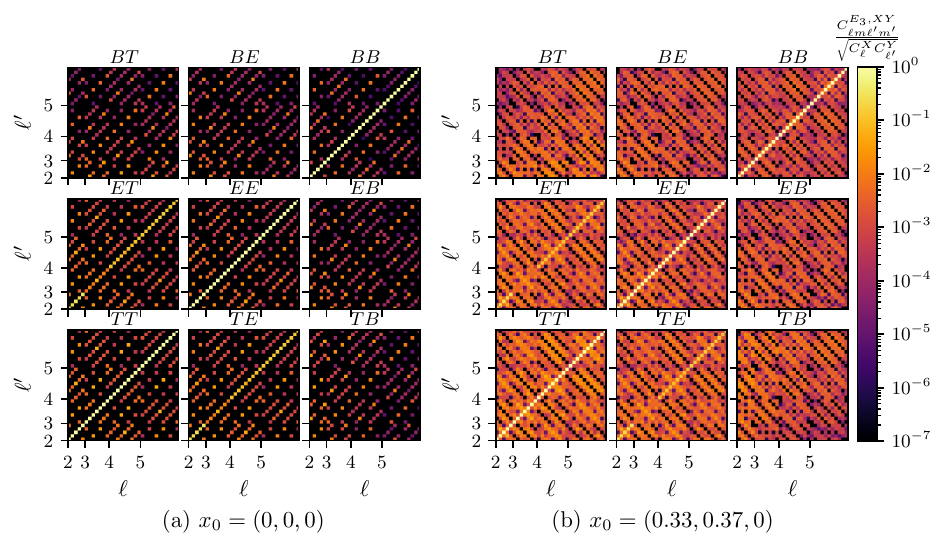}
    \end{subfigure}

    \begin{subfigure}[b]{\linewidth}
        \centering
        \includegraphics[width=\linewidth]{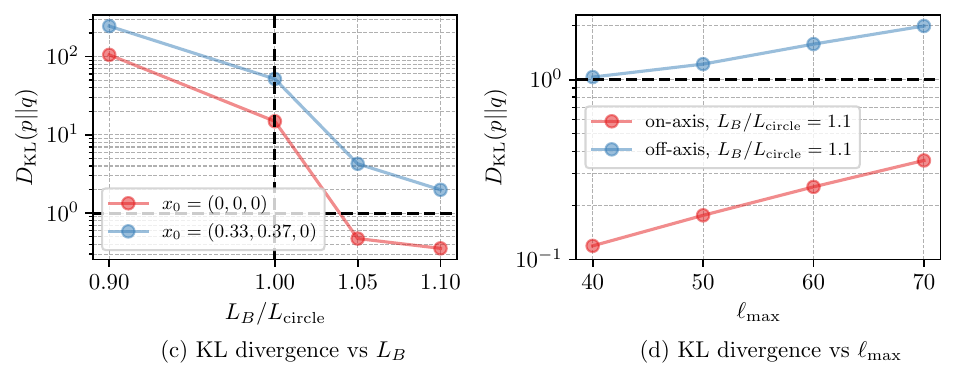}
    \end{subfigure}
    \caption{As in \cref{fig:cov_matrix_E2}, but for the \E{3} topology. Here, $L_A = 1.4 L_{\mathrm{LSS}}$, and the off-axis observer is located at $\vec{x}_0 = (0.33, 0.37, 0) \LSS$. The circle limit is $L_{\mathrm{circle}} = L_{\mathrm{LSS}}$ for the on-axis observer and $L_{\mathrm{circle}} \approx 0.71 \LSS$ for the off-axis observer. The left lower panel shows the KL divergence up to $\ell_{\mathrm{max}} = 70$.}
    \label{fig:cov_matrix_E3}
\end{figure}

\begin{figure}[h!]
    \centering
    \textbf{\E{4}}\par\medskip
    \begin{subfigure}[b]{\linewidth}
        \centering
        \includegraphics[width=\linewidth]{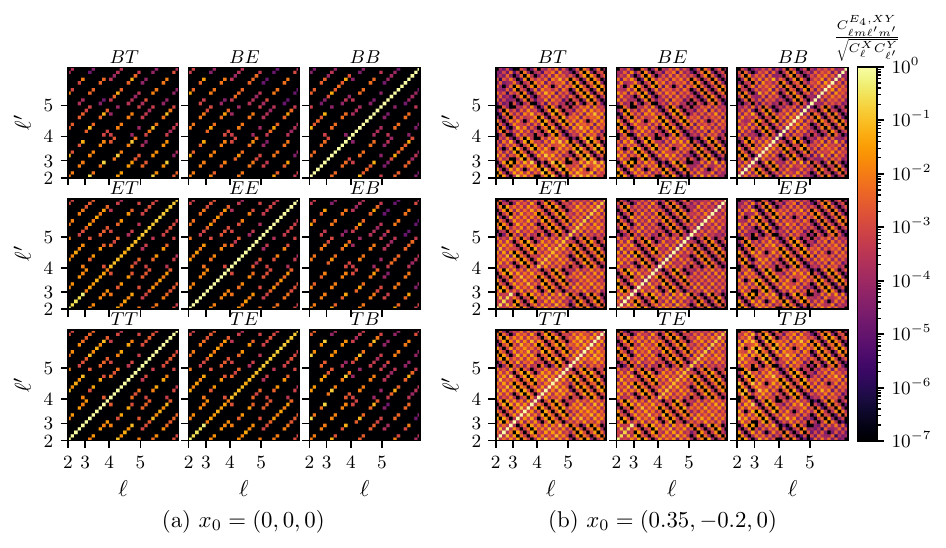}
    \end{subfigure}

    \begin{subfigure}[b]{\linewidth}
        \centering
        \includegraphics[width=\linewidth]{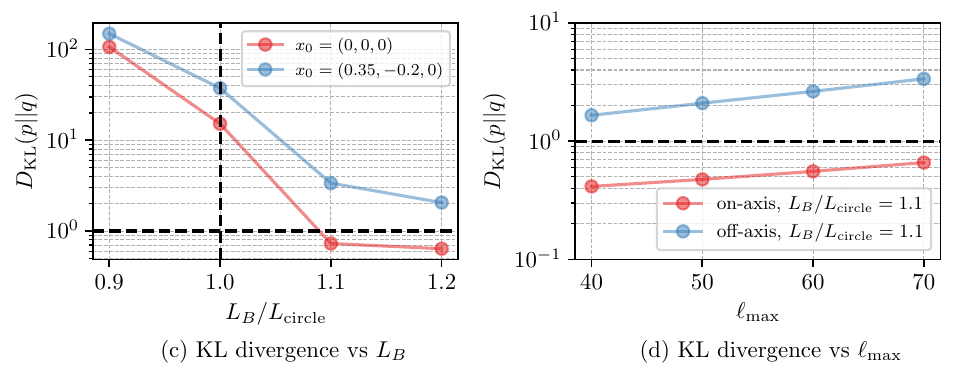}
    \end{subfigure}  
    \caption{As in \cref{fig:cov_matrix_E2,fig:cov_matrix_E3}, but for the \E{4} topology. Here, we set $L_A = 1.4\LSS$, and the off-axis observer is located at $\vec{x}_0 = (0.35, -0.2, 0)\LSS$. The circle limit remains $L_{\mathrm{circle}} = L_{\mathrm{LSS}}$ for the on-axis observer and $L_{\mathrm{circle}} \approx 0.71 \LSS$ for the off-axis observer. The bottom left panel shows the KL divergence up to $\ell_{\mathrm{max}} = 70$.}
    \label{fig:cov_matrix_E4}
\end{figure}

\begin{figure}[h!]
    \centering
    \textbf{\E{5}}\par\medskip
    \begin{subfigure}[b]{\linewidth}
        \centering
        \includegraphics[width=\linewidth]{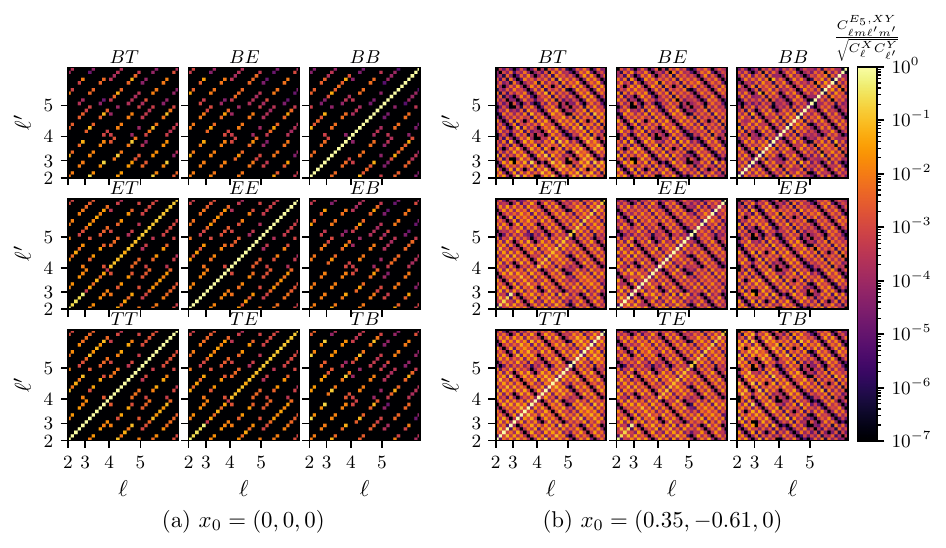}
    \end{subfigure}

    \begin{subfigure}[b]{\linewidth}
        \centering
        \includegraphics[width=\linewidth]{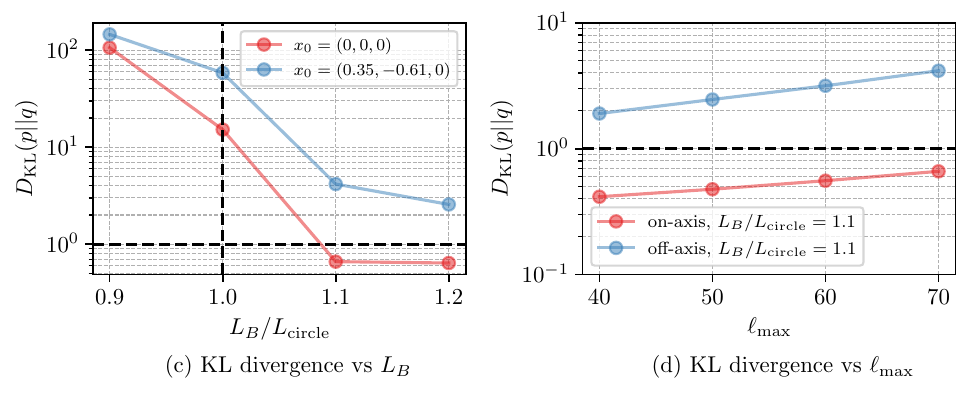}
    \end{subfigure}
    \caption{As in \cref{fig:cov_matrix_E2,fig:cov_matrix_E3,fig:cov_matrix_E4}, but for the \E{5} topology. Following the \E{4} topology, we use $L_A = 1.4 \LSS$, with the off-axis observer located at $\vec{x}_0 = (0.35, -0.61, 0) \LSS$. The circle limit remains $L_{\mathrm{circle}} = L_{\mathrm{LSS}}$ for the on-axis observer and $L_{\mathrm{circle}} \approx 0.71 \LSS$ for the off-axis observer. The bottom left panel shows the KL divergence up to $\ell_{\mathrm{max}} = 70$.}
    \label{fig:cov_matrix_E5}
\end{figure}

\begin{figure}[h!]
     \centering
    \textbf{\E{2} \hspace{6cm} \E{3}}\par\medskip
    \begin{subfigure}[b]{\linewidth}
        \centering
        \includegraphics[width=\linewidth]{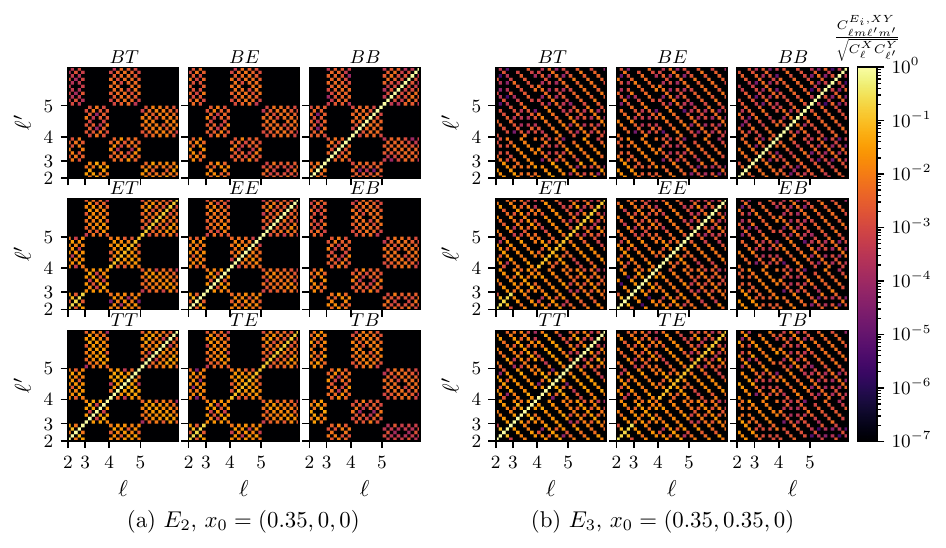}
    \end{subfigure}

    \begin{subfigure}[b]{\linewidth}
        \centering
        \includegraphics[width=\linewidth]{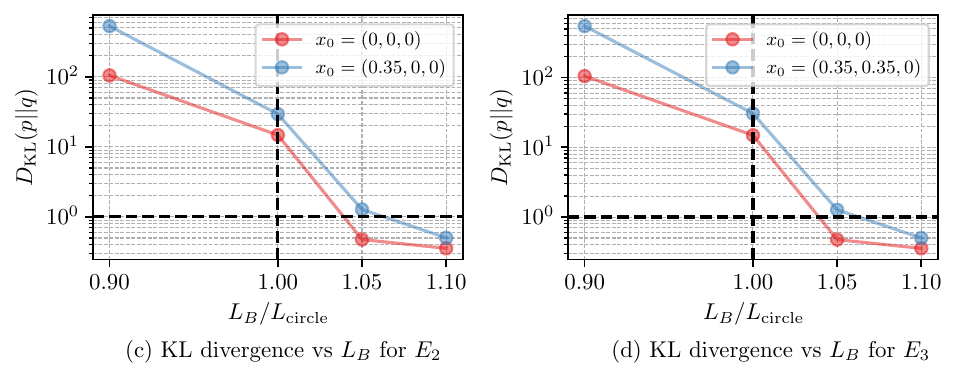}
    \end{subfigure}
    \caption{Rescaled pure tensor \emph{TEB} covariance matrices (upper panels) and corresponding KL divergences (lower panels) for the two off-axis observers in \E{2} and \E{3} used in \rcite{COMPACT:2023rkp} in comparison to the corresponding on-axis observers. Here, $L_A = 1.4 L_{\mathrm{LSS}}$ and $L_B$ is varied in the KL divergence plots. Both off-axis observers share the same volume with $L_{\mathrm{circle}} \approx 0.71$.
}
    \label{fig:cov_matrix_E2_E3}
\end{figure}

\begin{figure}[h!]
    \centering
    \textbf{\E{6}}\par\medskip
    \begin{subfigure}[b]{\linewidth}
        \centering
        \includegraphics[width=\linewidth]{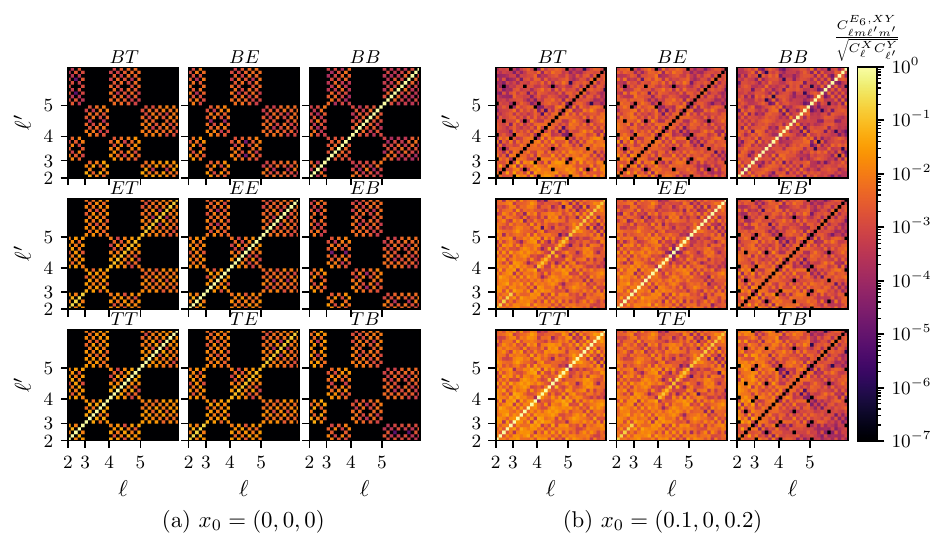}
    \end{subfigure}

    \begin{subfigure}[b]{\linewidth}
        \centering
        \includegraphics[width=\linewidth]{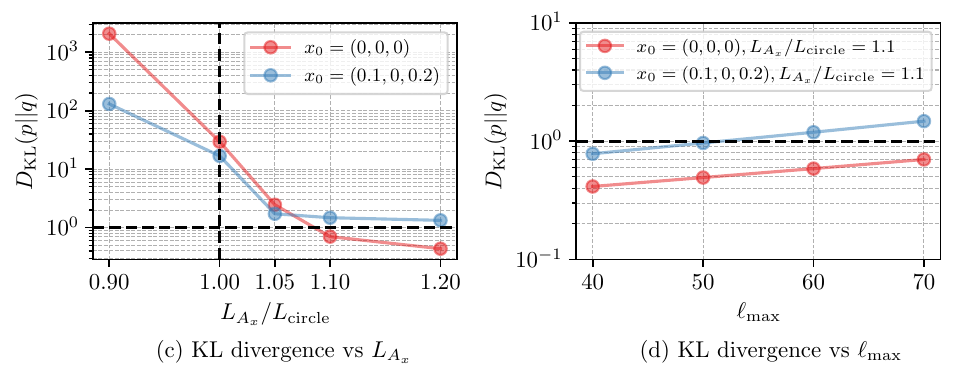}
    \end{subfigure}
    \caption{
    Upper panels: absolute values of the rescaled full \emph{TEB} CMB covariance matrix, $\Xi^{\E{6}; \mathrm{XY}}_{\ell m \ell' m'}$, for low multipoles ($\ell_{\rm max} = 5$) in the \E{6} topology. The results are shown for an untilted \E{6} topology with $r_x = r_y = r_z = \frac{1}{2}$ and $L_{A_x} = L_{\mathrm{circle}}$, $L_{B_y} = 0.8 \LSS$, and $L_{C_z} = \LSS$. The left and right panels correspond to an on-axis observer ($L_{\mathrm{circle}} = 0.6 \LSS$) and an off-axis observer ($L_{\mathrm{circle}} = 0.5 \LSS$, located at $\vec{x}_0 = (0.1, 0, 0.2)\LSS$), respectively.
    Lower panels: KL divergence as a function of $L_{A_x} / L_{\mathrm{circle}}$ up to $\ell_{\mathrm{max}} = 70$ (left panel) and as a function of $\ell_{\mathrm{max}}$ for $L_{A_x} = 1.1 L_{\mathrm{circle}}$ (right panel), highlighting differences between on-axis and off-axis observers. Circles denote computed data points, while solid lines connect these points for visual continuity.}
    \label{fig:cov_matrix_E6}
\end{figure}

In the upper panels of Figs.~\ref{fig:cov_matrix_E1}--\ref{fig:cov_matrix_E6}, we present the modulus of the rescaled pure-tensor \emph{TEB} covariance matrices, $\Xi^{\E{i}, XY}_{\ell m \ell' m'}$, and their corresponding KL divergence for the \E{1}--\E{6} topologies; however, we emphasize that these are inherently complex matrices. 
The covariance matrices are displayed in ``$\ell$ ordering,'' meaning that the entries are sequenced by increasing values of the multipole $\ell$ and, within each $\ell$, further organized by increasing \( m \). The $\{\ell, m\}$ elements are indexed as \( s = \ell(\ell+1) + m \), with \( m \) in the range \( -\ell \leq m \leq \ell \).
For each topology, we show the rescaled covariance matrices for two different observers, typically one located on the axis of rotation of the topology and one off-axis. For \E{2} and \E{3} we add a third observer.
The topological length scales and the off-axis positions for these correlation matrices
are selected to ensure that the observer cannot detect matched pairs of circles in the CMB, therefore satisfying current observational constraints on topology following the methodology described in \rcite{COMPACT:2022nsu}.

For \E{1} (upper panels of \cref{fig:cov_matrix_E1}), we set $L_1 = L_2 = 1.4 \LSS$ and $L_3 = \LSS$.
As \E{1} is homogeneous, the covariance matrices are independent of observer location.
We thus present the results for an untilted ($\beta=90^\circ$) configuration (top left panel) and a tilted ($\beta=75^\circ$) one (top right panel), with $\alpha = \gamma = 90^\circ$ in both cases. 
For the on-axis observer in \E{2}--\E{5} (top left panels of Figs.~\ref{fig:cov_matrix_E2}-\ref{fig:cov_matrix_E5}), we choose $L_A = 1.4 \LSS$ and $L_B = \LSS$. As the tilt and off-axis position of the observer are degenerate, we set the tilt parameters to zero. The off-axis observers in these topologies (top right panels of Figs.~\ref{fig:cov_matrix_E2}-\ref{fig:cov_matrix_E5}) are selected to lie between two excluded regions for $L_B < L_{\mathrm{LSS}}$, as illustrated in Fig.~2 of \rcite{COMPACT:2022nsu}. For these observers, we maintain $L_A = 1.4 \LSS$ and set $L_B = L_{\mathrm{circle}}$, where $L_\mathrm{circle}$ is defined as the smallest $L_B$ for which no pairs of identical circles appear on the CMB sky.
In the upper panels of \cref{fig:cov_matrix_E2_E3} we present the covariance matrices for \E{2} (left) and \E{3} (right) for $L_A = 1.4 \LSS$ and set $L_B = L_{\mathrm{circle}}$ (as in \cref{fig:cov_matrix_E2} and \cref{fig:cov_matrix_E3}) for the off-axis observer position considered for those manifolds in \rcite{COMPACT:2023rkp}.
Finally, for \E{6} (upper panels of \cref{fig:cov_matrix_E6}), one observer (top left panel) is located at the origin, with $L_{Ax} = 0.6 \LSS$, $L_{By} = 0.8 \LSS$, $L_{Cz} = \LSS$, and $r_x = r_y = r_z = 1/2$.
For an off-axis observer in \E{6} (top right panel), we set $L_{Ax}=L_{\mathrm{circle}}$, $L_{By} = 0.8 \LSS$, $L_{Cz} = \LSS$, and $r_x = r_y = r_z = 1/2$. We then choose the observer location as $\vec{x}_0 = (0.1, 0, 0.2)\LSS$. For this observer, $L_{\mathrm{circle}} = 0.5 \LSS$, where $L_{\mathrm{circle}}$ is the smallest value of $L_{A_x}$ with no matched circles on the sky.
These choices enable us to explore the allowed parameter space for smaller topology sizes, potentially enhancing the topological information encoded in the CMB.

The properties and symmetry violations in these topologies have been extensively discussed in \rcite{COMPACT:2023rkp,COMPACT:2024cud}. 
We briefly review these findings here and discuss new results where appropriate.

The \E{1} topology (\cref{fig:cov_matrix_E1}) is homogeneous and hence parity conserving for all observers, so
correlations vanish for odd $(\ell + \ell')$ in \( TT \), \( EE \), \( BB \), and \( TE \), and for even $(\ell + \ell')$ in \( EB \) and \( TB \) correlations. 
In the untilted \E{1} case (top left panel), correlations for the $m-m'\not\equiv 0\Mod{4}$ modes also vanish, in contrast to the tilted \E{1} case (top right panel), where such correlations do not necessarily vanish. 
This behavior arises as a consequence of the untilted manifold's accidental 
discrete 4-fold rotational symmetry about the $z$-axis
and our choice to orient the axes in the untilted \E{1} parallel to the translation vectors defining the generators.

For \E{2}--\E{6},  the manifold is no longer homogeneous: the axis of rotation introduces a preferred location that breaks statistical homogeneity. For most or all observers (as specified below) parity is then violated, potentially inducing non-zero elements in both even and odd  $(\ell + \ell')$ blocks across all \( TT \), \( EE \), \( BB \), \( TE \), \( EB \), and \( TB \) correlations in the \emph{TEB} covariance matrices. 
The patterns of these correlations of course vary depending on the specific values of the parameters of the manifold and the location of the observer.

For example, in \E{2} (\cref{fig:cov_matrix_E2}), when the observer is positioned along the axis of rotation (top left panel), the correlations exhibit symmetries similar to those in \E{1}, preserving parity in the 
$XY$ correlations. However, when the observer is off-axis in \E{2} (top right panel), parity is violated
except at certain special observer locations, such as shown in the top left panel of \cref{fig:cov_matrix_E2_E3}.
Meanwhile, in \E{6} (\cref{fig:cov_matrix_E6}) parity is preserved only when the observer is at the intersection of all three rotation axes (top left panel), where the clone pattern maintains reflection symmetry. Elsewhere (e.g., top right panel), this symmetry is lost and parity is violated.

In contrast, the \E{3}--\E{5} topologies (Figs.~\ref{fig:cov_matrix_E3}-\ref{fig:cov_matrix_E5}) exhibit parity violation even for on-axis observers (top left panels). Since rotations of $\pi/2$ (\E{3}), $2\pi/3$ (\E{4}), and $\pi/3$ (\E{5}) are not equivalent to their negative counterparts ($-\pi/2$, $-2\pi/3$, and $-\pi/3$, respectively), these spaces acquire a handedness, and parity is violated for all observer locations. Moreover, \E{3}--\E{5} induce non-zero diagonal elements in $(\ell, m)$ within the \emph{EB} and \emph{TB} covariance matrices for both on-axis (top left) and off-axis (top right) observers, while, \E{2} (\cref{fig:cov_matrix_E2}) and \E{6} (\cref{fig:cov_matrix_E6}) have zero diagonal elements in \emph{EB} and \emph{TB}, even for off-axis observers (top right) where parity is violated. 
This is easier to see in \E{6}, as shown in the top right panel of \cref{fig:cov_matrix_E6}. 
This behavior arises from the specific symmetry properties of \E{2} and \E{6}, which prohibit diagonal \emph{EB} and \emph{TB} correlations that are diagonal in $(\ell, m)$.

These effects underscore the importance of considering the full covariance matrix, particularly the off-diagonal elements, to discern the topological signature from other potential parity-violating signals, such as cosmic birefringence \cite{Diego-Palazuelos:2022dsq,Cosmoglobe:2023pgf}. By examining the \emph{TEB} correlations for \E{1}--\E{6}, we highlight the interplay between topology, parity violation, and tensor perturbations, offering a framework for distinguishing these effects in observational data.

In \rcite{COMPACT:2024cud}, we studied the behavior of the KL divergence for tensor-induced \emph{EE}, \emph{BB}, and \emph{EB} correlations as a function of topology scale for \E{2}. Here we offer much more complete results, extending the analysis to encompass the tensor-induced full \emph{TEB} correlations for \E{1}--\E{6}.

The lower left panels in \cref{fig:cov_matrix_E1,fig:cov_matrix_E2,fig:cov_matrix_E3,fig:cov_matrix_E4,fig:cov_matrix_E5,fig:cov_matrix_E2_E3,fig:cov_matrix_E6} 
show the KL divergence, $D_{\mathrm{KL}}(p || q)$, computed up to $\ell_{\mathrm{max}} = 70$, for the topology parameters and observer locations corresponding to the covariance matrices displayed (for $\ell=2-5$) in the top panels; however, one topology parameter is varied while the others are held fixed. 
Our analysis shows that the two statistics
$D_{\mathrm{KL}}(p || q)$ and 
$D_{\mathrm{KL}}(q || p)$, where $p$ represents the probability distribution for the non-trivial topology and $q$ corresponds to the probability distribution for the covering space,
exhibit similar behavior and convey comparable information. Consequently, we only present $D_{\mathrm{KL}}(p || q)$.

For \E{1}, we keep $ L_1 = L_2 = 1.4 \LSS $ and vary $L_3/L_{\mathrm{circle}}$ in the bottom left panel of \cref{fig:cov_matrix_E1}. Here, $L_{\mathrm{circle}}$ represents the smallest value of $L_3$ for which no matched circles appear in the CMB. In both the tilted and untilted \E{1} configurations, $ L_{\mathrm{circle}}$ is always equal to $L_{\mathrm{LSS}}$ for all observers. 
For $L_3<\LSS$, the CMB exhibits matched circles and the KL divergence is large, but 
as $L_3$ increases, the KL divergence gradually decreases. 
Once $L_3\geq\LSS$, the CMB no longer exhibits matched circles. 
As $L_3$ is increased further, the KL divergence eventually drops below the conventional model-distinguishability threshold (i.e., $D_{\mathrm{KL}}(p || q) = 1$) for both the tilted and untilted configurations. 
With further increases in the topology size, the KL divergence continues to decline at a slower rate and the difference between the tilted and untilted configurations diminishes, converging to nearly the same value by $L_3/\LSS = 1.1$, at least for $\ell_{\mathrm{max}}=70$.

\begin{figure}[h!]
    \centering
    \begin{subfigure}[b]{\linewidth}
        \centering
        \includegraphics[width=\linewidth]{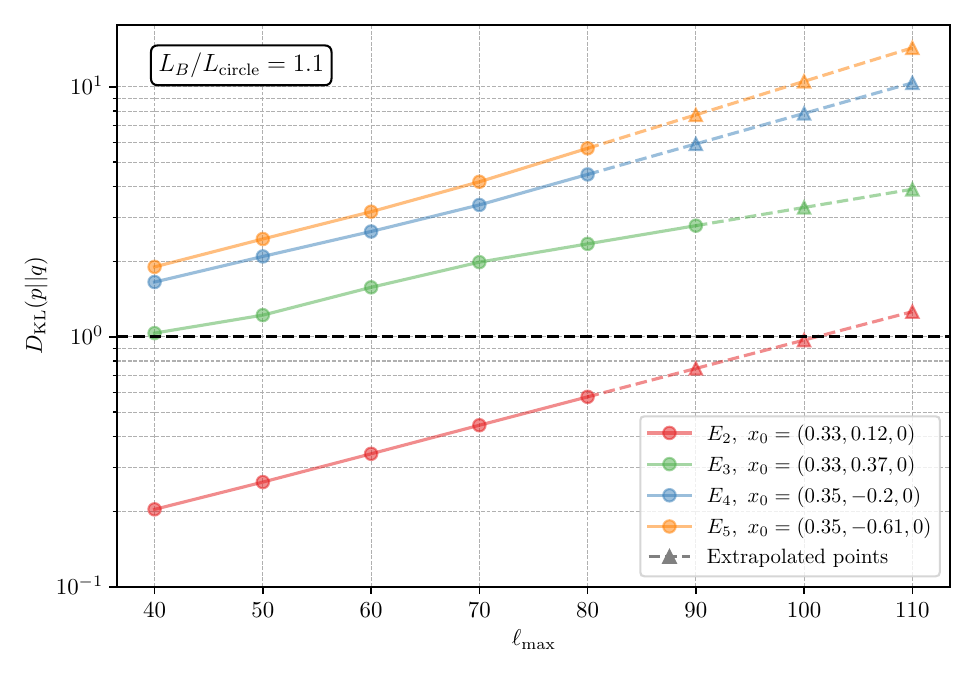}
    \end{subfigure}
 
    \caption{Comparison of $D_{\mathrm{KL}}(p|| q)$ as a function of $\ell_{\mathrm{max}}$ (from 40 to 110) for off-axis observers in \E{2}--\E{5} with $L_B=1.1L_{\mathrm{circle}}$. Circles represent computed data points, while triangles indicate linearly extrapolated values. Solid lines serve as connectors between computed points, and dashed lines represent extrapolated trends. Notably, for the \E{2} topology, the extrapolated KL divergence is above $D_{\mathrm{KL}}(p || q) = 1$ at $\ell_{\mathrm{max}} = 110$. }
    \label{fig:E2_E3_E4_E5_kl}
\end{figure}

In the bottom right panel of \cref{fig:cov_matrix_E1}, we plot the KL divergence as a function of $\ell_{\mathrm{max}}$ at $L_3/\LSS = 1.1$.  
We see that the KL divergence increases significantly with higher $\ell_{\mathrm{max}}$. 
This suggests that, unlike the scalar \emph{TT} case, where the KL divergence saturates at $\ell_{\mathrm{max}} \simeq 30$ (see, e.g., \rcite{Fabre:2013wia, COMPACT:2023rkp}), higher values of $\ell_{\mathrm{max}}$ are required to fully capture cosmic topology information in the tensor \emph{TEB} correlations.
Consequently, the KL divergence values shown in the bottom left panel should be interpreted as lower limits, reflecting only the information contained within tensor modes up to $\ell_{\mathrm{max}} = 70$. 
This raises the question of at what value of $\ell_{\mathrm{max}}$ the KL divergence saturates. 
We return to this question below where we discuss \cref{fig:E2_E3_E4_E5_kl}, showing the dependence of the KL divergence on $\ell_{\mathrm{max}}$ for \E{2}--\E{5}.

For \E{2}--\E{5}, we set $L_A = 1.4L_{\mathrm{LSS}}$ and vary $L_B$ in the KL divergence plots---subfigures (c) of \cref{fig:cov_matrix_E2,fig:cov_matrix_E3,fig:cov_matrix_E4,fig:cov_matrix_E5}.\footnote{To simplify the notation for \E{2}, where the length parameters are $ L_{A_1} $, $ L_{A_2} $, and $ L_B $, we assume $ L_{A_1} = L_{A_2} $ and refer to them collectively as $L_A$ .} The $x$-axis is labeled as $L_B / L_{\mathrm{circle}}$, where $L_{\mathrm{circle}}$ is defined as the smallest $L_B$ for which no identical pairs of circles appear in the CMB. For an on-axis observer, $L_{\mathrm{circle}}= L_{\mathrm{LSS}} $ as long as $ L_A > L_{\mathrm{LSS}}$, and so, in particular here.
For all off-axis observers in \E{2}--\E{5}, we locate the observers such that $L_{\mathrm{circle}} \approx 0.71 L_{\mathrm{LSS}}$.\footnote{Had we chosen $ L_A < L_{\mathrm{LSS}} $, identical circles would always appear in the CMB, regardless of the value of $ L_B $.}

As observed in \E{1}, for topologies \E{2}--\E{5}, when $L_B / L_{\mathrm{circle}} > 1$, the distinct circles in the CMB disappear, resulting in a decrease in the KL divergence and, consequently, a reduction in the information content related to cosmic topology. Notably, the KL divergence for \E{4} and \E{5} is generally larger than for \E{1}--\E{3}, both for on-axis and off-axis observers---this can be attributed to the higher degree of symmetry violation in \E{4} and \E{5}. Additionally, in \E{2}--\E{5}, the KL divergence for off-axis observers exceeds that for on-axis observers, although both converge to similar values as the topology scale increases.

As in the case of \E{1}, we plot the KL divergence, $D_{\mathrm{KL}}(p || q)$, as a function of $\ell_{\mathrm{max}}$ at $L_B / L_{\mathrm{circle}} = 1.1$. This is shown in the bottom left panel of  \cref{fig:cov_matrix_E2,fig:cov_matrix_E3,fig:cov_matrix_E4,fig:cov_matrix_E5}. As evident in these plots, the KL divergence increases with increasing $\ell_{\mathrm{max}}$.

In \cref{fig:cov_matrix_E2_E3}, where we explore the off-axis observers in \E{2} and \E{3} that were considered in \rcite{COMPACT:2023rkp}, we present 
in the lower panels the KL divergence for these observers in comparison to the on-axis observers. 
As before, we set $ L_A = 1.4 L_{\mathrm{LSS}} $ and vary $L_B$ in the KL divergence plots for both \E{2} and \E{3}.
These two off-axis observers share the same volume and have  
$ L_{\mathrm{circle}} \approx 0.71$, consistent with the previously analyzed off-axis observers in \cref{fig:cov_matrix_E2,fig:cov_matrix_E3}. 
However, as evident from the KL divergences, the results differ between the two cases. 
This discrepancy arises because the $XY$ correlations of this off-axis observer in \E{2} do not violate parity and the clone pattern remains symmetric around the observer---unlike the case in \cref{fig:cov_matrix_E2}. 
In contrast, for \E{3} the parity violation is solely due to chirality, whereas in \cref{fig:cov_matrix_E3} it results from both chirality and an asymmetric clone pattern around the observer.

Finally, in the bottom left panel of \cref{fig:cov_matrix_E6}, we display the KL divergence of \E{6}, where we set $L_{B_y} = 0.8 \LSS$ and $L_{C_z} = \LSS$ while varying $L_{A_x}/ L_{\mathrm{circle}}$ for both on-axis and off-axis observers. For the off-axis observer, $L_{\mathrm{circle}} = 0.5 \LSS$, whereas for the on-axis observer $L_{\mathrm{circle}} = 0.6 \LSS$. Additionally, the bottom right panel of \cref{fig:cov_matrix_E6} shows the KL divergence as a function of $\ell_{\mathrm{max}}$ (ranging from $40$ to $70$) with $L_{A_x}/L_{\mathrm{circle}} = 1.1$.

In \cref{fig:E2_E3_E4_E5_kl}, we compare the KL divergence as a function of $\ell_{\mathrm{max}}$ for the off-axis observers presented in \cref{fig:cov_matrix_E2,fig:cov_matrix_E3,fig:cov_matrix_E4,fig:cov_matrix_E5}. 
We extend the results to $\ell_{\mathrm{max}} = 80$ for \E{2}, \E{4}, and \E{5}, and to $\ell_{\mathrm{max}} = 90$ for \E{3}. 
Calculations for larger values of $\ell_{\mathrm{max}}$ would strain our current computational resources.
We then apply a linear extrapolation of the logarithm of the KL divergence values to estimate the KL divergence for all cases up to $\ell_{\mathrm{max}} = 110$, although the KL divergence may eventually converge at sufficiently high $\ell_{\mathrm{max}}$. These off-axis observers share the same volume, $L_{\mathrm{circle}} \approx 0.71$, making them a suitable case study. As shown in the figure, all these
topologies exhibit an increasing $D_{\mathrm{KL}}(p\|q)$ as $\ell_{\mathrm{max}}$ grows with nearly similar growth rates---a behavior consistent with that observed for both on-axis and off-axis observers.
This suggests that the growth in the KL divergence is a generic feature, largely independent of the specific topology parameters. In particular, for \E{2}, where $D_{\mathrm{KL}}(p\|q) < 1$ at $\ell_{\mathrm{max}} = 80$, the trend indicates that it will likely surpass the $D_{\mathrm{KL}}(p\|q) = 1$ threshold at higher \(\ell_{\mathrm{max}}\). Meanwhile, for \E{3}--\E{5}, which already have $D_{\mathrm{KL}}(p\|q) > 1$, the KL divergence is expected to increase further. Consequently, these findings hint towards the possibility that non-trivial topologies can be promising candidates for detecting tensor perturbations. This work, therefore, provides a window into further investigation of non-trivial topology and its potential signatures in the CMB polarization data.

\section{Conclusions}
\label{secn:conclusions}

Though the local geometry of the Universe appears to be well described by a homogeneous, isotropic FLRW metric, and specifically by the flat FLRW metric, it does not follow that the topology of spatial slices of the Universe is the trivial topology of the Euclidean 3-dimensional space (a.k.a.\ \E{18}, the covering space of the $E^3$ geometry).
In fact, there are $17$ other possible topologies that are compatible with this background metric, named \E{1}--\E{17}.
In this paper, we have focused on the orientable topologies (\E{1}--\E{6}, \E{11}, \E{12}, \slabh, and \slabi)---those on which there is a well-defined sense of chirality (or helicity, or clockwise and counterclockwise rotation).
In a future work we will consider the non-orientable manifolds, which exhibit a new phenomenon---helicity mixing for tensor modes.

In a previous work \cite{COMPACT:2023rkp}, we studied the implications of the orientable Euclidean topologies for scalar fields, with a focus on the CMB temperature anistropies. 
We concluded that all the orientable topologies imprint potentially observable effects on CMB temperature through their breaking of statistical isotropy.
This holds even when the topology scales are somewhat larger than the diameter of the LSS. 
In a later work \cite{COMPACT:2024cud}, we noted that most of these topologies also break statistical homogeneity, and therefore, for a typical observer, statistical parity. 
We pointed out that both the isotropy violation and the parity violation can induce significant CMB $TB$ and $EB$ correlations, forbidden in the usual case of the trivial topology. 
In this paper, we have described in detail how to compute the effects of tensor (spin-2) modes on CMB temperature and polarization for all orientable topologies. 
We have studied only the effects of tensor modes in order to fully understand their rich phenomenology.

In this work we have computed and presented for the first time the eigenmodes of the tensor Laplacian for all the orientable Euclidean topologies.
As in the scalar case, an immediate consequence of non-trivial topology is that only certain Fourier wavevectors $\vec{k}$ are allowed---in the compact manifolds \E{1}--\E{6} these allowed $\vec{k}$ form a lattice. 
For all but \E{1}, \E{11}, and \slabh, generic eigenmodes are no longer individual Fourier modes but are finite linear combinations of them.
Fields that are built out of these eigenmodes by the typical superposition, i.e., with amplitudes that are independent Gaussian random variables of zero mean, are 
not statistically isotropic. 
We have therefore presented for the first time analytic expressions for the correlation matrices $C^{\E{i};XY}_{iji'j', \vec{k} \vec{k}', \lambda \lambda'}$ of observables $X$ and $Y$ deriving from standard covering-space tensor modes, i.e., from  massless plane waves of definite polarization, in each topology \E{i}.
These correlation matrices fully characterize the statistical properties of the three-dimensional fields of the observables.

In the covering space, all values of the wave vector are allowed and the correlation matrices $C^{\E{i};XY}_{iji'j', \vec{k} \vec{k}', \lambda \lambda'}$ are fully diagonal in both wave vector (i.e., non-zero only for $\vec{k}'=\vec{k}$) and plane-wave polarization (i.e., non-zero only for $\lambda'=\lambda$).
In all the orientable topologies, only certain wave vectors are admissible.  This is precisely  the usual discretization of Fourier modes in a periodic box---and indeed for each topology the allowed wave vectors are those that are periodic on the fundamental domain of the associated \E{1} (for the compact topologies \E{1}--\E{6}), associated \E{11} (for \E{11}--\E{12}), and associated $\slabh$ for $\slabi$.
In all the orientable topologies  the correlation matrices remain diagonal in polarization;
however, in all except \E{1}, \E{11} and \slabh\ for almost every diagonal element of the correlation matrix there are  a finite topology-dependent number of off-diagonal terms  in wave vector.

These tensor eigenmodes have also been used to compute the effects of topology on the tensor modes of CMB temperature, and both $E$-mode and $B$-mode photon polarization.
Since the statistics of the tensor fluctuations are those of a zero-mean Gaussian field in three dimensions,
$T$, $E$, and $B$ are zero-mean Gaussian fields on the sphere of the sky, therefore
all the information in spherical-harmonic coefficients $a^X_{\ell m}$ is contained in the covariance between those coefficients: $C_{\ell m \ell' m'}^{XY}$, with $X, Y \in \{ T, E, B \}$. 
Analytic expressions for these covariance matrices for all the orientable Euclidean topologies have been provided.

In order to demonstrate some of the consequences of topology for CMB observables,
some example low-$\ell$ covariance matrices have been shown for both on-axis and off-axis observers (when applicable) for all $6$ compact orientable topologies.
These allowed us to see clearly that the violation of statistical isotropy inherent in all non-trivial topologies drastically alters the correlation matrices of CMB observables.
Whereas in the covering space these matrices were diagonal for pairs of like-parity observables ($TT$, $TE$, $EE$, $BB$) and zero for opposite-parity observables ($TB$, $EB$), now all elements of the correlation matrix can in principle be non-zero.
We emphasize that these parity-violating elements are non-zero despite the absence of microphysical parity violation---topology alone causes parity violation except in very specific cases.
Parity conservation in \E{1} (and \E{11} and \slabh) forces all $(\ell+\ell')$-odd elements of $C_{\ell m \ell' m'}^{XY}$ to be zero for $XY \in \{TT,TE,EE,BB\}$ and all $(\ell+\ell')$-even elements of $C_{\ell m \ell' m'}^{XY}$ to be zero for $XY \in \{TB,EB\}$, leaving about half of all the matrix elements non-zero.
Similarly residual symmetries can protect some of the matrix elements in \E{2} and \E{6}, but only at 
special locations that are sets of measure zero in the manifolds.  
Other accidental symmetries  can protect certain matrix elements, but  only for parameter values that are sets of measure zero in the parameter space.

These covariance matrices present distinct patterns and are significantly different from the ones obtained in a topologically trivial universe, where all $TT$, $EE$, $BB$, $TE$ blocks are diagonal and the entire $TB$ and $EB$ blocks are zero. 
They are also different in an important way from the correlation matrices due to scalar fluctuations which do not lead to $B$ modes, and thus do not source $TB$, $EB$ or $BB$ correlations.

While the patterns of the correlation matrix are interesting, what matters is whether they can be used to detect cosmic topology.
When considering tensor modes only, we have shown that the correlations between the different $a_{\ell m}^{X}$ ($X\in \{T,E,B\}$) would be sufficient to allow us to distinguish between the covering space and any of these non-trivial topologies if we could measure the tensor modes at high signal to noise.
We have demonstrated this by computing, for each topology, the KL divergences between the $T$, $E$, and $B$ joint  correlation matrices of the covering space and that topology as a function of an appropriate length scale of the topology. 
When the topologies are smaller than the diameter of the LSS, the topological information present in the CMB is very large, as expected. 
More significantly---since there is an existing CMB constraint that  the shortest distance around the Universe through us must exceed $98.5\%$ of the diameter of the LSS---the KL divergence remains much larger than $1$ for sizes somewhat larger than the diameter of the LSS. 
Depending on the topology, the CMB is still distinguishable from the trivial topology (i.e., the KL divergence between them is larger than $1$) up to sizes $1.1$ to $1.2$ times the diameter of the LSS. 

No substantial growth of topological information with $\ell$ has been found in the case of scalar modes, neither in temperature nor polarization, for $\ell\gtrsim 30$ \cite{Fabre:2013wia, COMPACT:2023rkp}. 
Unexpectedly, we have found that, for tensor modes, the KL divergences between topologically non-trivial manifolds and the covering space continue to increase up to whatever maximum $\ell$ ($\leq 90$) that we have calculated (due to computational limitations).  
In future studies, where we will incorporate both scalar and tensor modes, as well as noise, it may prove valuable to extend these calculations to still higher $\ell$.
This should be possible by making use of large memory nodes on appropriate machines, or with careful optimization.
The maximum $\ell_{\mathrm{max}}$  to which we can practically compute remains to be seen.

We foresee at least two important potential applications: first, the CMB polarization may have more information about topology than previously thought, thanks to the tensor modes at smaller scales;  second, it may be easier to measure the tensor modes (and therefore the tensor-to-scalar ratio $r$) in a Universe with a non-trivial topology, thanks to  the additional off-diagonal correlations. 
In future work, we plan to include the effects of both scalar and tensor modes in the full CMB $TEB$ signal in order to precisely quantify the interplay among non-trivial topology, tensor and scalar modes, and noise. 
We note that our work so far has focused on the theoretical limits of detectability and we have therefore considered perfect CMB observations---full-sky and noise-free. 
In the near future, we aim to drop these assumptions and reanalyze the \textit{Planck} data in search of the best-fit Euclidean topology.

\acknowledgments

Y.A. acknowledges support by the Spanish Research Agency (Agencia Estatal de Investigaci\'on)'s grant RYC2020-030193-I/AEI/10.13039/501100011033, by the European Social Fund (Fondo Social Europeo) through the  Ram\'{o}n y Cajal program within the State Plan for Scientific and Technical Research and Innovation (Plan Estatal de Investigaci\'on Cient\'ifica y T\'ecnica y de Innovaci\'on) 2017-2020, by the Spanish Research Agency through the grant IFT Centro de Excelencia Severo Ochoa No CEX2020-001007-S funded by MCIN/AEI/10.13039/501100011033, and by the Spanish National Research Council (CSIC) through the Talent Attraction grant 20225AT025.
C.J.C., D.P.M., G.D.S., and A.K.\ acknowledge partial support from NASA ATP grant RES240737; G.D.S. and Y.A.\ from the Simons Foundation; A.S. and G.D.S.\ from DOE grant DESC0009946; G.D.S., Y.A., and A.H.J.\ from the Royal Society (UK); and A.H.J.\ from STFC in the UK\@.
F.C.G\ is supported by the Presidential Society of STEM Postdoctoral Fellowship at Case Western Reserve University and by Ministerio de Ciencia, Innovaci\'on y Universidades, Spain, through a Beatriz Galindo Junior grant BG23/00061.
J.R.E. acknowledges support from the European Research Council under the Horizon 2020 Research and Innovation Programme (Grant Agreement No.~819478).
A.N.\ is supported by the Richard S.\ Morrison Fellowship.
T.S.P is supported by FAPERJ (grant E26/204.633/2024), CNPq (grant 312869/2021-5) and Funda\c{c}\~ao Arauc\'aria (NAPI de Fen\^omenos Extremos do Universo, grant 347/2024 PD\&I). J.C.D. and A.T. are supported by CSIC through grant No. 20225AT025. A.T.\ was also supported by the Richard S.\ Morrison Fellowship. 
This work made use of the High-Performance Computing Resource in the Core Facility for Advanced Research Computing at Case Western Reserve University and the facilities of the Ohio Supercomputing Center.

\appendix
\section{Non-trivial topology and Laplacian eigenmodes}
\label{app:topology_details}
The topologically non-trivial manifolds we are considering are obtained as a quotient of the covering space manifold of a homogeneous geometry (in particular the FLRW geometries $E^3$, $S^3$, or $H^3$, but in principle also one of the other homogeneous but anisotropic geometries) by a discrete, freely acting subgroup of the isometry group of the covering space. These are exactly the smooth manifolds that preserve the local geometry, i.e., the metric, of the covering space but have a different global topological structure.

The eigenmodes of the Laplacian operator acting in the topologically non-trivial manifold are then precisely those eigenmodes of the covering space that remain invariant under the action of the elements of the isometry group defining the non-trivial manifold. 
This is because the Laplacian operator involves only partial derivatives and the metric tensor, which are both local notions. 
Since locally the manifold with non-trivial topology and the covering space are identical by construction,
a function  will be acted upon by the Laplacian in the same way in both manifolds. 
Thus the only condition needed to ``recycle'' a covering-space eigenmode into a non-trivial manifold eigenmode is to ensure that it satisfies the boundary conditions, or equivalently, that it is invariant under the aforementioned isometries.

The discrete nature of the group quotienting the covering space implies that any three-dimensional (possibly continuous) parametrization of the eigenmode family will be partially or fully discretized. 
The discretization will be complete when the manifold under consideration is fully compact, while in other cases the discretization is only \emph{partial}, meaning that the parametrization of the eigenmode basis will necessarily include one or more continuous parameters alongside any discrete ones. 
In the fully compact case, the eigenspectrum itself will be discrete and the multiplicity of each eigenvalue finite.

In the most general case, when an isometry acts on a covering-space eigenmode the resulting function will be a (possibly infinite or continuous) linear combination of covering-space eigenmodes of the same eigenvalue. 
However, sometimes, for carefully chosen linear combinations of covering space eigenmodes of a given eigenvalue, the action of the isometry on an eigenmode will yield the eigenmode itself times a phase factor. 
If that phase factor is (i.e., can be arranged to be) $1$, then that linear combination of covering-space eigenmodes is  an eigenmode of the Laplacian on both the covering space and the topologically non-trivial manifold. 
There may be more than one such function of a given eigenvalue, but at least this gives us a selection criterion for  the parametrization of the eigenmode basis on the topologically non-trivial manifold. 
Typically, the dimensionality of the eigensubspace of each eigenvalue will be reduced compared to the covering space. 
In the non-compact case this is because it will require fewer continuous parameters to characterize the eigenmodes, while in the compact case the number of eigenmodes of a given eigenvalue will be finite.
In the case of a compact covering space (e.g., $S^3$), the eigensubpaces are already finite, but the multiplicity of eigenmodes in each eigensubspace will typically be reduced in the topologically non-trivial manifold.

Some covering-space eigenmodes will not contribute to any non-trivial-manifold eigenmode.
A simple, and well-known, example is the solution of the scalar Laplacian on a periodic cube of side length $L$ in flat (Euclidean) geometry.
A simple parametrization of the eigenmodes is that they are plane waves 
$\exp(i\vec{k}\cdot\vec{x})$, but only for $\vec{k}=2\pi(n_1,n_2,n_3)/L$ for arbitrary integers $n_i$, 
whereas in the covering space all values of the three-vector $\vec{k}$ would qualify.

\section{Notational conventions}

\subsection{Polarization tensors}
\label{app:polzn}
For the specific case where $\unitvec{k} = \unitvec{z}$ the helicity tensors are given as
\begin{equation}
    \label{HelicityMat}
    \mat{e}(\unitvec{z},\pm2)=\frac{1}{\sqrt{2}}\begin{pmatrix}
         1 & \pm i & 0\\
         \pm i & -1 & 0\\
         0 & 0 & 0
        \end{pmatrix}.
\end{equation}
To generalize these tensors to an arbitrary wavevector direction, one can apply a sequence of rotations
\begin{equation}
    \label{eqn:ArbitraryHelicityTensor}
    \mat{e}(\unitvec{k},\lambda) = \mat{R}_{\unitvec{z}}(\phi) \mat{R}_{\unitvec{y}}(\theta) \mat{e}(\unitvec{z},\lambda) \transpose{\mat{R}}_{\unitvec{y}}(\theta) \transpose{\mat{R}}_{\unitvec{z}}(\phi),
\end{equation}
where $\theta$ and $\phi$ are defined as
\begin{equation}
    \label{theta_phi_definition}
    \cos(\theta) \equiv \frac{k_z}{\sqrt{k^2_x+k^2_y+k^2_z}},\quad \tan(\phi) \equiv \frac{k_y}{k_x}.
\end{equation}
The rotation matrices used in the transformation are explicitly
\begin{align}
    \label{rotationMatrix}
    \mat{R}_{\unitvec{y}}(\theta) = \begin{pmatrix}
     \cos(\theta) & 0 &  \sin(\theta) \\
     0 & 1 & 0\\
     -\sin(\theta) & 0 & \cos(\theta) 
    \end{pmatrix}, 
    \quad \mat{R}_{\unitvec{z}}(\phi) = \begin{pmatrix}
     \cos(\phi) & -\sin(\phi) & 0 \\
     \sin(\phi) & \cos(\phi) & 0 \\
     0 & 0 & 1
    \end{pmatrix}.
\end{align}
For a generic rotation operator $\mathcal{R}$ acting on the helicity tensors we have 
\begin{equation}\label{eq:genericrotation}
\mathcal{R}[\mat{e}(\unitvec{k},\lambda)]=\transpose{\mat{R}}\mat{e}(\unitvec{k},\lambda)\,\mat{R}\,,
\end{equation}
where $\mat{R}$ is the matrix representation of the rotation.

\subsection{Spin-weighted spherical harmonics}
\label{app:2Ylm}
Spin-weighted spherical harmonics \({}_{\pm 2}Y_{\ell m}\) are obtained by applying the spin-raising (\(\eth\)) and spin-lowering (\(\bar{\eth}\)) operators
\begin{align}
 \eth = - (\sin{\theta})^s \left( \frac{\partial}{\partial \theta} + i \csc{\theta} \frac{\partial}{\partial \phi}\right)  (\sin{\theta})^{-s}, \ \ \ \
\bar{\eth} = - (\sin{\theta})^{-s} \left( \frac{\partial}{\partial \theta} - i \csc{\theta} \frac{\partial}{\partial \phi}\right)  (\sin{\theta})^{s},
\end{align}
on the standard (spin-0) spherical harmonics. Here, \(s\) represents the spin of the function to which the operator is applied. When applied to spin-weighted spherical harmonics, these operators yield the following identities:
\begin{align}
\eth {}_sY_{\ell m} = \sqrt{\frac{\ell - s}{\ell +s+1}} \ {}_{s+1}Y_{\ell m}, \ \ \ \
 \bar{\eth} {}_sY_{\ell m} = -\sqrt{\frac{\ell + s}{\ell -s+1}} \ {}_{s-1}Y_{\ell m}\,.
\end{align}
Notably, for spin-0 and spin-2 harmonics
\begin{align}
    {}_2Y_{\ell m} = \sqrt{\frac{(\ell - 2)!}{(\ell + 2)!}} \ \eth\eth Y_{\ell m}, \ \ \ {}_{-2}Y_{\ell m} = \sqrt{\frac{(\ell - 2)!}{(\ell + 2)!}} \ \bar{\eth}\bar{\eth} Y_{\ell m}.
\end{align} 
For generic spin-weighted harmonics
\begin{equation}
    {}_{s}Y^* _{\ell m}=(-1)^{s+m} {}_{-s}Y_{\ell m}.
\end{equation}
Additionaly, we have that
\begin{equation}
    \label{eqn:SHzaxis}
    (-1)^m{}_{s}Y_{\ell -m} (\unitvec{z})=\sqrt{\frac{2\ell+1}{4\pi}}\,\Kdelta_{ms}\,.
\end{equation}
Finally, these harmonics can also be expressed in terms of the reduced Wigner rotation matrices $d$ as \cite{Boyle:2013nka}
\begin{equation}
    {}_{s}Y_{\ell m}(\theta, \phi) = \sqrt{\frac{2\ell+1}{4\pi}} d^{\ell}_{m -s}(\theta) \mathrm{e}^{-im\phi} .
\end{equation}
In fact, for computational purposes it proves to be more efficient to work directly with the Wigner D-matrices.

\bibliographystyle{utphys}
\bibliography{topology, additional}

\end{document}